\documentclass[noinfoline,twoside]{imsart} 

\RequirePackage[OT1]{fontenc}
\RequirePackage{amsthm,amsmath}
\RequirePackage[numbers]{natbib}
\RequirePackage[colorlinks,citecolor=blue,urlcolor=blue]{hyperref}

\usepackage[ruled]{algorithm2e}

\SetAlFnt{\small}
\SetAlCapFnt{\small}
\SetAlCapNameFnt{\small}
\SetAlCapHSkip{0pt}
\IncMargin{-\parindent}

\usepackage[cmex10]{amsmath}
\usepackage{amssymb}
\usepackage{algorithmic}
\usepackage{array}
\usepackage{fixltx2e}
 \usepackage{dblfloatfix}
 \usepackage{url}

\usepackage[]{graphicx}
\usepackage{booktabs}
\usepackage{rotating}
\usepackage{xspace}
\usepackage{multirow}
\usepackage{array}

\usepackage{lineno}

\newcommand {\ADD}[1]{{#1}}

\newcommand {\quot}[1]{``#1''}
\newcommand{\vect}[1]{\mathbf{#1}}

\newcommand{\burstF}{\emph{Burst of Color III}\xspace}
\newcommand{\burst}{Burst\xspace}

\newcommand{\fashionF}{\emph{Fashion II}\xspace}
\newcommand{\fashion}{Fashion\xspace}

\newcommand{\landscapeF}{\emph{Landscape V}\xspace}
\newcommand{\landscape}{Landscape\xspace}

\newcommand{\selfF}{\emph{Self Portrait VII}\xspace}
\newcommand{\self}{Self\xspace}

\newcommand{\zenF}{\emph{Zen Photography III}\xspace}
\newcommand{\zen}{Zen\xspace}

\newcommand{\primaryF}{\emph{Primary Colors II}\xspace}
\newcommand{\primary}{Primary\xspace}
\newcommand{\silhouettesF}{\emph{Silhouettes VI}\xspace}
\newcommand{\silhouettes}{Silhouettes\xspace}
\newcommand{\selfieF}{\emph{Selfie!}\xspace}
\newcommand{\selfie}{Selfie\xspace}
\newcommand{\pixelsF}{\emph{160 Pixels}\xspace}
\newcommand{\pixels}{Pixels\xspace}
\newcommand{\redF}{\emph{Red V}\xspace}
\newcommand{\red}{Red\xspace}
\newcommand{\shallowF}{\emph{Shallow DOF VI}\xspace}
\newcommand{\shallow}{Shallow\xspace}

\newcommand{\dummytab}{\parbox{18pt}{\vspace{18pt}}}

\newcommand{\fparamsfull}  {\mathbf{S};\boldsymbol{\mathcal{I}},\boldsymbol{\mathcal{M}},\mathcal{C}}
\newcommand{\fparams}  {\cdot}
\newcommand{\fparamssfull} {\mathbf{S};\boldsymbol{\mathcal{I}},\boldsymbol{\mathcal{M'}},\boldsymbol{\mathcal{F}},\mathcal{C}}
\newcommand{\fparamss} {\cdot}

\newcommand{\fone}{sc}
\newcommand{\imaps}{\{sal, qua, har\}}
\newcommand{\dsets}{\{\burst,\fashion,\landscape,\self,\zen\}}

\newcommand{\pirlingurl}{\protect\url{http://www.ivl.disco.unimib.it/research/collage/}}

\graphicspath{{./images2/}}

\newcolumntype{C}[1]{>{\centering\arraybackslash}p{#1}}

\def \collagewidth {1.0in}
\def \collagewidthhh {0.90in}
\def \collagewidthhhh {0.83in}

\def \imagesizee {28pt}
\def\std{\mathop{std}}

\def \tbl {}

\begin{document}

\begin{frontmatter}
\title{User Preferences Modeling and Learning for Pleasing Photo Collage Generation}
\runtitle{User Preferences Modeling and Learning}

\begin{aug}
\author{\fnms{Simone} \snm{Bianco}\ead[label=e1]{bianco@disco.unimib.it}}
\and
\author{\fnms{Gianluigi} \snm{Ciocca}\ead[label=e2]{ciocca@disco.unimib.it}}

\address{Dipartimento di Informatica Sistemistica e Comunicazione\\
University of Milano-Bicocca\\
Viale Sarca 336, 20126, Milano, Italy\\
\printead{e1,e2}}

\runauthor{S. Bianco and G. Ciocca}
\end{aug}

\begin{abstract}
In this paper we consider how to automatically create pleasing photo collages created by placing a set of images on a limited canvas area. The task is formulated as an optimization problem. Differently from existing state-of-the-art approaches, we here exploit subjective experiments to model and learn pleasantness from user preferences. To this end, we design an experimental framework for the identification of the criteria that need to be taken into account to generate a pleasing photo collage. Five different thematic photo datasets are used to create collages using state-of-the-art criteria. A first subjective experiment where several subjects evaluated the collages, emphasizes that different criteria are involved in the subjective definition of pleasantness. We then identify new global and local criteria and design algorithms to quantify them. The relative importance of these criteria are automatically learned by exploiting the user preferences, and new collages are generated. \ADD{To validate our framework, we performed several psycho-visual experiments involving different users. The results shows that the proposed framework allows to learn a novel computational model which effectively encodes an inter-user definition of pleasantness. The learned definition of pleasantness generalizes well to new photo datasets of different themes and sizes not used in the learning. Moreover, compared with two state of the art approaches, the collages created using our framework are preferred by the majority of the users.}
\end{abstract}


\end{frontmatter}

\markboth{S. Bianco and G. Ciocca}{User Preferences Modeling and Learning for Pleasing Photo Collage Generation}

\maketitle

\section{Introduction}

\label{sec:intro}

\begin{figure}[!tb]
\centering
\includegraphics[width=\textwidth]{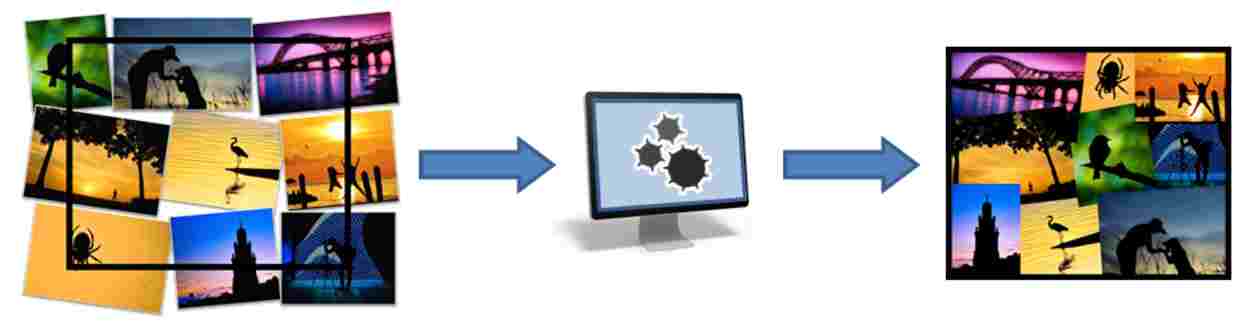}%
\caption{Example photo collage created by placing a number of photo images on a canvas area of limited size. }\label{fig:esempio}%
\end{figure}

Photo collages are created by placing a number of photo images on a canvas area of limited size. They are used to visually represent in an appealing and compact way events of interest. The images can be fitted on the canvas by simply scaling them at the risk of losing important details contained in them and making the collage dull. In this paper we consider the problem of how to automatically create pleasing photo collages: given a set of photo images and a canvas area, we want to arrange the photos on the canvas in a pleasant unsupervised manner and without scaling them (see Figure~\ref{fig:esempio}). Assuming that the size of the canvas area is smaller than the sum of the sizes of the photos to be displayed, two main issues arise. The first is that photos may occlude themselves, the second is that photos may be partially outside the canvas area. These issues must be addressed by taking into account the pleasantness of the resulting collage that is influenced by the order with which the photos are placed on the canvas and their spatial arrangement. Usually, the most important photos are placed at the top of the less important ones in order to minimize the risk of being severely occluded, and composition properties related to photo contents, geometric constraints and aesthetic consideration are taken into account to maximize the pleasantness of the resulting collage. The criteria that define what is important in a photo and what composition properties should be satisfied may vary from user to user. Moreover the single criteria may compete against each other. To be used in an automatic system for photo collage generation, the pleasantness criteria and their relative importance must be properly quantified using suitable algorithms. At the end of this process a fitness function can be defined whose value represents the overall degree of pleasantness of a photo collage. To obtain the most pleasant collage, an automatic algorithm must search the best arrangement of the photos by maximizing the value of the fitness function. For this purpose an optimization algorithm is usually exploited. Several formulations of some of the above criteria have been proposed in the literature but none of the existing works performed an user study in order to actually determine what are the criteria that made a photo important, what constraints must be satisfied in order to have a collage balanced, or what hints users pay attention to in judging the pleasantness of a photo collage. We argue that if we could elicit the criteria by modeling the preferences of the users, we would be able to create more pleasant photo collages.

\subsection{Related Work}
\label{subsec:related}

Previous works on photo collage can be categorized into two main groups depending on the processing applied to the photos. These two groups are: \emph{content-preserving} and \emph{non content-preserving}. 

In the non content-preserving group belong the photo collage methods that select relevant regions within the photos in order to maximize the information that is conveyed in the final collage. The methods ensure that these regions are made visible in the final collage while the less relevant regions are either removed (cropping) or hidden by other, more relevant, ones (hiding). In addition to scaling and translation, these methods usually perform a layering of the photos to decide the order with which they are positioned on the canvas and/or rotate the photos to further preserve their content as much as possible.

Among the methods that also apply a rotation operation on the photos, we find Picture Collage \cite{Wang2006,liu2009picture} that is the one of the first works that formalized the problem of photo collage as an optimization problem using different, competing collage criteria, namely image saliency, blank space, and saliency raio balance. Inspired by this work is the collage strategy proposed in \cite{Battiato2008} which uses the same criteria but images are firstly classified into three categories and then different relevant region detection strategies are adopted on the basis of the image category. Also inspired by the work of \cite{Wang2006} are the improved collage strategies proposed by \cite{Wei2009} and \cite{Yang2009} where the collages can be also interactively modified by the user. A recent photo collage approach \cite{Yu2014} uses a heuristic search process to ensure that salient information of each photo is displayed in the polygonal area resulting from a power-diagram-based circle packing algorithm. Most of the previous approaches use a saliency map, solely or coupled with other descriptors, as informativeness criteria. In \cite{Ekhtiyar2011} instead, the informativeness criteria corresponds to foreground objects detected on depth maps. Finally, differently from all the aforementioned approaches, the method proposed in \cite{Huang2011} creates Arcimboldo-like collages with multiple thematically-related cutouts from filtered Internet images.

The stained glass-like photo collage by \cite{Girgensohn2004} is one of the methods that preserve the photo orientation without rotating them. The photos are cropped with respect to the contained face regions. These cropped regions have straight edges that are used to arrange the photos on the canvas. Digital Tapestry \cite{Rother2005} subdivides the photo into a set of sub-blocks and from them, the relevant regions of the photo are reconstructed and merged together. A pixel-based variant of this approach, named AutoCollage, is described in \cite{Rother2006}. Here the relevant regions, with variable shapes are merged with a seamless blending that ensures that no sharp boundaries between them are formed in the final collage. A similar approach is the Mobile Photo Collage presented in \cite{Lee2010}. The Puzzle-Like collage \cite{Goferman2010} instead, cuts out from each photo an irregular shaped region which follows the area surrounding a relevant object within the image. Finally, we can cite the Dynamic Media Assemblage  \cite{Luo2013}, a photo collage approach that can be used to summarize video content as well as a photo collection in a stained glass-like collage.

In the content-preserving group belong those methods that arrange the photos according to the relevance of their content defined in some way. The only operations performed on them are scaling and translation. Usually the most relevant photos are scaled bigger than the less relevant ones, and they are positioned on the most salient regions of the canvas. Moreover the aspect ratio of the photos is preserved. These methods are also referred as photo layout methods. 

An example is the work of \cite{Chen2006} where the photo layout is constructed using a larger topic photo and several small-size supportive photos. The photos are selected and sized according to their temporal and content coherence. A similar approach is exploited in \cite{Calic2007} on video sequences where key-frames in a visual summary are arranged on the canvas using the narrative grammar of comics. Also within this group we can cite the work of \cite{Chao2010} where exclusion zones are used to layout a set of photos on a canvas using different spatial criteria. This method was further improved in \cite{fan2012}. In \cite{Sandhaus2011} spatial criteria are coupled with aesthetic principles to layout photos in a pleasant composition. Recently, taking advantage of information usually found in social networking, and building on the previous PicWall work \cite{Wu2013}, FriendWall (\cite{Wu2014}) uses social attributes (intrinsic labels) to create photo collage employing both image visual features and associated Metadata. As a final example, we can cite the interactive approach \cite{Diakopoulos2005} where pre-designed layout templates of annotated cells are used to arrange the photos according to their metadata, and focus area can be selected by the user.

\subsection{Paper Contribution and Organization}
\label{subsec:contrib}

\begin{figure}[!tb]%
\centering
\includegraphics[width=3.2in]{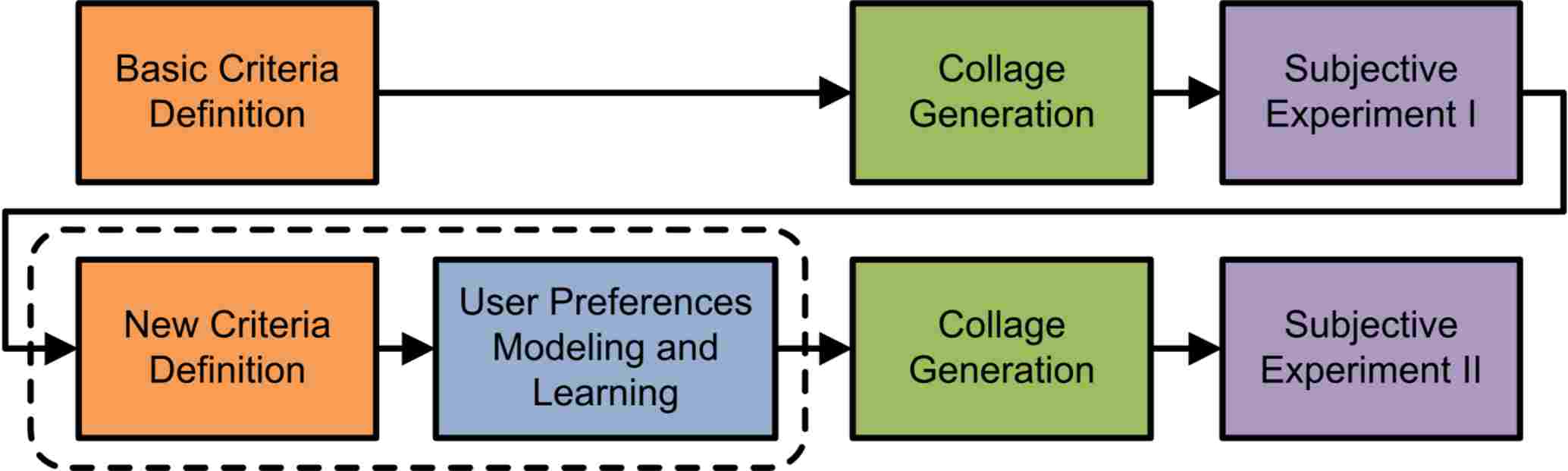}%
\caption{Experimental framework.}%
\label{fig:expflow}%
\end{figure}

The focus of this work is to exploit subjective experiments to model user preferences in order to learn what criteria (and to what extent) need to be taken into account to automatically generate a pleasing photo collage.  To this end we designed an experimental framework that incorporates the identification of the criteria via user preference modeling, the implementation of the corresponding computational algorithms, the learning of their relative importance, and the validation of the results. 
We applied our framework in the context of non content-preserving collages. We believe that this category permits to investigate more criteria underlying the definition of pleasantness as the associated problem  has more degrees of freedom than the one associated to content-preserving collages. However, our proposed framework can be adapted to these methods as well. 
The different steps of our framework are depicted in Figure \ref{fig:expflow}. A first subjective experiment is conducted to investigate how different criteria are involved in the user subjective definition of pleasantness. For this experiment, we redefine the three basic criteria (image informativeness, canvas area coverage, and information ratio balance) exploited in most of the works in the state-of-the-art (e.g. \cite{Wang2006,Battiato2008}). We evaluate three different representations of image informativeness: the first one, which is usually used in the state-of-the-art, is based on saliency; the other two are based on quality and color harmony respectively, and are here introduced. Collages are created by exploiting a Direct Search optimization algorithm. Since user image collections are of very different contents, and different contents may lead to different pleasantness criteria, we considered five thematic image datasets. The results obtained from this experiment are used to identify new criteria both at global and local level. The new global criteria are: face ratio, axis alignment, centrality, and  convexity; the new local criteria are: color similarity, orientation diversity and minimum orientation difference. After having developed algorithms to compute these new criteria, their relative importance is learned by exploiting user rankings on the previously created collages. The identified criteria and their learned importance are then used to generate new sets of collages that are evaluated by a new panel of users. 
\ADD{To further validate the proposed framework, we performed three additional experiments. In order to verify if the identified criteria and their learned relative importance generalize well, that is, if they can be used to create collage on unseen image sets, we performed a subjective experiments on six other image collections of different contents with respect to the ones used in the previous experiments. We also tested the generalizability of the learned definition of pleasantness by creating collages varying the number of images in the set and the canvas size. Moreover, we compared the performance of our proposal against two state-of-the-art algorithms.} 
To the best of our knowledge this is the first work which extensively exploits subjective experiments within the collage generation process to learn user preferences, and that uses datasets of images of different contents to validate the proposed approach.

The rest of the paper is organized as follows. The problem formulation is mathematically described in Section \ref{sec:problem} along with the description of the basic criteria. Section \ref{sec:generation} illustrates the collage generation by describing the three different importance maps considered in our experiments, the photo datasets used, and the optimization algorithm responsible for the collage creation. The first subjective experiment and its outcomes are described in Section \ref{sec:exp1}. The set of the new criteria derived from the first experiment is described in Section \ref{sec:newcriteria}, while  the user preferences modeling and learning strategy is detailed in Section \ref{sec:modeling}. Results of the second subjective experiment performed on the newly created collages are illustrated in Section \ref{sec:exp2}. \ADD{The generalizability of the learned definition of pleasantness and the comparison with state of the art methods on new datasets are reported in Section \ref{sec:further}.} Finally Section \ref{sec:conclusion}, concludes the paper.

\section{Problem Formulation and Basic Criteria Definition}
\label{sec:problem}
Given $N$ input photo images $\vect{I}=\{I_i\}_{i=1}^{N}$ and their corresponding importance maps $\vect{M}=\{M_i\}_{i=1}^{N}$ (importance map representations will be discussed in the next section), a photo collage algorithm must arrange all the images on a canvas area $\mathcal{C}$. In a photo collage, each image $I_i$ is characterized by its state $\vect{s_i}=(\vect{t_i},\theta_i,l_i)$, where $\vect{t_i}=(t_{i,x},t_{i,y})$ is the 2D translation vector (w.r.t. the canvas origin),  $\theta_i$ is the orientation angle (w.r.t. the x-axis), and $l_i$ is the layering index used to determine the placement order of the image. The state is used in a roto-translation transformation $T(\cdot,\vect{s_i})$ to position the image (and its importance map) on the canvas area: 
\begin{equation}
\mathcal{I}_i=T(I_i,\vect{s_i}) \hspace{0.5in} \mathcal{M}_i=T(M_i,\vect{s_i})
\end{equation}

\begin{figure}[!tb]%
\centering
\includegraphics[width=\columnwidth]{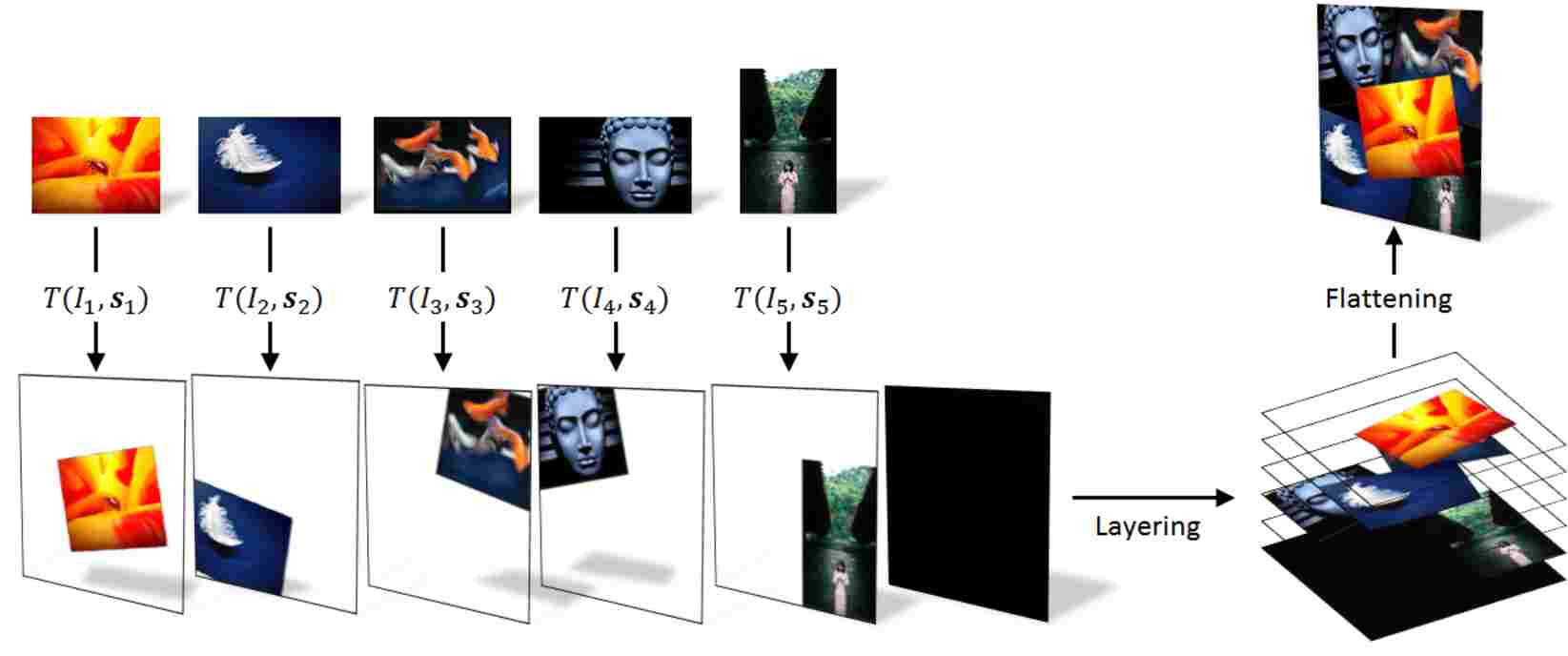}%
\caption{Photo collage layering and compositing}%
\label{fig:creazioneMosaico}%
\end{figure}

The layering indexes can be manually or automatically assigned according to some heuristics. We compute the layering indexes $l_i$ on the basis of the 2D integrals of the importance maps $M_i$: images with higher importance maps are placed on top layers, while images with lower importance maps are placed on bottom layers.
An example of the procedure used for photo collage layering and compositing is reported in Figure~ \ref{fig:creazioneMosaico}.

The picture collage creation is formulated as an optimization problem in order to find the best configuration of states 
$\vect{S}=~\{\vect{s_i}\}_{i=1}^N$ which optimizes all the criteria considered. 

\subsection{Basic Criteria Definition}
\label{subsec:basiccriteria}

\begin{table}[!tb]
\tbl{Basic criteria used in most photo collage formulations\label{tab:cues_base}}{
\begin{tabular}{p{1.5in}p{3in}c}
\toprule 
Criterion & Description & Function \\
\midrule
\ADD{Visibility} & Visible image content (based on importance map) & $C_{1}$ \\
Canvas coverage & Canvas area covered by the photos & $C_{2}$ \\
\ADD{Visibility ratio balance} & Visible image region w.r.t. image size  & $C_{3}$ \\
\bottomrule
\end{tabular}
}
\end{table}

Most of the existing photo collage methods (e.g. \cite{Wang2006}) exploit the three \quot{\emph{basic criteria}} listed in Table \ref{tab:cues_base}. These criteria are quantified by the functions $C_i(\fparamsfull)$. The functions are parametrized by the configuration of states $\mathbf{S}$, and take as data  the set of transformed images $\boldsymbol{\mathcal{I}}$, the set of transformed importance maps $\boldsymbol{\mathcal{M}}$, and the canvas $\mathcal{C}$. 
In the following we write the functions $C_i(\fparamsfull)$  as $C_i(\fparams)$ dropping the dependencies for a more compact notation.

\vspace{0.5truecm}
\noindent
{\ADD{\bf{Visibility}}} The overall collage \ADD{visibility} 
is the average of all information ratios (based on an importance map) computed on the visible regions of the images:

\begin{equation}
C_1(\fparams)=\frac{1}{N}\sum_{i=1}^{N} \frac{sum2(vis(\mathcal{M}_i))}{sum2(M_i)}
\label{eq:c1}
\end{equation}

\noindent where $vis(\cdot)$ is a function that computes the visible parts (taking into account clipping and overlapping) of the given map, and $sum2(\cdot)$ is a function that computes the 2D integrals of the map.

\vspace{0.5truecm}
\noindent
{\bf{Canvas coverage}} The canvas coverage is defined as the ratio of canvas area covered by the arranged photos:

\begin{equation}
C_2(\fparams)=\frac{1}{area(\mathcal{C})}\sum_{i=1}^{N}area(vis(\mathcal{M}_i))
\label{eq:c2}
\end{equation}

\noindent where $area(\cdot)$ is a function that computes the area corresponding to the given input.
 
\vspace{0.5truecm}
\noindent
{\ADD{\bf{Visibility ratio balance}}} The \ADD{visibility} 
ratio balance is computed as the standard deviation of the information ratios:

\begin{equation}
C_3(\fparams)=1-\std_{i=1\dots N} \left\lbrace\frac{sum2(vis(\mathcal{M}_i))}{sum2(M_i)}\right\rbrace
\label{eq:c3}
\end{equation}

\noindent where $std\{\cdot\}$ computes the standard deviation of the given values. 

\vspace{0.5truecm}
The values obtained are combined into a fitness function $f$ that must be maximized:

\begin{equation}
f(\fparamsfull)=\sum_{i=1}^{3}\lambda_iC_i(\fparamsfull)
\label{eq:fitness1}
\end{equation}

\noindent with $\lambda_i$, $i=1,\dots,3$, a weight used to define the contribution of the $i$-th criterion (usually fixed to $1$). This fitness function is at the basis of most of the photo collage algorithms in the state-of-the-art. 

\section{Collage Generation}
\label{sec:generation}
In the following subsections, assuming that a proper dataset of images is available, we describe three different approaches to compute the image importance map: the first approach is inspired by \cite{Ma2003}; the other two are here introduced. We also describe the algorithm used to place the images on the canvas area by searching the best configuration of states. The algorithm optimizes the fitness function defined in Equation~ \ref{eq:fitness1}.

\subsection{Photo Datasets}
\label{subsec:datasets}

\begin{figure*}[!Htbp]
\tabcolsep=1pt
\centering
\small
\resizebox{\columnwidth}{!}{
\begin{tabular}{cccccccccccccc}
\includegraphics[height=\imagesizee]{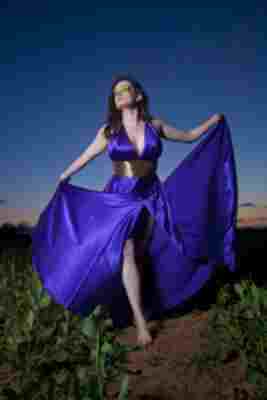} &
\includegraphics[width=\imagesizee]{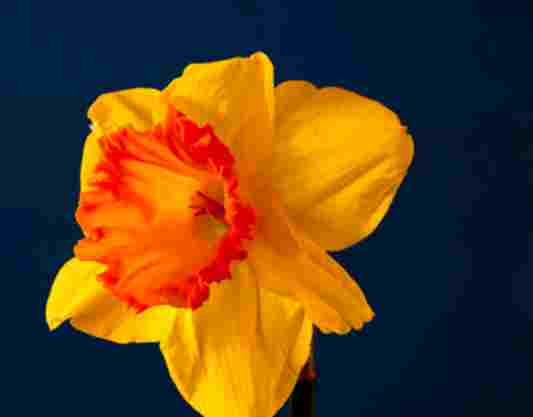} &
\includegraphics[height=\imagesizee]{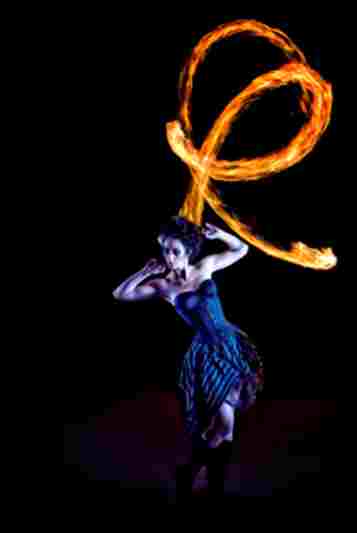} &
\includegraphics[width=\imagesizee]{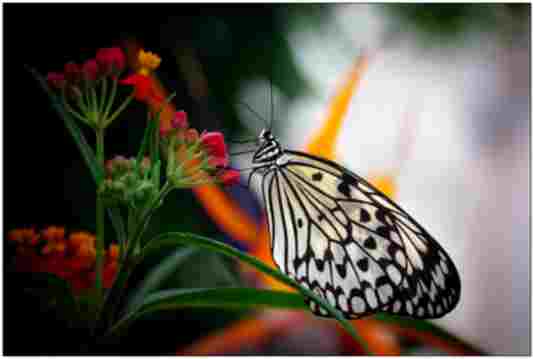} &
\includegraphics[height=\imagesizee]{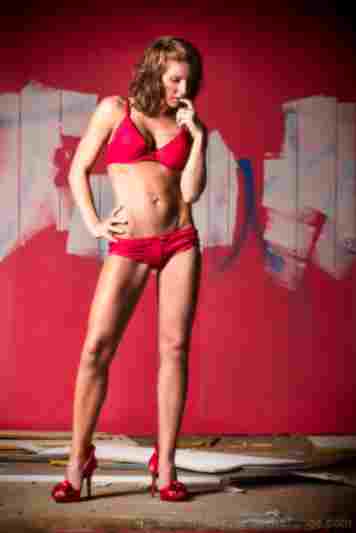} &
\includegraphics[height=\imagesizee]{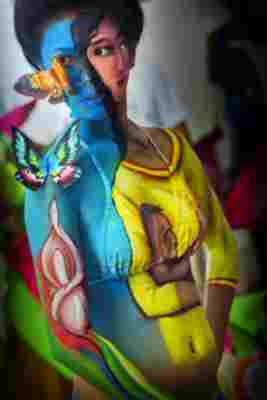} &
\includegraphics[width=\imagesizee]{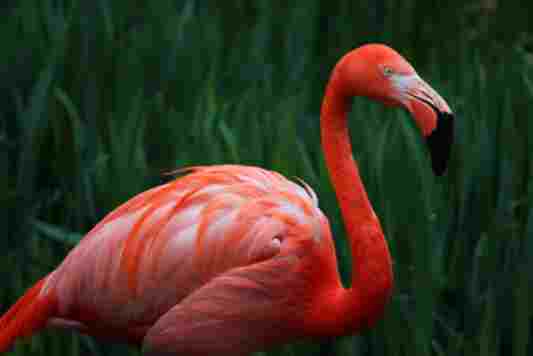} &
\includegraphics[height=\imagesizee]{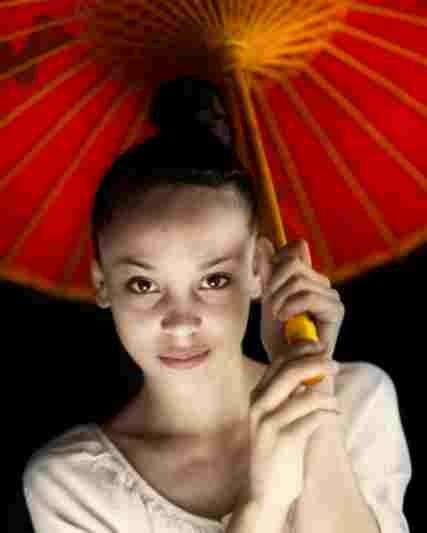} &
\includegraphics[width=\imagesizee]{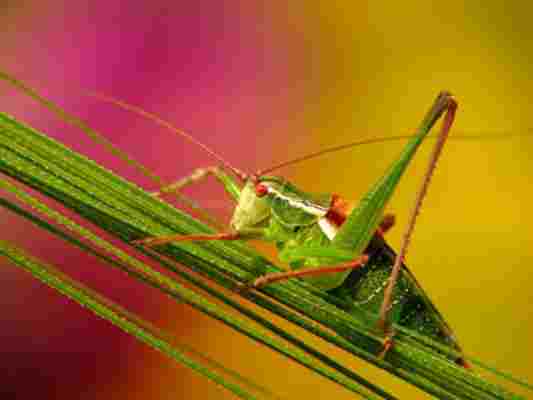} &
\includegraphics[height=\imagesizee]{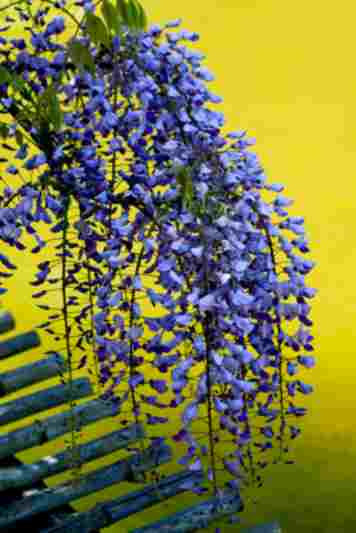} &
\includegraphics[width=\imagesizee]{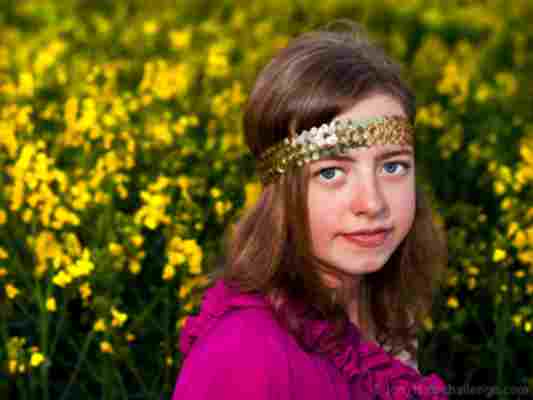} &
\includegraphics[height=\imagesizee]{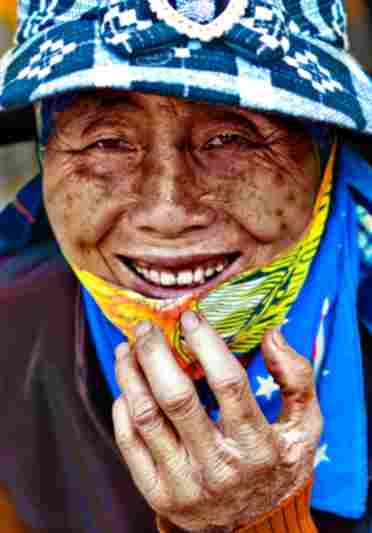} &
\includegraphics[height=\imagesizee]{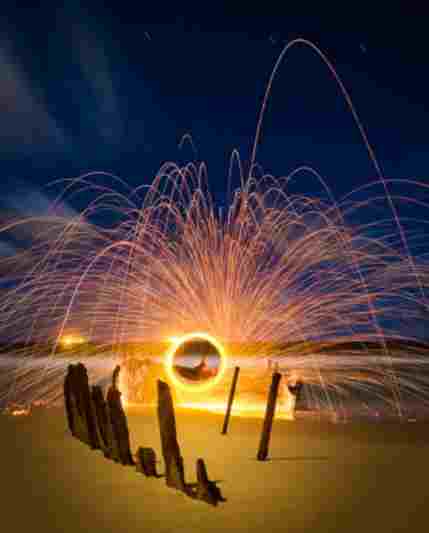} &
\includegraphics[height=\imagesizee]{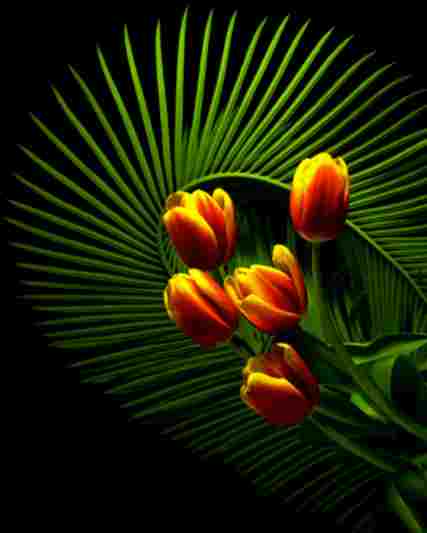} \\
\includegraphics[height=\imagesizee]{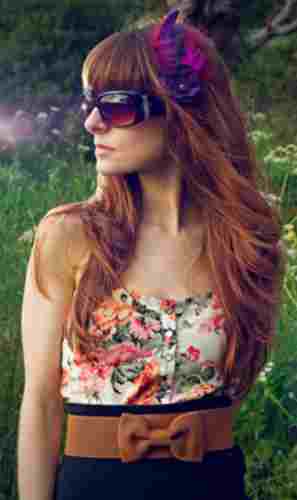} &
\includegraphics[height=\imagesizee]{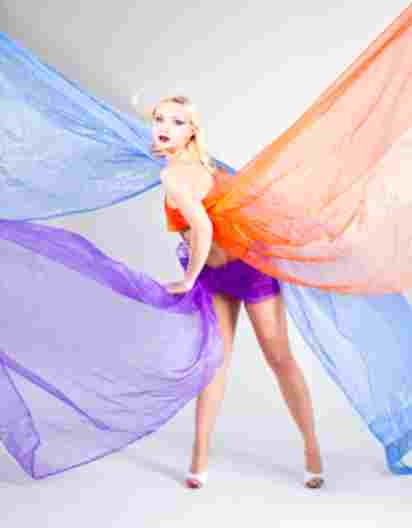} &
\includegraphics[width=\imagesizee]{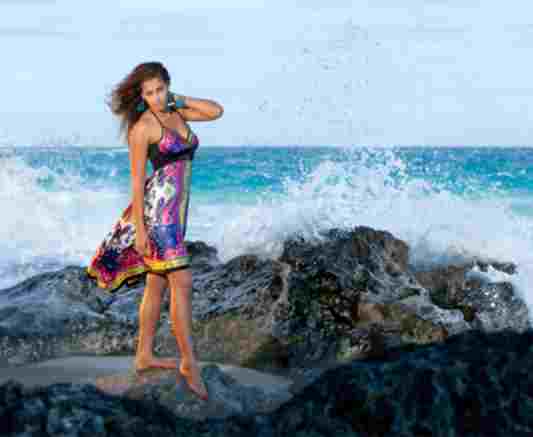} &
\includegraphics[width=\imagesizee]{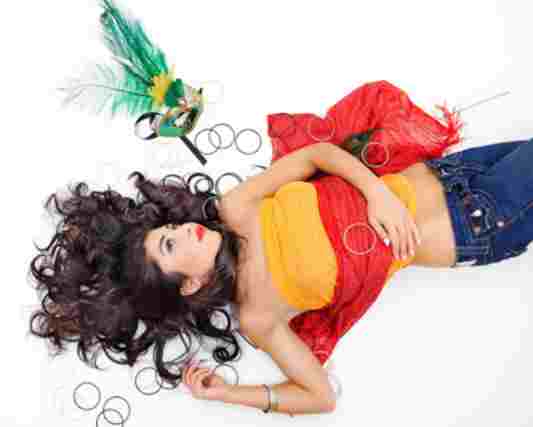} &
\includegraphics[width=\imagesizee]{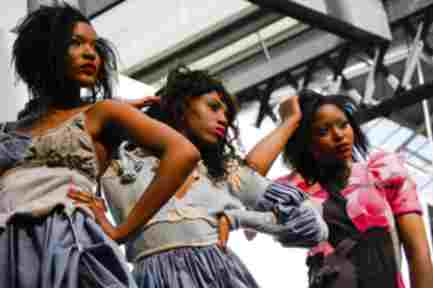} &
\includegraphics[width=\imagesizee]{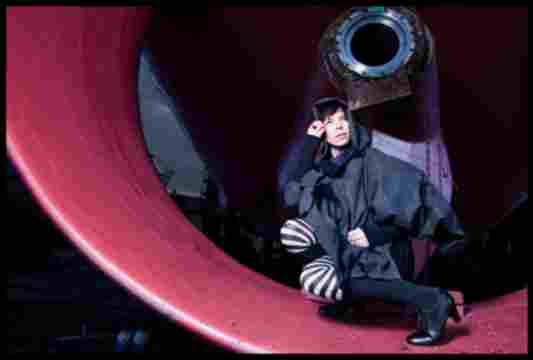} &
\includegraphics[width=\imagesizee]{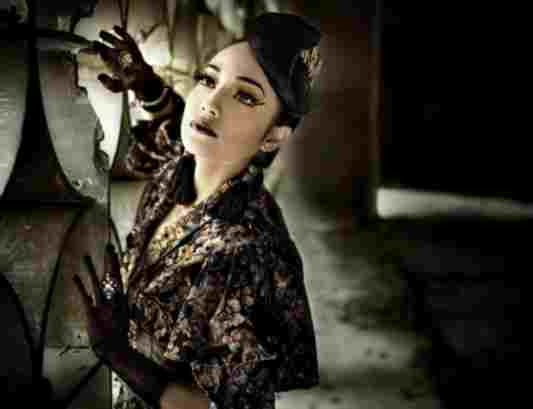} &
\includegraphics[height=\imagesizee]{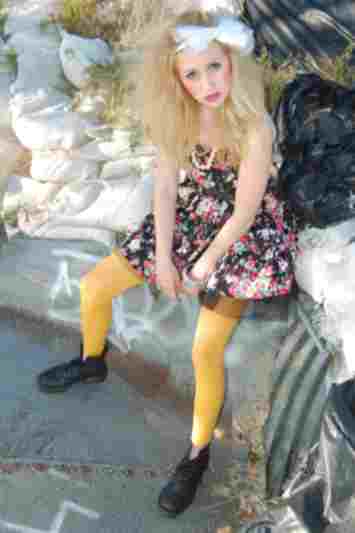} &
\includegraphics[height=\imagesizee]{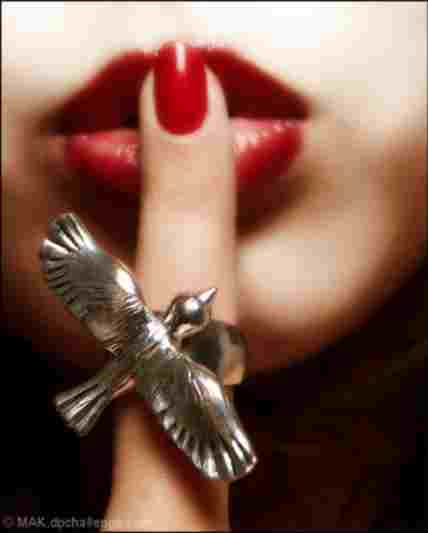} &
\includegraphics[width=\imagesizee]{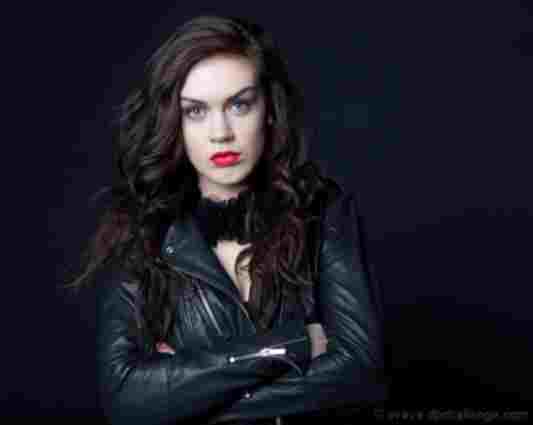} &
\includegraphics[width=\imagesizee]{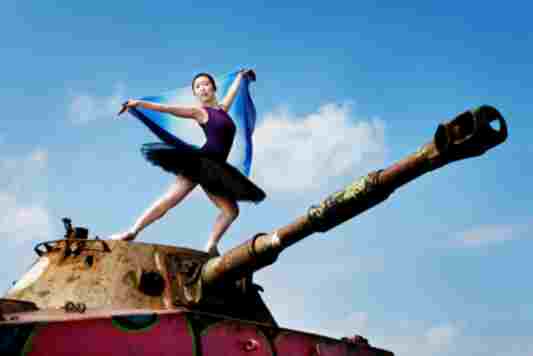} &
\includegraphics[width=\imagesizee]{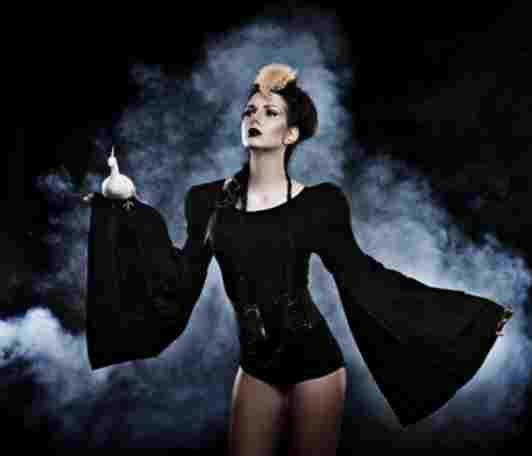} &
\includegraphics[width=\imagesizee]{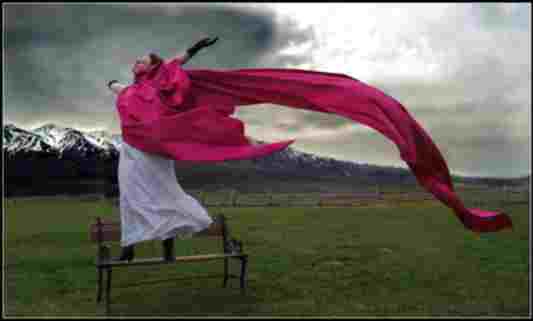} &
\includegraphics[height=\imagesizee]{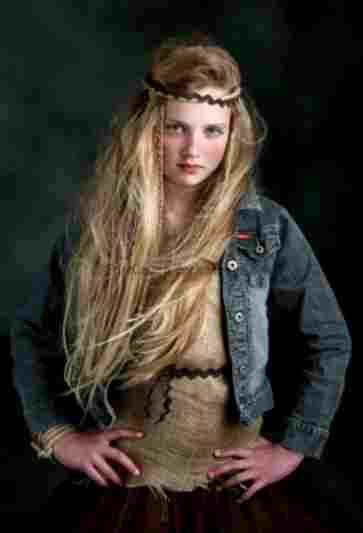} \\
\includegraphics[height=\imagesizee]{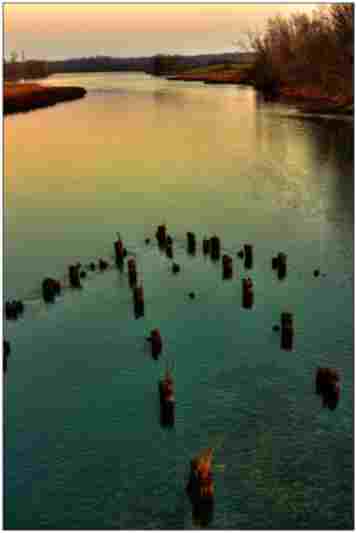} &
\includegraphics[width=\imagesizee]{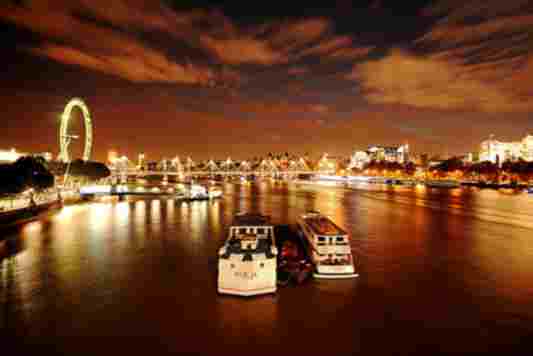} &
\includegraphics[width=\imagesizee]{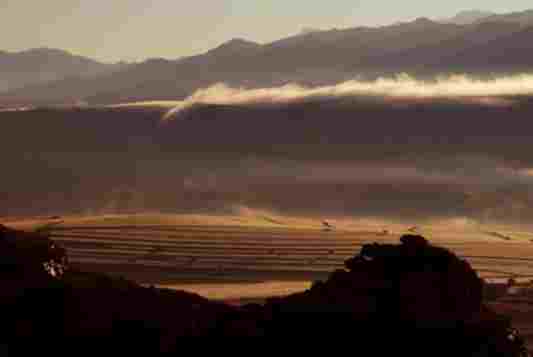} &
\includegraphics[width=\imagesizee]{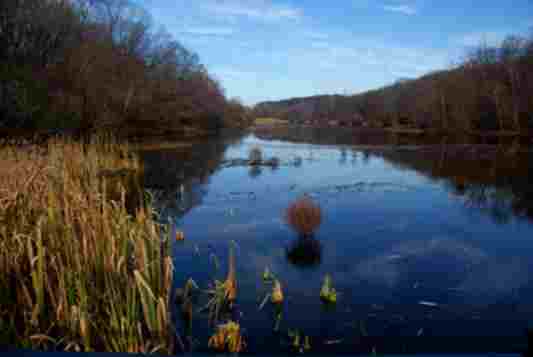} &
\includegraphics[width=\imagesizee]{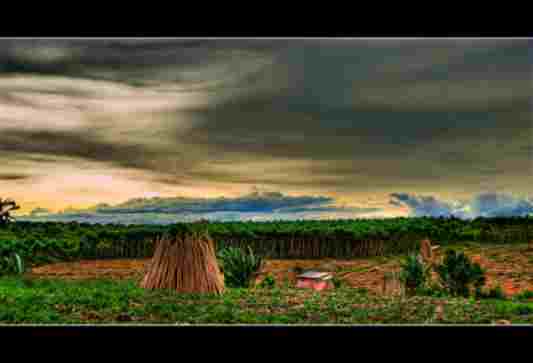} &
\includegraphics[width=\imagesizee]{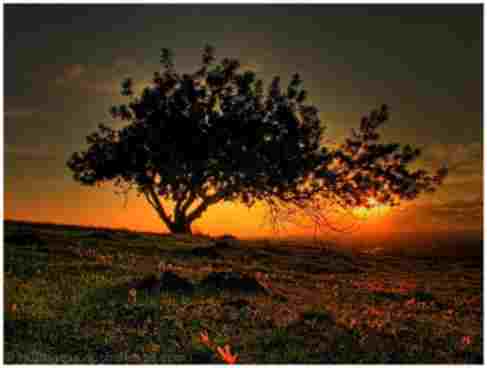} &
\includegraphics[height=\imagesizee]{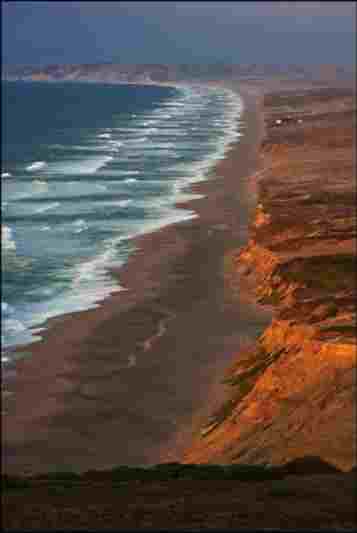} &
\includegraphics[height=\imagesizee]{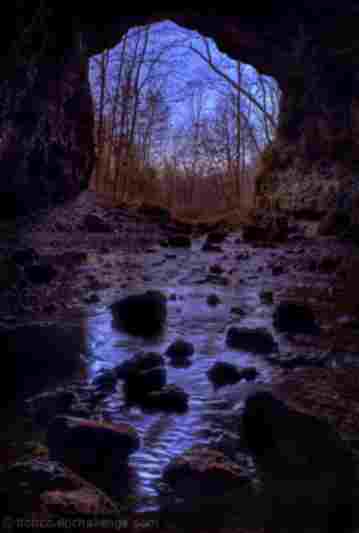} &
\includegraphics[width=\imagesizee]{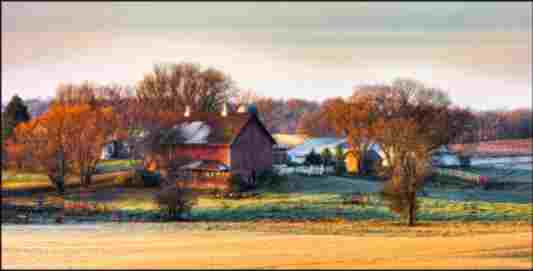} &
\includegraphics[width=\imagesizee]{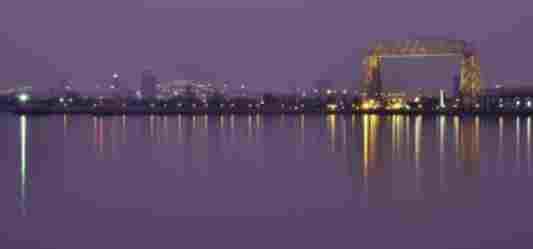} &
\includegraphics[width=\imagesizee]{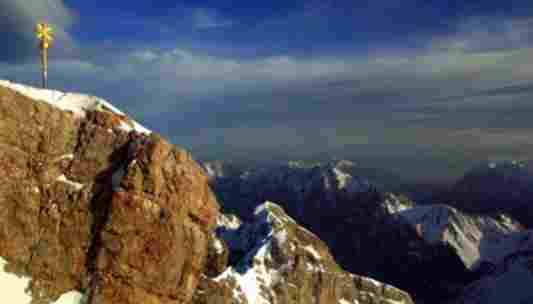} &
\includegraphics[width=\imagesizee]{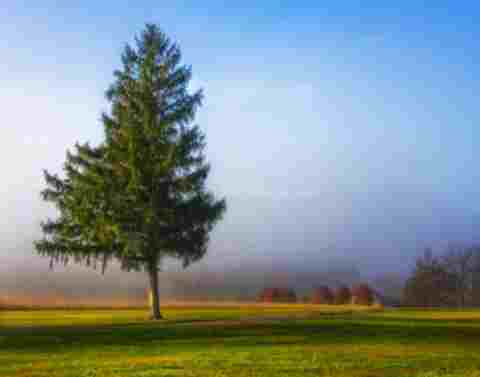} &
\includegraphics[width=\imagesizee]{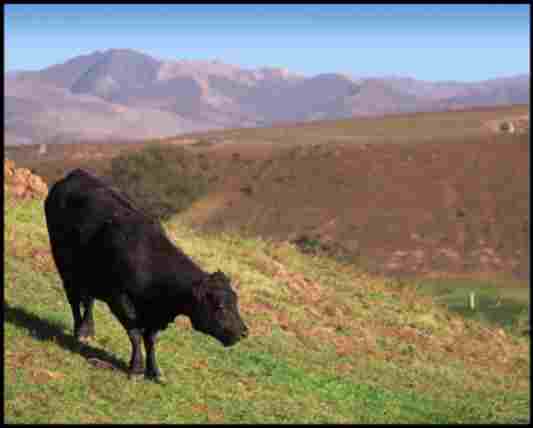} & 
\includegraphics[height=\imagesizee]{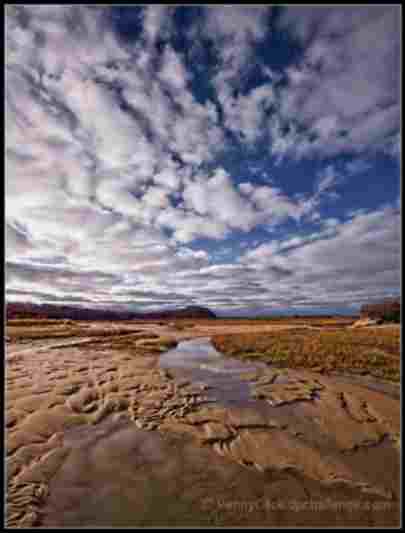} \\
\includegraphics[height=\imagesizee]{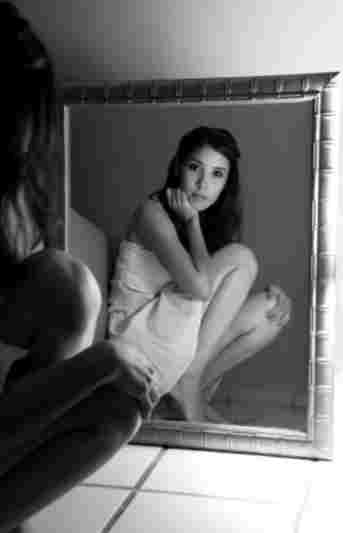} &
\includegraphics[width=\imagesizee]{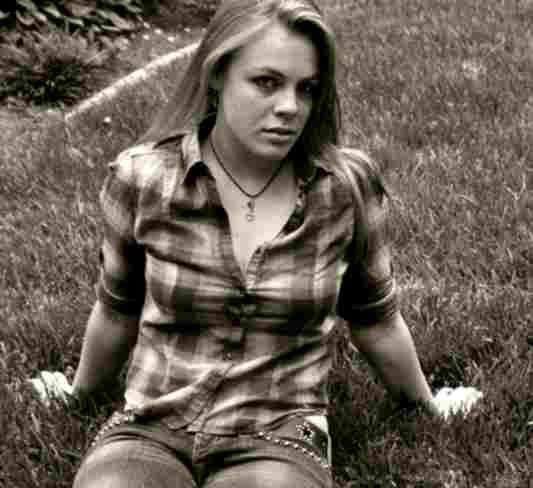} &
\includegraphics[height=\imagesizee]{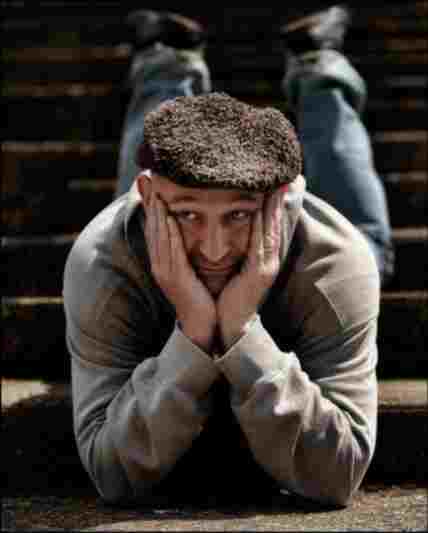} &
\includegraphics[height=\imagesizee]{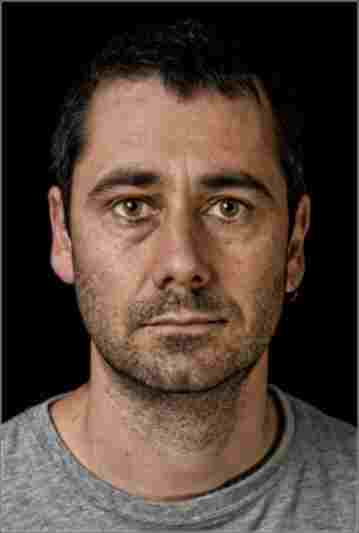} &
\includegraphics[height=\imagesizee]{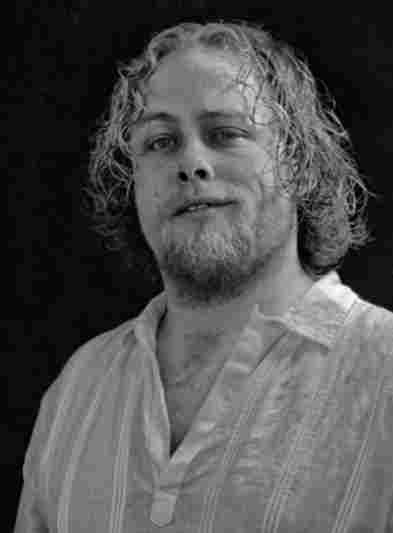} &
\includegraphics[width=\imagesizee]{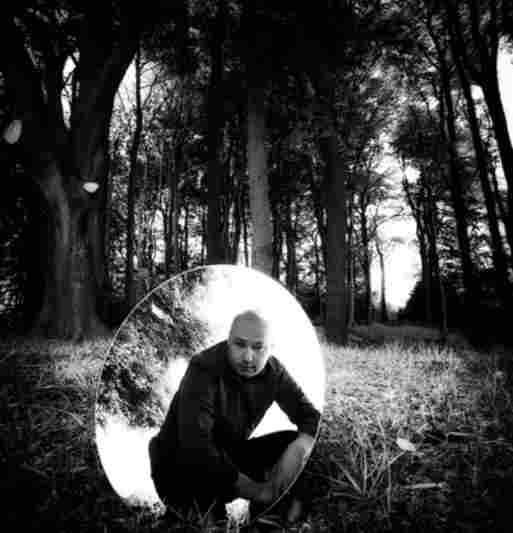} &
\includegraphics[width=\imagesizee]{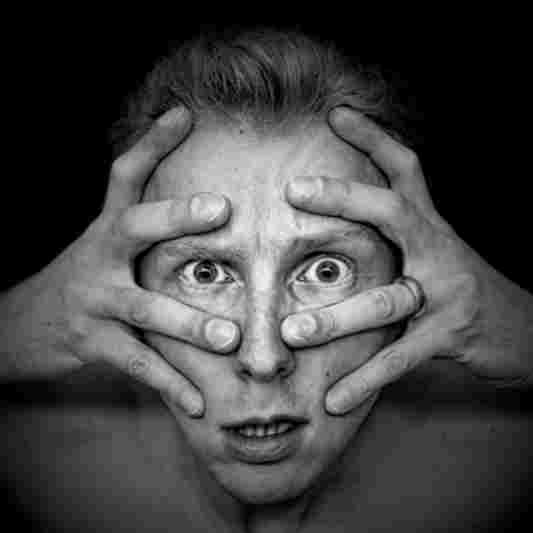} &
\includegraphics[height=\imagesizee]{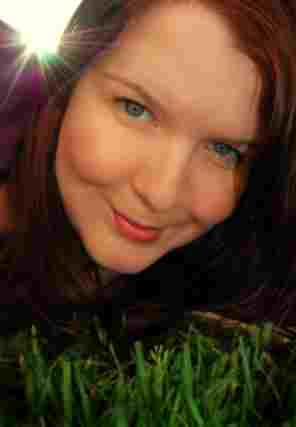} &
\includegraphics[width=\imagesizee]{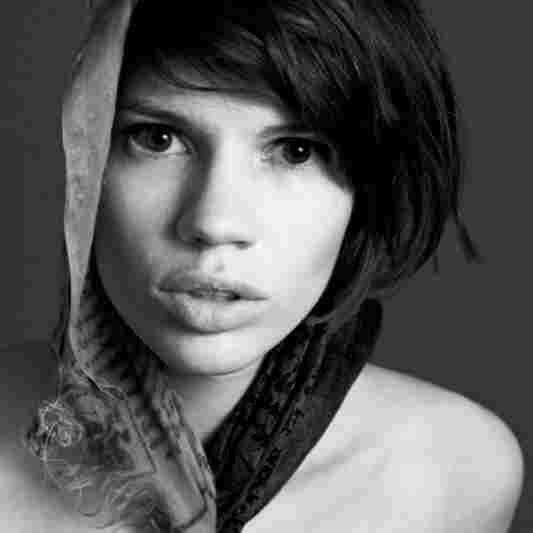} &
\includegraphics[width=\imagesizee]{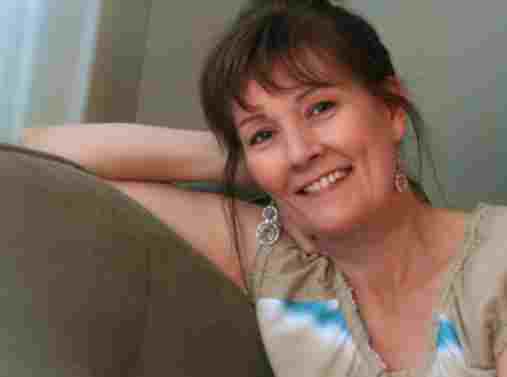} &
\includegraphics[width=\imagesizee]{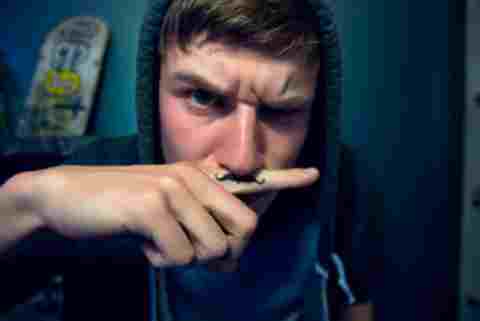} &
\includegraphics[height=\imagesizee]{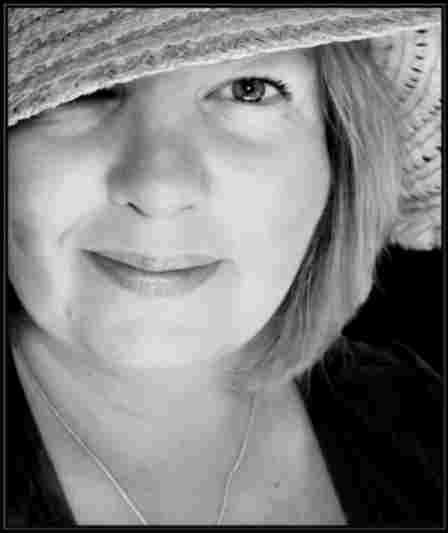} &
\includegraphics[height=\imagesizee]{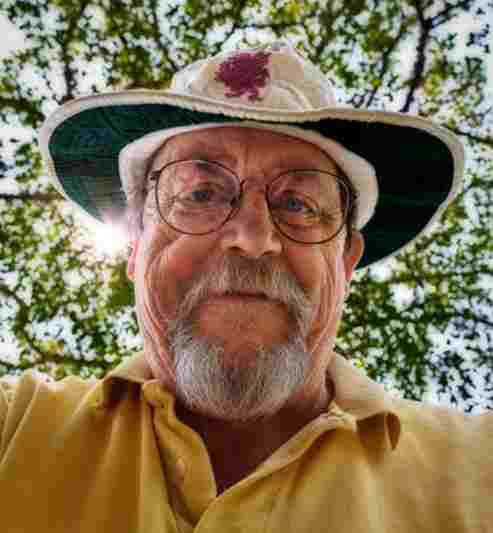} &
\includegraphics[width=\imagesizee]{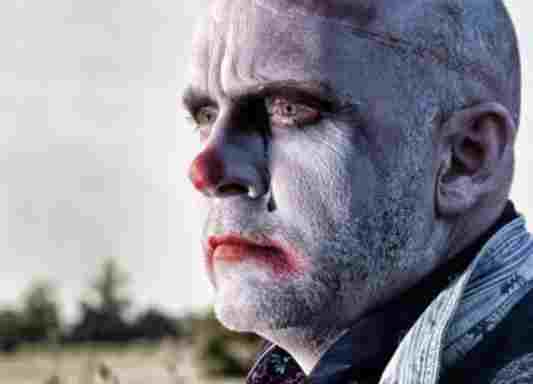} \\
\includegraphics[height=\imagesizee]{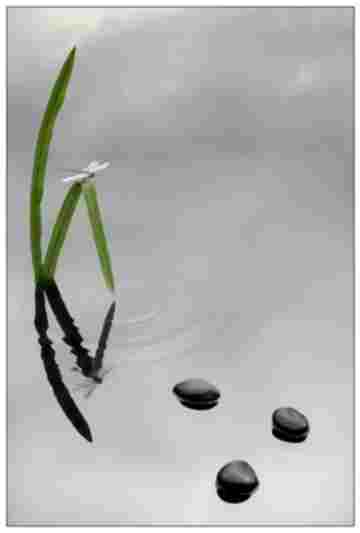} &
\includegraphics[width=\imagesizee]{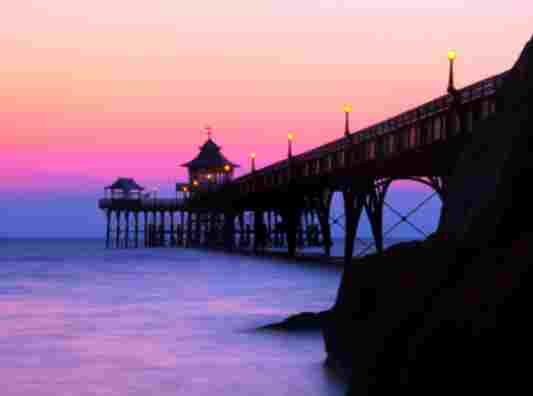} &
\includegraphics[width=\imagesizee]{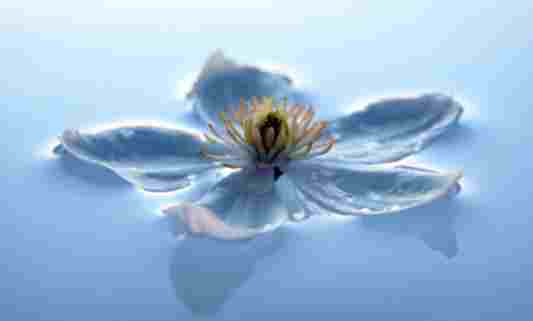} &
\includegraphics[width=\imagesizee]{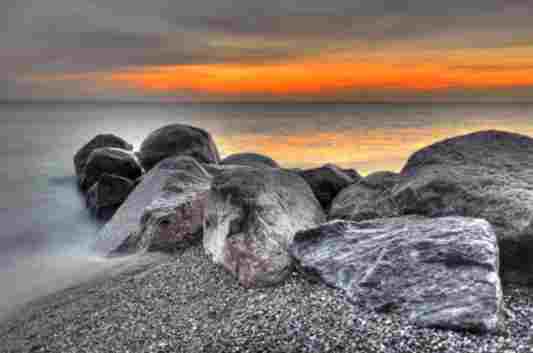} &
\includegraphics[width=\imagesizee]{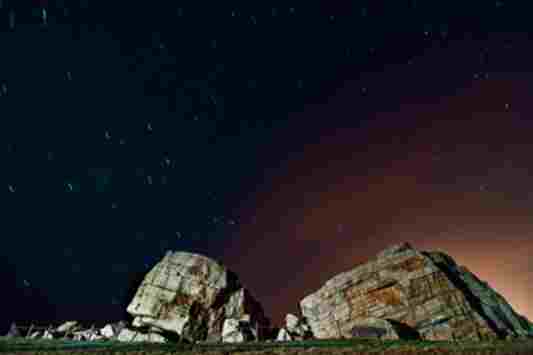} &
\includegraphics[width=\imagesizee]{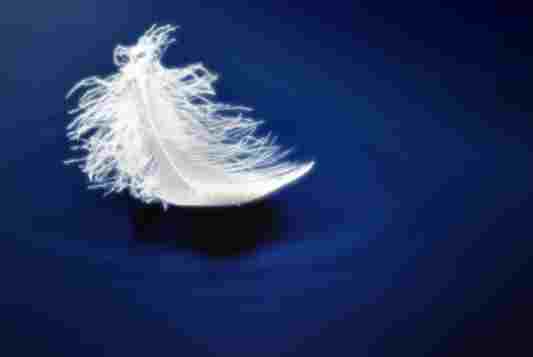} &
\includegraphics[width=\imagesizee]{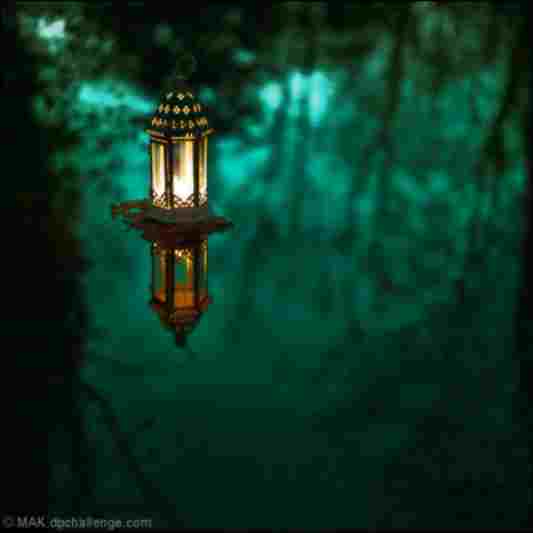} &
\includegraphics[height=\imagesizee]{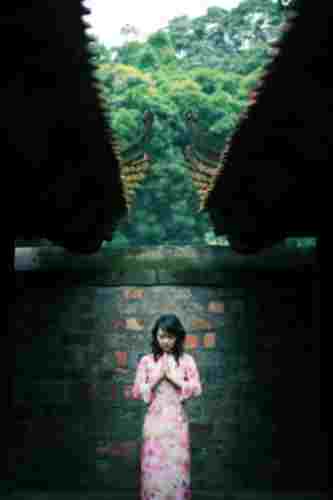} &
\includegraphics[width=\imagesizee]{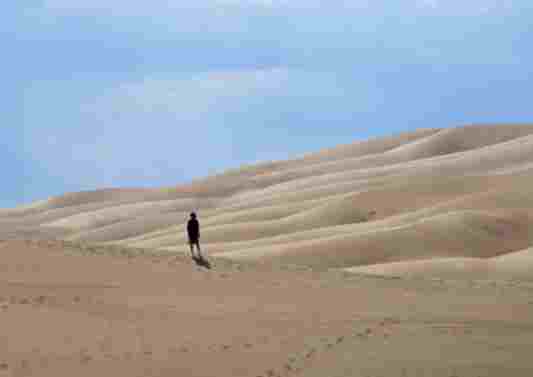} &
\includegraphics[width=\imagesizee]{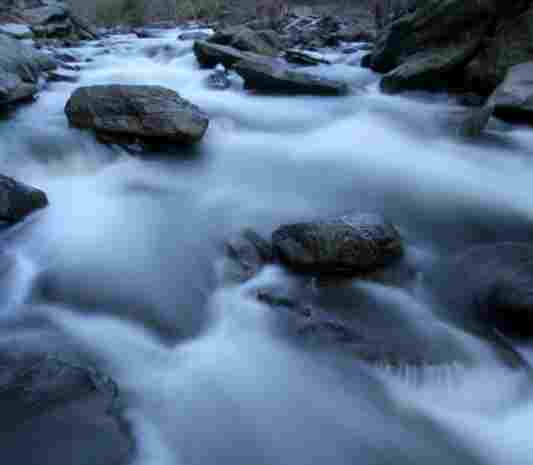} &
\includegraphics[width=\imagesizee]{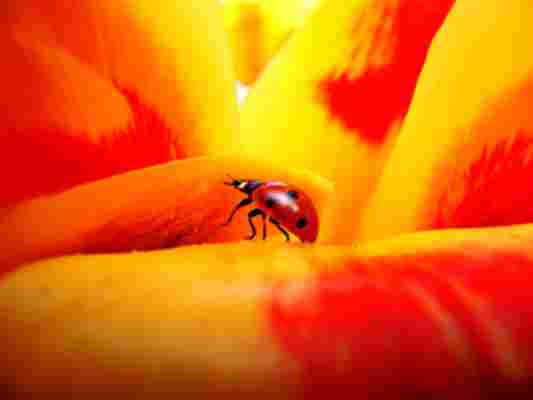} &
\includegraphics[width=\imagesizee]{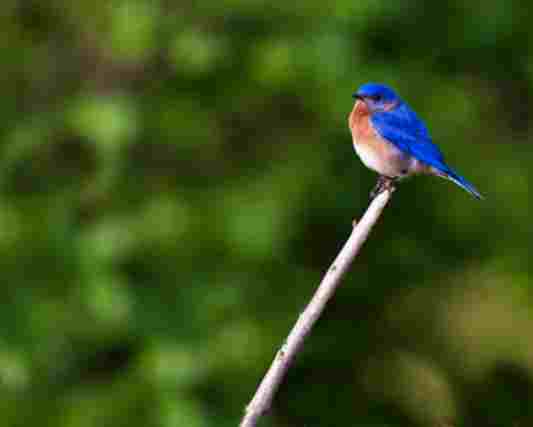} &
\includegraphics[width=\imagesizee]{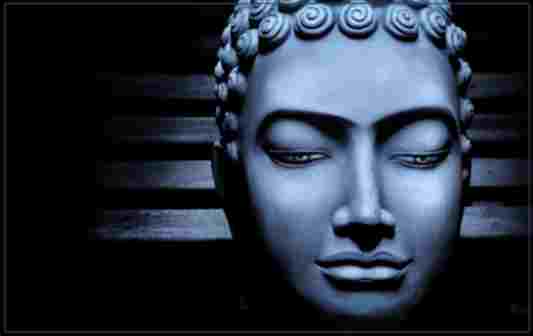} &
\includegraphics[width=\imagesizee]{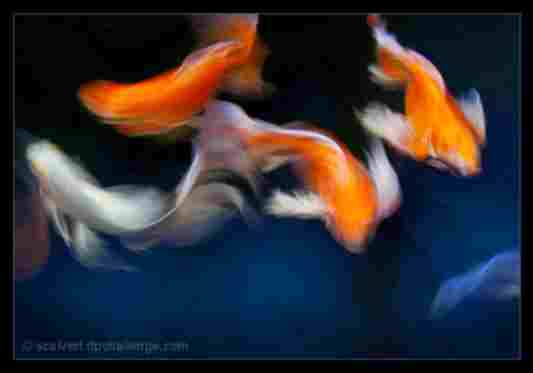} \\
\end{tabular}
}
\caption{The five thematic datasets. From top to bottom: \burstF (\burst),\fashionF (\fashion), \landscapeF (\landscape), \selfF (\self), and \zenF (\zen).}
\label{fig:datasets}
\end{figure*}

A collage is usually created from a set of images sharing a common underlying theme. To create our dataset, we downloaded the images from the DPChallenge\footnote{http://www.dpchallenge.com/} web site. The site collects photos of both amateur and professional photographers that participate to digital photography challenges. Each challenge has a main theme that the participants must follow. All the submitted photos are then judged by other participants by giving a numerical score. We selected five photo challenges among the hundreds published and for each of them we collected the 14 best rated photos. The challenges have been chosen to include diverse subjects of generic themes. The chosen challenges are: \burstF (\burst for brevity), \fashionF (\fashion), \landscapeF (\landscape), \selfF (\self), and \zenF (\zen). 
The \burst dataset is composed of images with a single subject; the \fashion dataset contains images of people and accessories; the \landscape dataset is composed of mostly horizontal images; on the contrary, the \self dataset contains mostly portrait images both in colors and black and white; finally, the \zen dataset is composed of heterogeneous images and in most cases it is not easy to identify the subject. This diversity makes it possible to investigate if people use different criteria in the creation of photo collages for different themes. Figure \ref{fig:datasets} shows the five sets of photos.

\subsection{Importance Maps}
\label{subsec:maps}

Two of the basic criteria used in Equation \ref{eq:fitness1} require the computation of importance maps to locate the most informative regions in an image. The underlying idea is that the most informative regions should not be hidden by other images thus maximizing the information displayed. Since there is no a unique definition of what is important in an image, in our investigation we tested different importance maps exploiting three different image properties: saliency, color harmony, and quality. Each importance map is plugged in turn into Equation~\ref{eq:fitness1} obtaining three different collages for each photo dataset. 

\begin{figure}[tb]
\centering
\begin{tabular}{cccc}
\small{Original} & \small{Saliency} & \small{Harmony} & \small{Quality} \\
\includegraphics[width=40pt]{land/estratte/834855.jpg} &
\includegraphics[width=40pt]{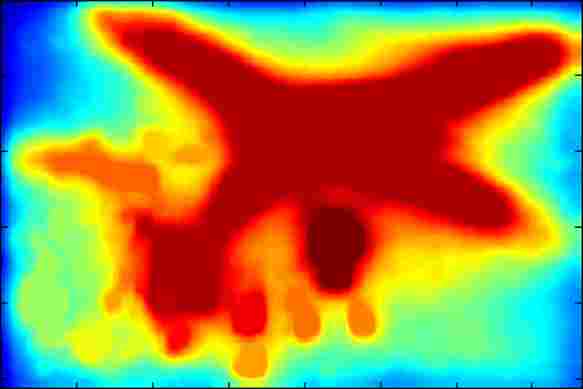}  &
\includegraphics[width=40pt]{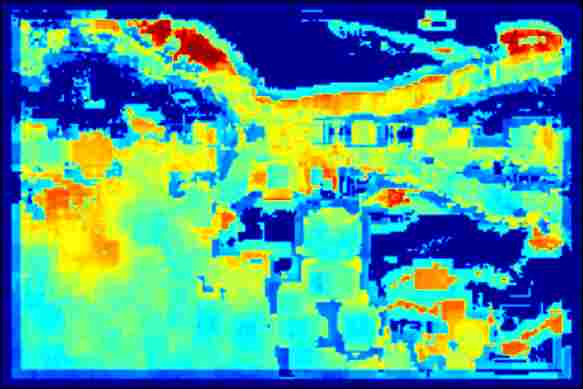}  &
\includegraphics[width=40pt]{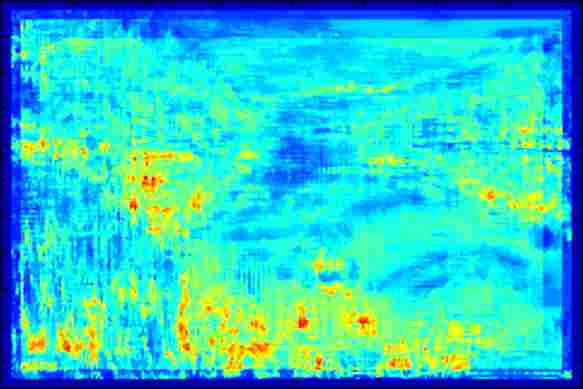}  \\
\includegraphics[width=40pt]{self/estratte/887356.jpg} &
\includegraphics[width=40pt]{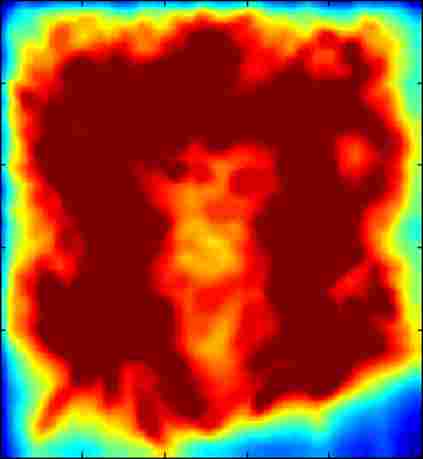}   &
\includegraphics[width=40pt]{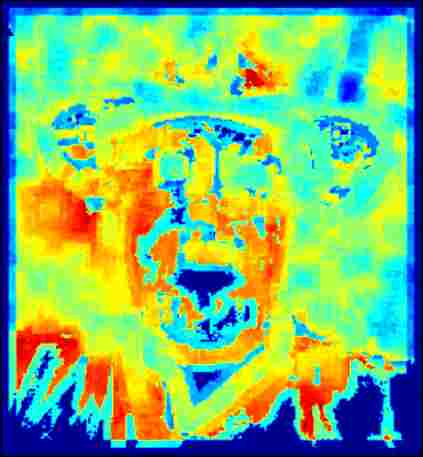}   &
\includegraphics[width=40pt]{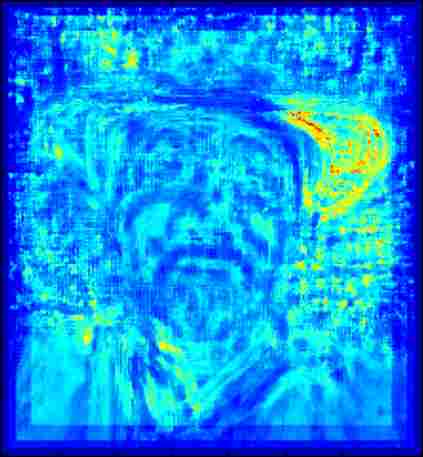}  \\
\includegraphics[width=40pt]{burst/estratte/945921.jpg} &
\includegraphics[width=40pt]{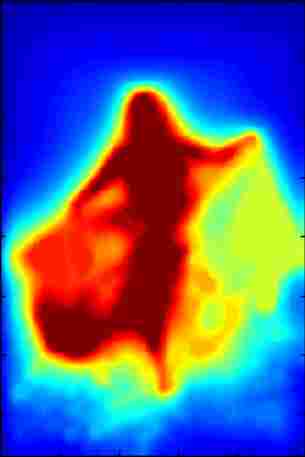} &
\includegraphics[width=40pt]{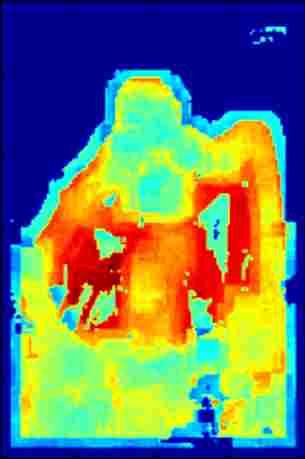} &
\includegraphics[width=40pt]{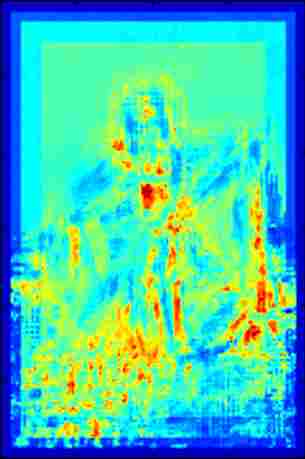} \\
\end{tabular}
\begin{tabular}{c}
\\
\includegraphics[height=148pt]{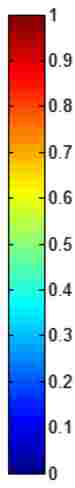}
\end{tabular}
\caption{Examples of importance maps.}
\label{fig:importance_maps}
\end{figure}

\noindent\textbf{Saliency} The first importance map is based on saliency and uses an approach to compute it similar the one presented in \cite{Ma2003}. \ADD{We used this approach in a previous work on image thumbnailing \cite{ciocca2010multiple-image} and the resulting saliency maps show that, on the overall, a compact set of salient regions are produced. We considered these results reasonable for our purposes. Other, more recent and precise saliency methods can be exploited. The recent paper \cite{cheng2015globalcontrast} shows the performances of several algorithms on reference datasets that can be used as alternative ones. For surveys related to saliency see \cite{duncan2012,kimura2013computational,ali2014}. To compute the saliency map}, the image is divided into small rectangular tiles. On each tile, a contrast score is computed by comparing its average color with the average colors of the neighbor's tiles. The contrast score is assigned to each pixel in the tile. The basic algorithm has been extended by computing three different saliency maps in the LUV color space using neighborhoods of increasing size. Each map captures the saliency at a different scale. These saliency maps are then filtered and combined together into a single normalized map of values in the range $[0,1]$.  We denote the importance map of the i-th image computed using saliency as $M_{i,sal}$. Examples of saliency maps are shown in the second column of Figure~\ref{fig:importance_maps}. 

\noindent\textbf{Harmony} Since color combinations are related to the pleasantness of an image, for the second importance map, we used the method proposed in \cite{Solli2009} to evaluate color harmony of the image locally by creating a color harmony map. We choose to use this approach because, in contrast to other approaches (e.g. \cite{Ou2006}), it computes an image color harmony score by considering the distribution and spatial relationship between color regions found by the MeanShift segmentation algorithm. In order to have a color harmony map we computed the harmony score on pixel's neighborhoods (i.e. pixels in a square region surrounding a given pixel's location) of different sizes. The harmony map is obtained by summing all the scores and by normalizing them in the $[0,1]$ range.  We denote the importance map of the i-th image computed using color harmony as $M_{i,har}$. The third column of Figure \ref{fig:importance_maps} shows some examples of color harmony maps.

\noindent\textbf{Quality} Image quality approaches model how an image is perceived if affected by different image distortions. We cannot predict what kind of image distortions are present, nor we have a reference image to which compare our photos, thus we must consider generic (or \quot{universal}) no references image quality approaches. We exploited the BRISQUE (Blind/Referenceless Image Spatial Quality Evaluator) computational model described in \cite{Mittal2012}. The model uses different image features in order to quantify the image quality. Since BRISQUE computes a single quality index for an image, we implemented a neighborhood-based strategy in order to obtain a quality map. We considered the quality index computed on the whole image and on three pixel's neighborhoods. The indexes are summed and normalized in the $[0,1]$ range. We denote the importance map of the i-th image computed using image quality as $M_{i,qua}$. The fourth column of Figure \ref{fig:importance_maps} shows some examples of image quality maps.

\subsection{Optimization Algorithm}
\label{subsec:optalgo}
Let us consider, for now, a generic photo dataset and a generic importance map definition. Under these assumptions, the optimal collage is generated by finding the best configuration of states $\vect{S}$ which maximizes Equation \ref{eq:fitness1}. The solution space of this maximization problem is of mixed type: in fact for each state $\vect{s}_i=(\vect{t}_i,\theta_i,l_i)$ we have $\mathbf{t}_i \in \mathbb{Z}^2$, $\theta_i \in \mathbb{R}$, and $l_i \in \mathbb{N}$. In order to uniform the state variables types, and since small variations of $\theta_i$ do not affect the final collage, the allowed orientations are uniformly quantized in the range $[-\theta_{max},\theta_{max}]$.

The chosen optimization method is an extension of a Direct Search algorithm (DS) modified to deal with discrete solution spaces \cite{bianco2012sensor,bianco2012sampling}. DS is a derivative-free method for solving optimization problems \cite{hooke1961direct,kolda2003optimization}. Since the focus of this paper is not on the optimization algorithm used, any non-gradient method could be used \cite{nelder1965simplex} as well as stochastic ones \cite{Goldberg1989,eberhart1995new}. 

The algorithm is initialized with a random configuration of states: the i-th image is placed on the canvas at a random position $\vect{t}_i$ and with a random orientation $\theta_i$. Its layering index instead is determined by the importance map as previously described. 
At each iteration of the implemented DS algorithm, the algorithm finds the best configuration of states by testing the current best configuration against all those obtained by varying the position and orientation in each image's state. The position of each image is then updated and a new iteration is started. The algorithm terminates when the maximum number of iterations has been reached.

\ADD{We used the modified Direct Search algorithm without any heuristic whose computational cost is $\mathcal{O}(gan)$ with $g$ being the number of grid points on the canvas, $a$ being the number of allowed orientation on ecah grid point, and $n$ being the number of images to be placed. Other, more efficient, optimization algorithms can be used. Here we are interested in the effects of using different criteria in creating the photo collages, not the most efficient way to create them.}

\subsection{Experimental Setup}
In our implementation the size of the canvas $\mathcal{C}$ is set 400$\times$400 pixels and all the images have been resized such that $\min\{\mbox{width},\mbox{height}\}=128$ pixels maintaining the same aspect ratio. With these constraints, the ratios between the sum of the areas of the images in a given dataset and the area of the canvas are $2.0096$ for the \burst dataset, $2.0024$ for the \fashion dataset, $2.2008$ for the \landscape dataset, $1.8168$ for the \self dataset, and $2.0144$ for the \zen dataset. In practice this means that we need to hide about $50\%$ of the pixels of the images to fit them on the canvas in a pleasant manner. Or, conversely, we need to retain the most informative and pleasant $50\%$ of the pixels. 
The canvas and image dimensions have been chosen solely for the purpose of evaluating the performance of our framework under typical constraints usage. \ADD{We are here interested more in the ratio of size between images and canvas than in the absolute dimensions themselves and we wanted the placement problem to be hard. A larger canvas and/or larger images can be used in an actual application. It should be pointed out that while optimization has been done on canvas of 400$\times$400 pixels, the subjective tests have been done on their 1600$\times$1600 versions.}

Top left image corners were allowed to be placed on a regular grid from $-2g$ to 400 in both canvas directions, with a step of $g=50$ pixels. 
The set of allowed orientations is defined in the range $[-\theta_{max},\theta_{max}]=[-\frac{\pi}{3}, \frac{\pi}{3}]$ in $\frac{\pi}{18}$ steps. 
For Experiment I, the values of $\lambda_1$, $\lambda_2$, and $\lambda_3$ in Equation \ref{eq:fitness1} have been set to $1$, while for Experiment~II, these have been learned from the users.

\begin{figure*}[!tb]
\small
\tabcolsep=2pt
\centering
\begin{tabular}{cccccc}
 & \burst & \fashion & \landscape & \self & \zen \\
\begin{sideways}Saliency\end{sideways} &
\includegraphics[width=\collagewidth]{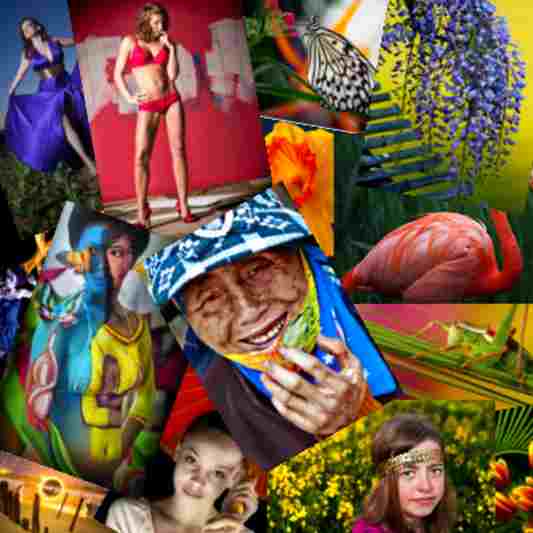} &
\includegraphics[width=\collagewidth]{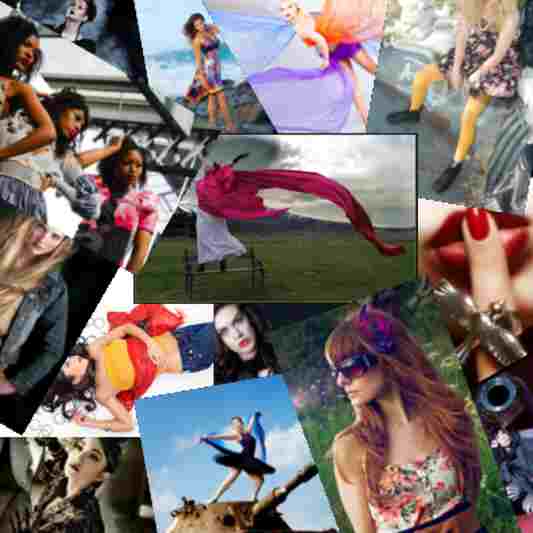} &
\includegraphics[width=\collagewidth]{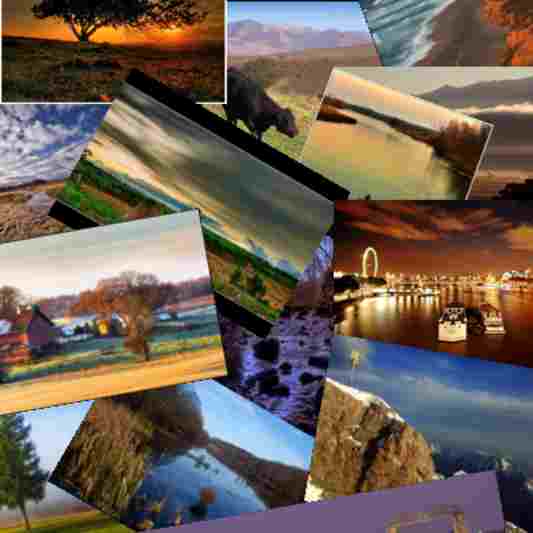} &
\includegraphics[width=\collagewidth]{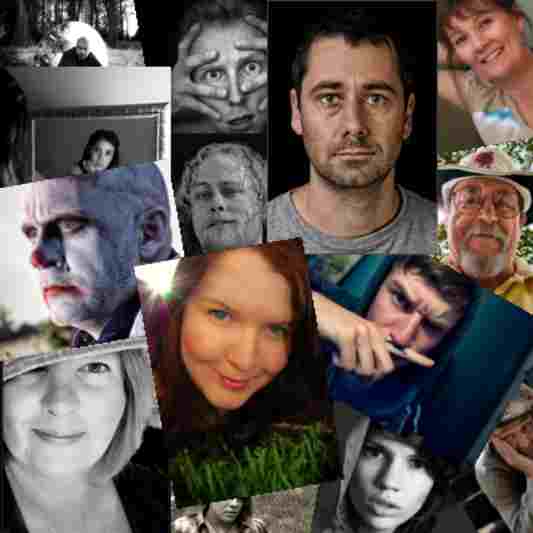} &
\includegraphics[width=\collagewidth]{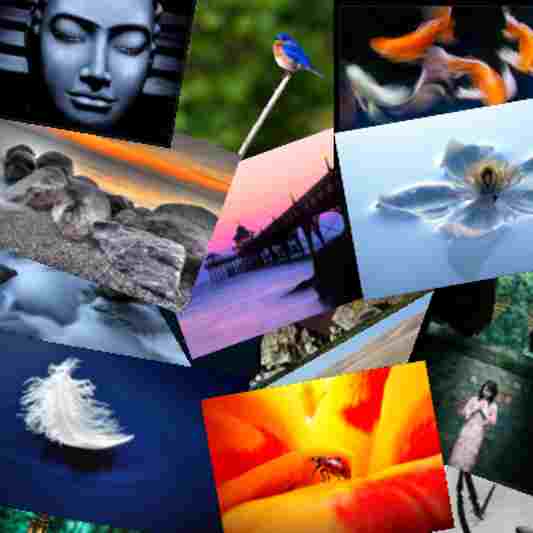} \\

\begin{sideways}Harmony\end{sideways} &
\includegraphics[width=\collagewidth]{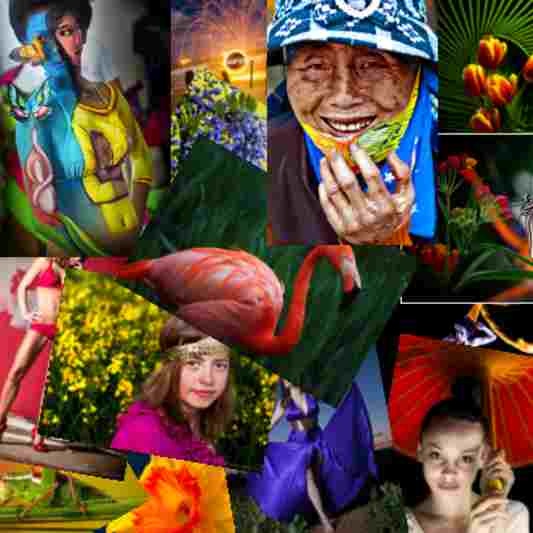} &
\includegraphics[width=\collagewidth]{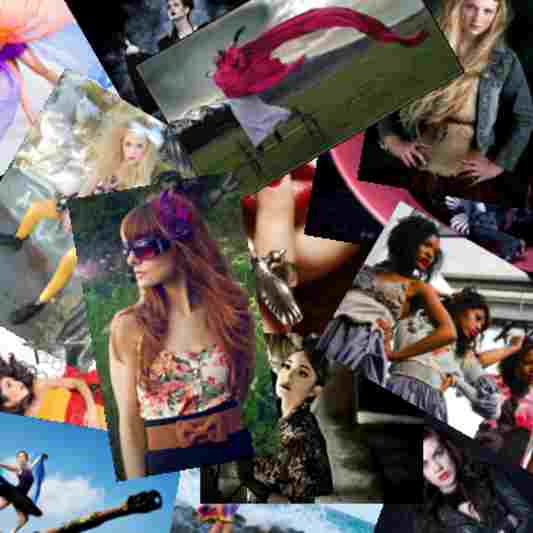} &
\includegraphics[width=\collagewidth]{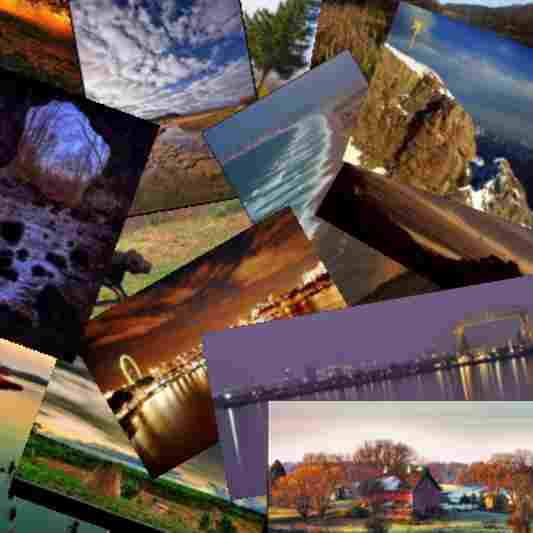} &
\includegraphics[width=\collagewidth]{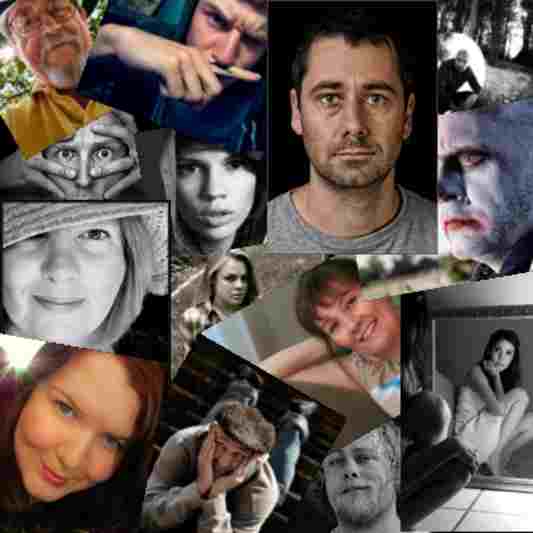} &
\includegraphics[width=\collagewidth]{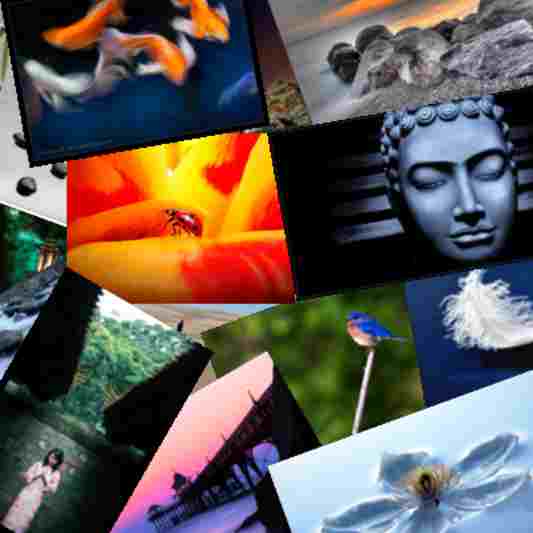} \\

\begin{sideways}Quality\end{sideways} &
\includegraphics[width=\collagewidth]{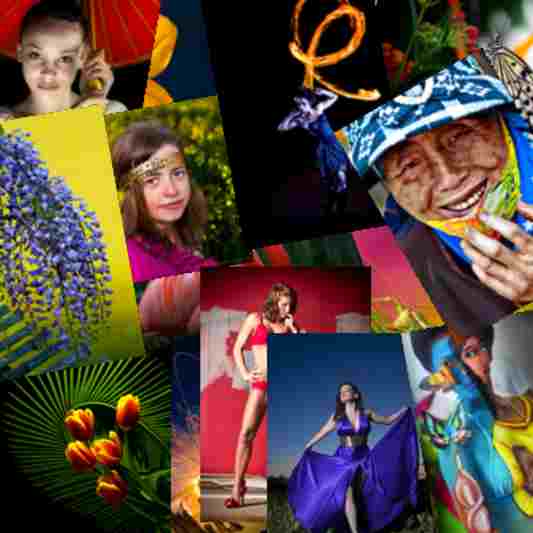} & 
\includegraphics[width=\collagewidth]{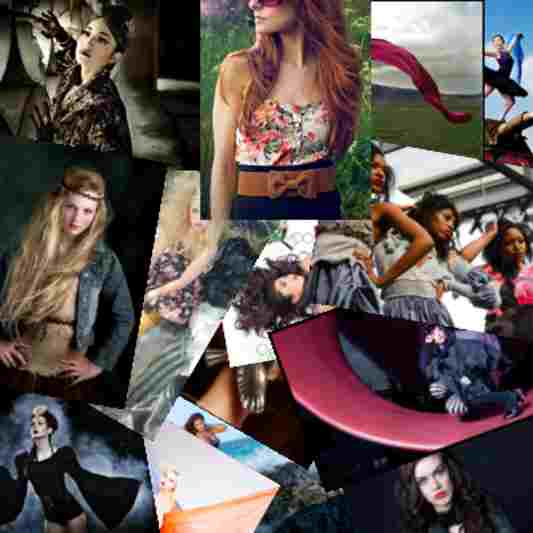} & 
\includegraphics[width=\collagewidth]{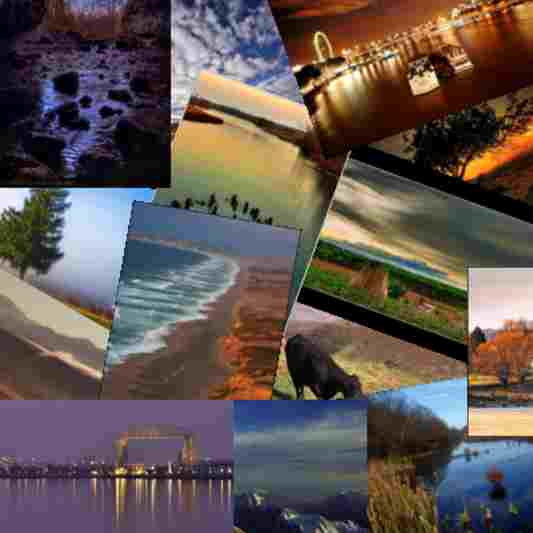} & 
\includegraphics[width=\collagewidth]{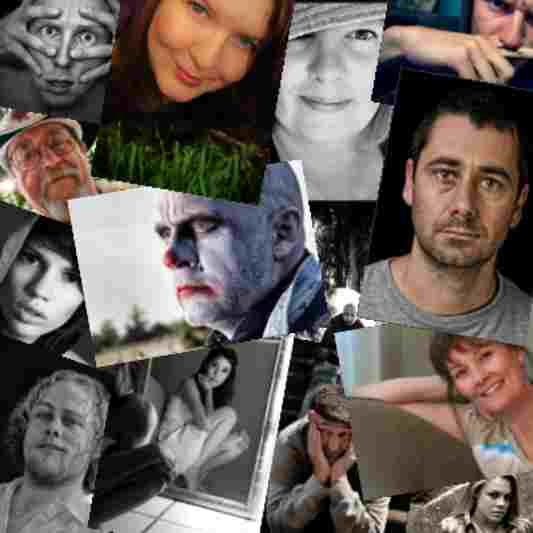} & 
\includegraphics[width=\collagewidth]{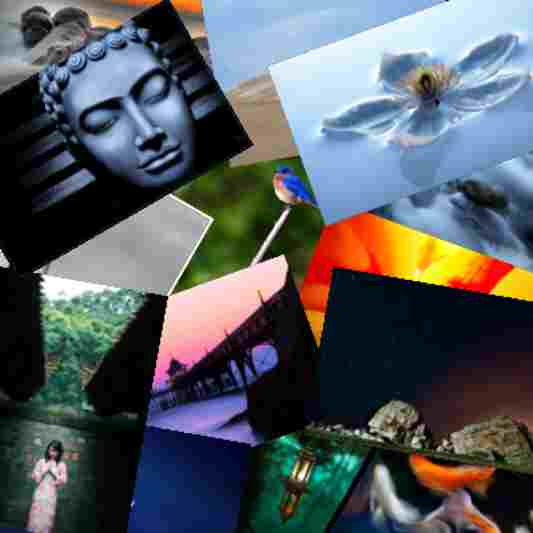} \\
\end{tabular}
\\
\normalsize
{\begin{tabular}{|C{0.7in}|C{0.7in}|C{0.7in}|C{0.7in}|C{0.7in}|}
\hline
\dummytab $\vect{S}_{\burst}^{sal}$ & $\vect{S}_{\fashion}^{sal}$ & $\vect{S}_{\landscape}^{sal}$ &
$\vect{S}_{\self}^{sal}$ & $\vect{S}_{\zen}^{sal}$ \\ \hline
\dummytab $\vect{S}_{\burst}^{har}$ & $\vect{S}_{\fashion}^{har}$ & $\vect{S}_{\landscape}^{har}$ & $\vect{S}_{\self}^{har}$ & $\vect{S}_{\zen}^{har}$ \\ \hline
\dummytab $\vect{S}_{\burst}^{qua}$  & $\vect{S}_{\fashion}^{qua}$ & $\vect{S}_{\landscape}^{qua}$ & $\vect{S}_{\self}^{qua}$ & $\vect{S}_{\zen}^{qua}$  \\ \hline
\end{tabular}}
\caption{Photo collages created on the five datasets using the three importance maps. Color and full size images can be found at {\pirlingurl}.}
\label{fig:collages1}
\end{figure*}

\section{Subjective Experiment I}
\label{sec:exp1}
The above algorithm has been applied to each dataset using the three importance maps yielding a total of 15 photo collages as shown in Figure \ref{fig:collages1}. Let us denote each collage with the corresponding configuration of states $\vect{S}_{m}^{ds}$: $ds \in \dsets$ denotes the photo dataset, and $m \in \imaps$ the importance map used.

In order to identify the criteria to be used to create pleasing photo collages, we performed a subjective test involving several users. Test subjects were selected taking into account age, gender and expertise in photography. Specifically, 16 subjects (Italian native speakers) were enlisted. Subjects are between 21 and 41 years old, three females and 13 males. Only one of the subjects can be considered an expert photographer (although not professional) while the others consider themselves amateurs. Half of the subjects stated that they shoot an average of 3,000-4,000 photos/year. The remaining subjects shoot an average of 100-300 photos/year. All of them have a certain knowledge about digital image processing. No relation exists between subjects and images in the photo datasets.
In this first experiment, we showed to each subject the five sets of three photo collages, one set at time, and asked him/her to rank the three collages according to his/her liking without judging the semantic of the scenes depicted. The subjects were aware that the collages have been generated by different algorithms but no technical information and no hints  about the underlying criteria were given. This was done in order to not bias their choices. The sets of photo collages, as well as the collages within each set, were presented in a random order. The evaluation of all the collages and the related interviews took on average 30 minutes per subject.

After all the test sessions have been performed, we counted the number of times that each collage was ranked at the first (i.e. best), second, or third position in its photo dataset. \ADD{During the counting, we checked for noisy user feedback that, in the pairwise experiment, manifests in the form of circular preferences (e.g. A$>$B, B$>$C, and C$>$A). We planned to remove these subjects from the analysis, but at the end of the experiment no one of the subjects showed this behavior.}

Table \ref{tab:counts1} shows the detailed results. As it can be seen, in the case of the \zen photo dataset, the results are quite polarized. Almost all the subjects have judged the collages in a similar manner ranking first the collage created with the Saliency map, then the one using the Harmony map, and lastly the collage using the Quality map. The same ranking, although with a less polarization effect, can be observed for the \burst, \fashion and \landscape sets. In all the three sets, the collages created with the Saliency map is clearly the preferred one. The one using the Harmony map is the second best since it has been selected second or third a fewer number of times than the one using the Quality map. The only set displaying a different ranking is \self. In this case, the ranking is the opposite of the ones obtained from the other four sets. 

In Table \ref{tab:ranks1} the final ranking of the importance maps for the five photo collage sets are reported. The ranking is determined by applying the Formula One World Championship points scoring system: each collage receives 25, 18, or 15 points each time that it is selected respectively first, second, or third. The numbers in parenthesis are the computed scores.


\begin{table}[!tb]
\tbl{Number of times that each collage was ranked fist, second or third\label{tab:counts1}}{
\begin{tabular}{ccc}
\begin{tabular}{lccc}
\toprule
\burst & 1st & 2nd & 3rd \\
\midrule
Saliency & 9 & 3 & 4 \\
Harmony & 5 & 6 & 5 \\
Quality & 2 & 7 & 7 \\
\bottomrule
\end{tabular}
&
\begin{tabular}{lccc}
\toprule
\fashion & 1st & 2nd & 3rd \\
\midrule
Saliency & 9 & 4 & 3 \\
Harmony & 6 & 8 & 2 \\
Quality & 1 & 4 & 11 \\
\bottomrule
\end{tabular}
&
\begin{tabular}{lccc}
\toprule
\landscape & 1st & 2nd & 3rd \\
\midrule
Saliency & 9 & 6 & 1 \\
Harmony & 6 & 5 & 5 \\
Quality & 1 & 5 & 10 \\
\bottomrule
\end{tabular}
\\ \\
\multicolumn{3}{c}{\begin{tabular}{cc}
\begin{tabular}{lccc}
\toprule
\self & 1st & 2nd & 3rd \\
\midrule
Saliency & 2 & 5 & 9 \\
Harmony & 5 & 7 & 4 \\
Quality & 9 & 4 & 3 \\
\bottomrule
\end{tabular}
&
\begin{tabular}{lccc}
\toprule
\zen & 1st & 2nd & 3rd \\
\midrule
Saliency & 11 & 3 & 2 \\
Harmony & 2 & 11 & 3 \\
Quality & 3 & 2 & 11 \\
\bottomrule
\end{tabular}
\end{tabular}}
\end{tabular}
}
\end{table}


\begin{table}[!tb]
\tbl{Importance maps ranking, according to Table \ref{tab:counts1}. The numbers in parenthesis are the scores computed using the Formula One World Championship points scoring system\label{tab:ranks1}}{
\begin{tabular}{lccc}
\toprule
 Set & 1st & 2nd & 3rd \\
\midrule
 \burst & Saliency (339) & Harmony (309) & Quality (281)\\
 \fashion & Saliency (342) & Harmony (324)& Quality (262)\\
 \landscape & Saliency (348) & Harmony (314) & Quality (265)\\
 \self & Quality (342) & Harmony (324) & Saliency (225)\\
 \zen & Saliency (359) & Harmony (293) & Quality (276)\\ 
\bottomrule
 \end{tabular}
}
\end{table}

After each test, we also interviewed each subject about the reasons of his/her choices, what factors have influenced the selection of a photo collage over the others, and what criteria they used. In the following, for each photo dataset, we report a summary of the answers given by the users during the interviews.

\subsection{Experiment I: Results}
\label{subsec:results1}
The {\bf{\burst}} photo dataset is composed of images with bright colors. Many of these images are close-ups. It is not surprising that most subjects indicated color as a primary feature in collage evaluation. In particular, several subjects suggested that the images should have been positioned in the canvas by taking into account the color similarity. Very dissimilar colors among neighbor images were considered disturbing. One subject suggested to hide very dark regions preferring to have a collage with bright colors. Most subjects preferred images placed with randomized orientations. Collages containing images with their borders parallel to the canvas borders were penalized. Most of the images in this dataset contain a single object of interest. Collages where this object was fully visible were thus preferred, in particular in the case of faces.  

The {\bf{\fashion}} photo dataset is mainly composed of images of full-body women models. Only one image is a close-up. These images are less colorful than the \burst dataset but they contain high contrast regions. The main criterion used in evaluating the collages was the visibility of the models. Several subjects also indicated that having the top layer image in a central position makes the collage more pleasing. Secondary criteria include the visibility of a (impossible to model) favourite image, and loss of bright colored regions. No other criteria were suggested on this dataset.

The {\bf{\landscape}} photo dataset contains images with mostly dull colors if compared against the previous ones. No people are visible and the scenes depicted are mostly natural scenes. Several shots have a panoramic aspect ratio. For these reasons, according to the subjects, the collages created on this dataset resulted among the most difficult to be evaluated. Images arranged in a regular way were considered disturbing. If an image was mostly covered by the others (as for example the violet sunset in the collage created with the Saliency Map), it was considered acceptable by many users. On the overall, the collages were often considered equivalent.

The {\bf{\self}} photo dataset was the easiest to evaluate probably because contains self portraits. As expected, the criteria arisen from the interviews referred mostly to the visibility of the faces. One interesting insight on this dataset is that, even though we encouraged the subjects to avoid judging the image content from the semantic point of view, many choices were made based on the appealing of the faces depicted. For example, some subjects considered the photo of the clown unpleasant and thus a photo that could be covered before others. On the contrary others considered this photo very artistic. It seems that when human  are depicted, personal preferences are difficult to ignore. This is the only dataset containing both gray-scale and color type of images. Some test subjects did not appreciate collages with spatial clusters of images of the same type.

The {\bf{\zen}} photo dataset should inspire peace and tranquillity. It contains photos with very few colors and details. They are mostly close-ups, and some of the photos show soft-focus effects. Most of the subjects found it difficult to judge the collages and express the rationale behind their choices. However, color composition and harmony were the most important criteria. The best collages were those where the relevant objects were visible. One interesting criterion emerged on this dataset is that the shape of the visible image regions should not be jagged. Regular (i.e. convex) shapes are considered more appealing.

\section{New Criteria Definition}
\label{sec:newcriteria}
{From the results reported in the previous section, we can see that the users evaluated the collages using different criteria. These criteria are both local and global. Local criteria refer to either properties of single images or of their neighborhoods, while global criteria refer to properties of the collage seen as a whole. The three basic criteria reported in  Table \ref{tab:cues_base} and exploited in previour works are not enough to capture the different nuances of pleasantness expressed by the users.
}
Thus, on the basis of the insights obtained from Experiment~I, and taking into account that we need to model them with  computational algorithms, we have selected the criteria in Table~\ref{tab:newcriteria} to be used in the generation of pleasing collages. {The first three criteria are extensions of the ones in Table~\ref{tab:cues_base}, where now the importance map is computed by using a combination of the three importance maps described in Section~\ref{subsec:maps}. The other seven criteria have been defined following the results of Experiment I.} 
Since the results showed that we also need to take into account the presence of faces within the images, it is necessary to consider, for each image, a binary mask $F_i$ containing the face regions. These masks undergo the same geometric transformations as the importance maps:

\begin{equation}
\mathcal{F}_i=T(F_i,\vect{s}_i)
\end{equation}

We indicate with $\boldsymbol{\mathcal{F}}=\{\mathcal{F}_i\}_{i=1}^{N}$ the set of transformed masks which is passed along with the other data to the criteria functions.
In the following we write the functions $C_i(\fparamssfull)$  as $C_i(\fparamss)$ dropping the dependencies for a more compact notation.

%
%

\begin{table}
\tbl{New criteria for a pleasing photo collage\label{tab:newcriteria}}{
\begin{tabular}{p{1.5in}p{3in}c}
\toprule
Criterion & Description & Function \\ 
\midrule

\ADD{Visibility} & Visible image content (based on the three importance maps) & $C'_{1}$ \\
Canvas coverage & Canvas area covered by the photos & $C'_{2}$ \\
\ADD{Visibility ratio balance} & Visible image region w.r.t. image size  & $C'_{3}$ \\

Face Ratio & Percentage of visible faces & $C'_4$ \\ 
Axis Alignment & Percentage of images having sides parallels to the canvas & $C'_5$ \\ 
Centrality & Position of the top-level image & $C'_6$ \\ 
Convexity &  Measure of jaggedness of the shape of the visible regions & $C'_7$\\
Color Similarity & Measure of color similarity between neighbor images & $C'_8$ \\ 
Orientation Diversity & Measure of variability in the image orientations & $C'_9$ \\ 
Minimum Orientation Difference & Minimum orientation difference between neighbor photos & $C'_{10}$\\ 
\bottomrule
\end{tabular}
}
\end{table}

\vspace{0.5truecm}
\noindent
{\ADD{\bf{Visibility}}} For each image we combined the three importance maps computed on saliency, quality and harmony, in order to obtain a global importance map:

\begin{equation}
M'_i=\sum_{k\in\imaps}\alpha_k{M}_{i,k}
\label{eq:newImportanceMap}
\end{equation}

where $\alpha_k$ are found as described in the next section.
\ADD{Visibility} is thus computed as in Equation \ref{eq:c1} by substituting the set of transformed importance maps $\boldsymbol{\mathcal{M}}$ with the new one $\boldsymbol{\mathcal{M}'}$: 

\begin{equation}
C'_1(\fparamss)=\frac{1}{N}\sum_{i=1}^{N} \frac{sum2(vis(\mathcal{M'}_i))}{sum2(M'_i)}
\label{eq:c1b}
\end{equation}

\vspace{0.5truecm}
\noindent
{\bf{Canvas coverage}} The definition of the canvas coverage is identical to the definition of $C_2$ in Equation \ref{eq:c2}:

\begin{equation}
C'_2(\fparamss)=\frac{1}{area(\mathcal{C})}\sum_{i=1}^{N}area(vis(\mathcal{M'}_i))
\label{eq:c2b}
\end{equation}

\vspace{0.5truecm}
\noindent
{\ADD{\bf{Visibility ratio balance}}} The ratio balance is computed as in Equation \ref{eq:c3}:

\begin{equation}
C'_3(\fparamss)=1-\std_{i=1\dots N}\left\lbrace\frac{sum2(vis(\mathcal{M'}_i))}{sum2(M'_i)}\right\rbrace
\label{eq:c3b}
\end{equation}

\vspace{0.5truecm}
\noindent
{\bf{Face ratio}} A face detector is run on each image $I_i$ to find the mask containing the face regions (i.e. face bounding boxes). Let the mask $F_i$ be
\begin{equation}
F_i(x,y)= \left\{ 
\begin{array}{ll}
1 & \mbox{if $(x,y)\in$ face region} \\
0 & \mbox{otherwise}	\\ 
\end{array}
\right.
\end{equation}

\noindent the face ratio feature is then defined as follows:
\begin{equation}
C'_4(\fparamss)= \frac{\sum_{i=1}^{N}sum2(vis(\mathcal{F}_i))}{\sum_{i=1}^{N}sum2(F_i)}
\label{eq:c4b}
\end{equation}

\vspace{0.5truecm}
\noindent
{\bf{Axis alignment}} This feature measures the ratio of images with orientation parallel to the axis, i.e. $0$ given that $\theta_{max}=\frac{\pi}{3}$.
\begin{equation}
C'_5(\fparamss) = \frac{\# \{ \vect{s_i} \in \vect{S} : \theta_i=0 \}}{N}
\label{eq:c5b}
\end{equation} 

\vspace{0.5truecm}
\noindent
{\bf{Centrality}} The centrality feature measures how central is the image in the first layer, i.e. the top-most image. Let us call $\mathbf{c_1}=(x_1,y_1)$ the centroid of the visible part of the image in the top layer and $\mathbf{c_0}=(x_0,y_0)$ the centroid of the canvas $\mathcal{C}$.
The centrality is defined as:
\begin{equation}
C'_6(\fparamss)=1-\frac{{\|\mathbf{c_1}-\mathbf{c_0}\|}_2}{hdiag(\mathcal{C})}
\label{eq:c6b}
\end{equation}
where $hdiag(\cdot)$ is used to compute the half diagonal length.

\vspace{0.5truecm}
\noindent
{\bf{Convexity}} For each transformed image $\mathcal{I}_i$ the convexity ratio is defined as ratio between the area corresponding to the image's visible region and the area of its convex hull. The convexity feature is computed as the minimum convexity ratio over all the transformed images:
\begin{equation}
C'_7(\fparamss) = \min_{i=1,\ldots,N} \left\{ \frac{area(vis(\mathcal{I}_i))}{area(convex(vis(\mathcal{I}_i)))} \right\}
\label{eq:c7b}
\end{equation}

\vspace{0.5truecm}
\noindent
{\bf{Color similarity}} This feature is computed by evaluating the color histogram similarity of each image on the canvas with respect to its neighbors. For each image we first compute:

\begin{equation}
d_i=\sum_{j \in neigh(\mathcal{I}_i)} \chi^2(hist(vis(\mathcal{I}_i)),hist(vis(\mathcal{I}_j)) )
\end{equation}

\noindent where  $hist(vis(\mathcal{I}_i))$ is the color histogram computed on the visible portion of $\mathcal{I}_i$, $\chi^2$ is the chi-squared distance, and $neigh(\mathcal{I}_i)$ represents the set of the indexes of the images neighbors of $\mathcal{I}_i$.  Color similarity is then computed as:
\begin{equation}
C'_8(\fparamss)=\frac{1}{N} \sum_{i=1}^{N}  \frac{{d_i}}{\# \{ neigh(\mathcal{I}_i) \} }
\label{eq:c8b}
\end{equation}

\vspace{0.5truecm}
\noindent
{\bf{Orientation diversity}} This feature measures the average of the variance in orientation in each set of neighbor images:
\begin{equation}
C'_9(\fparamss)=\frac{1}{N} \sum_{i=1}^{N}\std_{j \in \vect{NI}}\left\{\frac{\theta_{j}}{\theta_{max}}\right\}
\label{eq:c9b}
\end{equation}
where $\vect{NI}=neigh(\mathcal{I}_i) \cup \{i\}$ and $\theta_{max}$ is the maximum rotation angle allowed.

\vspace{0.5truecm}
\noindent
{\bf{Minimum orientation difference}} This feature measures the average of the minimum orientation differences between each image $\mathcal{I}_i$ and its neighboring set $neigh(\mathcal{I}_i)$:
\begin{equation}
C'_{10}(\fparamss)= \frac{1}{N} \sum_{i=1}^N{\min_{j \in neigh(\mathcal{I}_i)} \left\{\frac{|\theta_i - \theta_j|}{\theta_{max}}\right\}}
\end{equation}

\noindent The new fitness function $f'(\cdot)$ to be optimized in the generation of pleasing photo collages, can be compactly written as:

\begin{equation}
f'(\fparamssfull)=\sum_{i=1}^{10}{\lambda'}_i{C'}_i(\fparamssfull)
\label{eq:fitness2}
\end{equation}

where each $\lambda'_i$ weights the contribution of criterion $C'_i$, and are found as described in the next section. Please recall that the fitness function also depends on the three weights $\alpha_k$ introduced in Equation \ref{eq:newImportanceMap}, and that are used to compute the new importance maps.


\section{User Preferences Modeling and Learning}
\label{sec:modeling}
Given as input the values $C'_i$, $i=1,\ldots,10$, we want to learn \ADD{a single set of} 
optimal weights $[\boldsymbol{\lambda'},\boldsymbol{\alpha}]=\left[\{\lambda'_i \}_{i=1}^{10},\{ \alpha_i \}_{i\in\imaps}\right]$ to be plugged into Equation \ref{eq:fitness2} so that they produce fitness values in accordance with user preferences emerged from Experiment~I \ADD{on all the datasets considered}. 
To this end, for each dataset, the fitness values obtained for the collages created using the saliency, harmony, and quality importance maps must be in the same order reported in Table~\ref{tab:ranks1}. Taking as example the Burst dataset, where the user preferences were Saliency~$\succ$~Harmony~$\succ$~Quality, we want that $f'(\vect{S}_{\burst}^{sal})>f'(\vect{S}_{\burst}^{har})>f'(\vect{S}_{\burst}^{qua})$.
Furthermore, the relative distances between the normalized scores obtained by the different maps and reported in Table \ref{tab:scores1} should be preserved as much as possible. 
\begin{table}[!tb]
\tbl{Normalized scores obtained by scaling the scores in Table \ref{tab:ranks1} for the maximum score in each dataset\label{tab:scores1}}{
\begin{tabular}{lrrr}
\toprule
 Set & Saliency & Harmony & Quality \\
\midrule
 \burst 		& 1.00 & 0.91 & 0.83 \\
 \fashion 	& 1.00 & 0.95 & 0.77 \\
 \landscape & 1.00 & 0.91 & 0.76 \\
 \self 			& 0.80 & 0.91 & 1.00 \\
 \zen 			& 1.00 & 0.82 & 0.77 \\ 
\bottomrule
\end{tabular}
}
\end{table}
As an example, let us indicate with $\fone(\cdot)$ the function that computes the Formula One score. Taking again as example the Burst dataset, we want that the fitness $f'(\cdot)$ satisfies
\begin{equation}
\frac{f'(\vect{S}_{\burst}^{har})}{f'(\vect{S}_{\burst}^{sal})}=\frac{\fone(\vect{S}_{\burst}^{har})}{\fone(\vect{S}_{\burst}^{sal})}=0.91
\end{equation}
and
\begin{equation}
\frac{f'(\vect{S}_{\burst}^{qua})}{f'(\vect{S}_{\burst}^{sal})}=\frac{\fone(\vect{S}_{\burst}^{qua})}{\fone(\vect{S}_{\burst}^{sal})}=0.83 
\end{equation}

\ADD{Similar constraints come from the other four datasets considered, giving a total of ten simoultaneous contraints that Equation \ref{eq:fitness2} has to satisfy.}

The optimal weights $[\boldsymbol{\lambda'},\boldsymbol{\alpha}]$ 
are found by solving the following optimization problem: 


\begin{align}
\label{eq:userfitting}
[\boldsymbol{\lambda'},\boldsymbol{\alpha}]=
 \mbox{arg} \!\!\!\!\!\!\!\!\! \max_{ \substack{\lambda'_1,\ldots,\lambda'_{10} \\ \alpha_{sal},\alpha_{qua},\alpha_{har} }} \sum_{ds \in \mathcal{D}}  \tau(\mbox{ord}_{ds}^{E_1},\mbox{ord}_{ds}^{f'}) - 
\eta \sum_{k=\{2,3\}}
\left|\left|
\frac{f'(\vect{S}_{ds}^{j_k})}{f'(\vect{S}_{ds}^{j_1})}-\frac{\fone(\vect{S}_{ds}^{m_k})}{\fone(\vect{S}_{ds}^{m_1})}
\right|\right|_1
\end{align}

\ADD{\noindent where $\mathcal{D}=\dsets$,} $\mbox{ord}_{ds}^{E_1}$ and $\mbox{ord}_{ds}^{f'}$ are respectively the per-dataset user rankings computed using $\fone(\cdot)$ and the rankings induced by $f'(\cdot)$: 

\begin{align}
\mbox{ord}_{ds}^{E_1}=[m_1,m_2,m_3 ], m_k \in \imaps : 
\fone(\vect{S}_{ds}^{m_1}) > \fone(\vect{S}_{ds}^{m_2}) >\fone(\vect{S}_{ds}^{m_3})
\label{eq:ordE}
\end{align}

\begin{align}
\mbox{ord}_{ds}^{f'}=[j_1,j_2,j_3 ], j_k \in \imaps : 
 f'(\vect{S}_{ds}^{j_1}) > f'(\vect{S}_{ds}^{j_2}) > f'(\vect{S}_{ds}^{j_3}) 
\label{eq:ordf}
\end{align}

\noindent $\tau(\cdot,\cdot)$ is the Kendall tau rank correlation coefficient \cite{kendall1938new}, $||\cdot||_1$ is the $1-$norm, and $\eta$ is a weight term that balances the relative contributions of the two parts of which Equation \ref{eq:userfitting} is made of. In this work, $\eta$ is heuristically set to $1$. 

\ADD{The rationale behind the optimization is that we want to automatically find the best set of weights $[\boldsymbol{\lambda'},\boldsymbol{\alpha}]$ that, plugged into Equation \ref{eq:userfitting}, produce a fitness function $f'(\cdot)$ in maximum accordance with user rankings on all the datasets used for training.}


\ADD{
The Kendall tau rank correlation coefficient in the first term is used to measure if, and to what extent, a given set of weights $[\boldsymbol{\lambda'},\boldsymbol{\alpha}]$ produces a fitness $f'(\cdot)$ in accordance with user rankings. This means that when the fitness function $f'(\cdot)$ is valued on the set of collages, its outputs should be in the same order in which the users judged them. The second term is introduced to avoids the scores to be too close to each other and to have a meaningful ranking.}

\ADD{Having such a fitness function permits, given a new set of images for which we want to build a collage, to have a measure of how good is a certain configuration of image states. Furthermore, maximizing $f'(\cdot)$ we are confident that we are generating a collage that the users will judge good on the overall.
The optimization to solve Equation \ref{eq:userfitting} has to be performed just once and offline so the computational time required to solve it is of secondary importance. However, it requires a bunch of seconds to run, since all the inputs are computed offline and the only operations involved are the computation of the Kendall tau rank (first term of Equation \ref{eq:userfitting}) and the relative distances between user scores and induced scores (second term of Equation \ref{eq:userfitting}).}

The optimization problem in Equation \ref{eq:userfitting} is solved using the continuous-space implementation of the DS algorithm. The signs of the criteria weights are reported in Table \ref{tab:pesiCues} together with a brief explanation of their effect in the creation of the collage.


\begin{table}[!tb]
\tbl{Signs of the criteria weights learned by the optimization algorithm, and a description of their effects on the creation of the collage\label{tab:pesiCues}}{
\begin{tabular}{ccp{3.8in}}
\toprule
 Criterion & Sign & Interpretation\\
\midrule
$C'_1$ 		& + & Promotes the visibility of the \ADD{image informativeness} \\
 $C'_2$ 		& + & Promotes the covering of the whole canvas, demoting holes\\
 $C'_3$ 		& - & Demotes large variations in the size of the visible parts of the images\\
 $C'_4$ 		& + & Promotes faces to be visible\\
 $C'_5$ 		& + & Promotes images to be placed aligned with canvas axis\\ 
 $C'_6$ 		& + & Promotes image in the top layer to be placed in the center\\
 $C'_7$ 		& + & Promotes visible parts of the images to be convex\\
 $C'_8$ 		& - & Promotes proximity of images with similar color histograms\\
 $C'_9$ 		& - & Promotes small variation of orientations among neighboring images\\
 $C'_{10}$ & + & Demotes neighboring images to have the same orientation\\ 
\bottomrule
\end{tabular}
}
\end{table}

Once the weights $[\boldsymbol{\lambda'},\boldsymbol{\alpha}]$ are learned, a new set of collages is generated by maximizing Equation \ref{eq:fitness2}. For each dataset, the layering order of the images is induced by the weigths $\boldsymbol{\alpha}=[\alpha_{sal},\alpha_{qua},\alpha_{har}]$. More in details, for each image $i$ a new importance map $M'_i$ is created using Equation \ref{eq:newImportanceMap}. The layering order is then obtained by sorting in decreasing order the 2D integrals of the importance maps $M'_i$. The collages are generated using the discrete version of DS algorithm introduced in Section \ref{subsec:optalgo}. 

\section{Subjective Experiment II}
\label{sec:exp2}

\begin{figure*}[!tb]
\tabcolsep=2pt
\begin{center}
\small
\begin{tabular}{cccccc}
& \burst & \fashion & \landscape & \self & \zen \\
\begin{sideways}Exp. I - Best coll.\end{sideways} &
\includegraphics[width=\collagewidth]{collages1/burst_saliency_DS14.jpg} &
\includegraphics[width=\collagewidth]{collages1/fashion_saliency_DS14.jpg} &
\includegraphics[width=\collagewidth]{collages1/land_saliency_DS14.jpg} &
\includegraphics[width=\collagewidth]{collages1/self_quality_DS14.jpg} &
\includegraphics[width=\collagewidth]{collages1/zen_saliency_DS14.jpg} \\

\begin{sideways}Exp. II - New coll.\end{sideways} &
\includegraphics[width=\collagewidth]{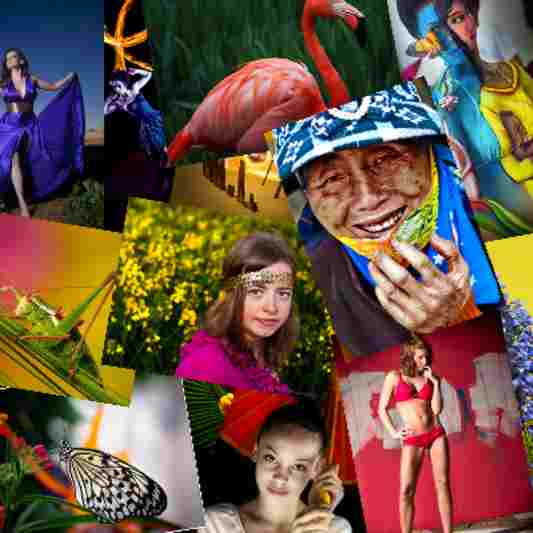} &
\includegraphics[width=\collagewidth]{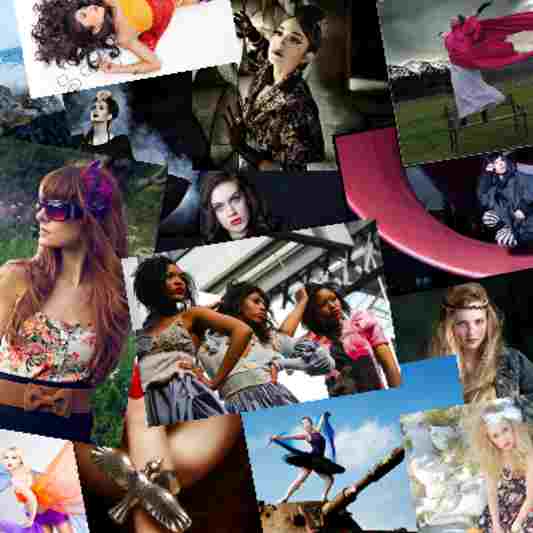} &
\includegraphics[width=\collagewidth]{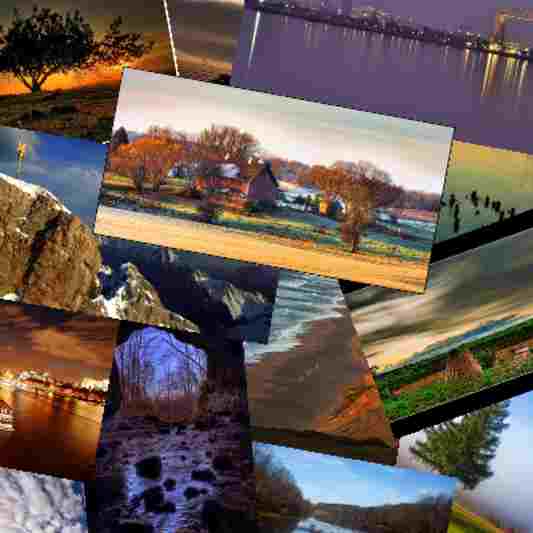} &
\includegraphics[width=\collagewidth]{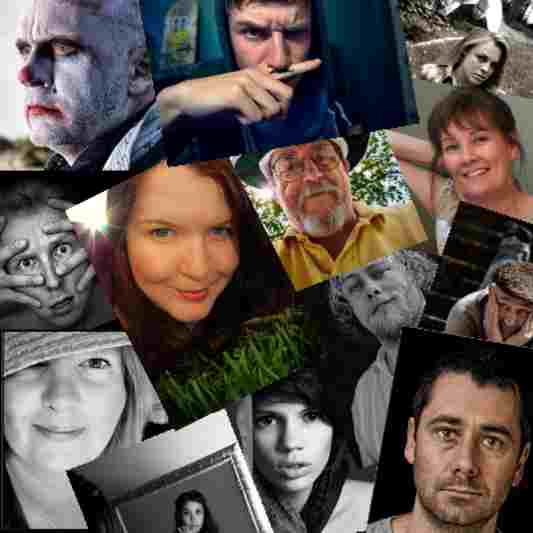} &
\includegraphics[width=\collagewidth]{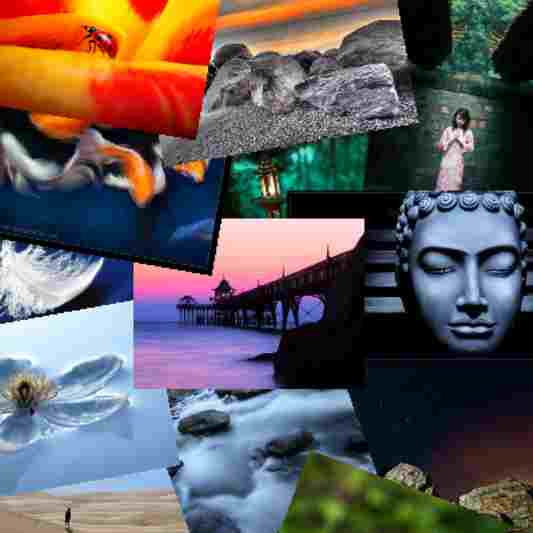}  \\
\end{tabular}

\normalsize{\begin{tabular}{|C{0.7in}|C{0.7in}|C{0.7in}|C{0.7in}|C{0.7in}|}
\hline
\dummytab $\vect{S}_{\burst}^{sal}$ & $\vect{S}_{\fashion}^{sal}$ & $\vect{S}_{\landscape}^{sal}$ &
$\vect{S}_{\self}^{qua}$ & $\vect{S}_{\zen}^{sal}$ \\ \hline
\dummytab $\vect{S}_{\burst}^{\prime}$ & $\vect{S}_{\fashion}^{\prime}$ & $\vect{S}_{\landscape}^{\prime}$ & $\vect{S}_{\self}^{\prime}$ & $\vect{S}_{\zen}^{\prime}$ \\ \hline
\end{tabular}}
\end{center}
\caption{Collages comparison between the best ranked collages in Experiment I (top) and the collages created with the user preference modeling and learning procedure for Experiment~II (bottom). Color and full size images can be found at {\pirlingurl}.}
\label{fig:collages2}
\end{figure*}

The final collages obtained on each dataset using the above described procedure, are reported in Figure~\ref{fig:collages2}. We denote each new collage with the corresponding configuration of states $\vect{S}^{\prime}_{ds}$. For each dataset we also report the best ranked collage from Experiment I according to the scores in Table~\ref{tab:scores1}. In this experiment we wanted to understand if the new collages were judged better than the previous ones. To this end, we performed a pairwise subjective test. For each dataset, users were presented with the two collages in Figure \ref{fig:collages2} and they were asked to choose the preferred one. A total of 39 subjects participated to this experiment: 26 males and 13 females. 


Results of Experiment II are reported in Table \ref{tab:counts2}. In three datasets (\burst, \landscape, and \self) the new collages were preferred by over 64\% of the subjects. In particular, the \self dataset exhibits the higher percentage of preferences with about 72\% of the subjects choosing the new collage. 
The \fashion dataset shows about 56\% of preference for the new collage. For this dataset, the new criteria seem to be marginally effective. This is due to the artistic nature of the photos that makes them good-looking regardless of their positioning. The \zen dataset continues to be the most problematic to be evaluated. Substantially, the subjects split in half in judging the collages due to the particular nature of the photo's content.
On average 62\% of the subjects preferred the new photo collages.

%
%

\begin{table}[!tb]
\tbl{Number of times (\#) and percentage (\%) that a collage was preferred in Experiment~II\label{tab:counts2}}{
\begin{tabular}{c}
\begin{tabular}{ccc}

\begin{tabular}{lcc}
\toprule
\burst & \# & \% \\
\midrule
Saliency & 13 & 33.3 \\
Final collage & 26 & 66.7 \\
\bottomrule
\end{tabular}
&
\begin{tabular}{lcc}
\toprule
\fashion & \# & \% \\
\midrule
Saliency & 17 & 43.6 \\
Final collage & 22 & 56.4 \\
\bottomrule
\end{tabular}
&
\begin{tabular}{lcc}
\toprule
\landscape & \# & \%\\
\midrule
Saliency & 14 & 35.9 \\
Final collage & 25 & 64.1 \\
\bottomrule
\end{tabular}
\end{tabular}

\\ \\

\begin{tabular}{cc}
\begin{tabular}{lcc}
\toprule
\self & \# & \% \\
\midrule
Quality & 11 & 18.2 \\
Final collage & 28 & 71.8 \\
\bottomrule
\end{tabular}
 &
\begin{tabular}{lcc}
\toprule
\zen & \# & \% \\
\midrule
Saliency & 19 & 48.7 \\
Final collage & 20 & 51.3 \\
\bottomrule
\end{tabular}
\end{tabular}
\end{tabular}
}
\end{table}

\section{Further experiments}
\label{sec:further}
\ADD{In this section further experiments are carried out to verify the generalization ability of the identified criteria and their learned relative importance. Three different experiments are presented: i) the learned definition of pleasantness is used to create collages on unseen image sets; ii) results obtained by our proposal are compared against two state-of-the-art algorithms; iii) the behavior of the learned definition of pleasantness is also tested by varying the number of images in the dataset and the canvas size.}

\subsection{Generalization to new photo themes}
\label{sec:generalization}
In order to test how the learned definition of pleasantness generalizes to photo collages not seen during the training phase, a further experiment has been done. 
\ADD{The optimal set of weights learned on the five training datasets in the previous section is used \emph{as-is} to create photo collages on the new datasets. }
For this experiment, six new challenges have been selected from the DPChallenge web site. The chosen challenges are: \shallowF (\shallow for brevity), \redF (\red), \primaryF (\primary), \silhouettesF (\silhouettes), \selfieF (\selfie), and \pixelsF (\pixels). 

\begin{figure*}[!tb]
\tabcolsep=2pt
\begin{center}
\small

\resizebox{\textwidth}{!}{
\setlength{\tabcolsep}{1pt}
\begin{tabular}{cccccc}
 \shallow & \red & \primary & \silhouettes & \selfie & \pixels \\
\includegraphics[width=\collagewidthhh]{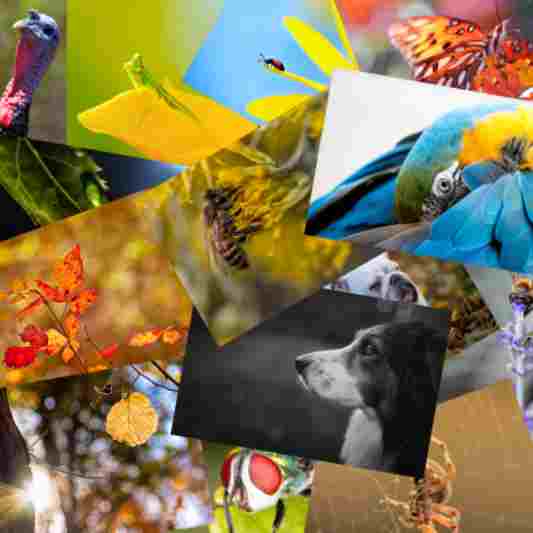} &
\includegraphics[width=\collagewidthhh]{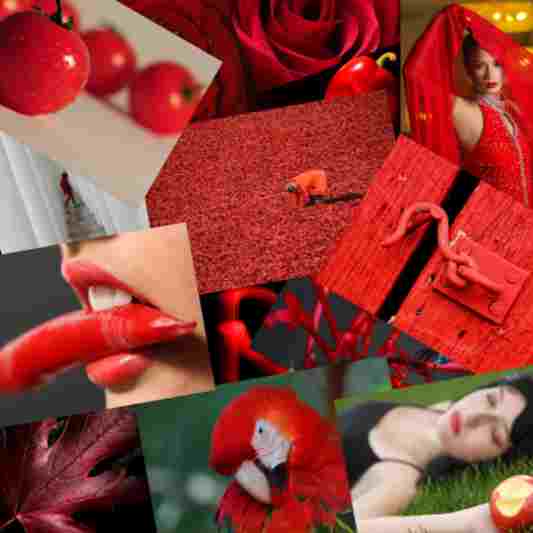} &
\includegraphics[width=\collagewidthhh]{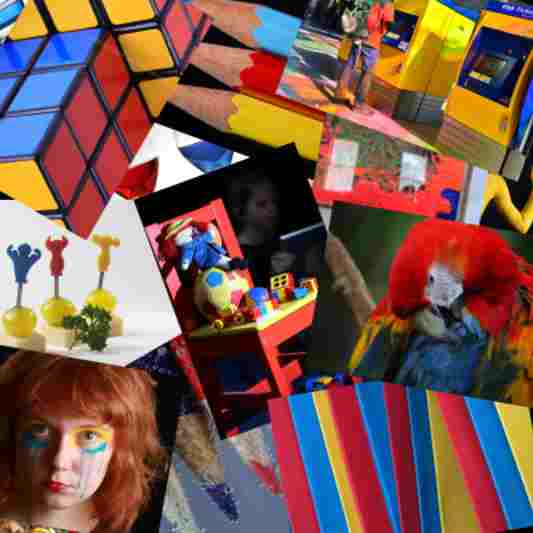} &
\includegraphics[width=\collagewidthhh]{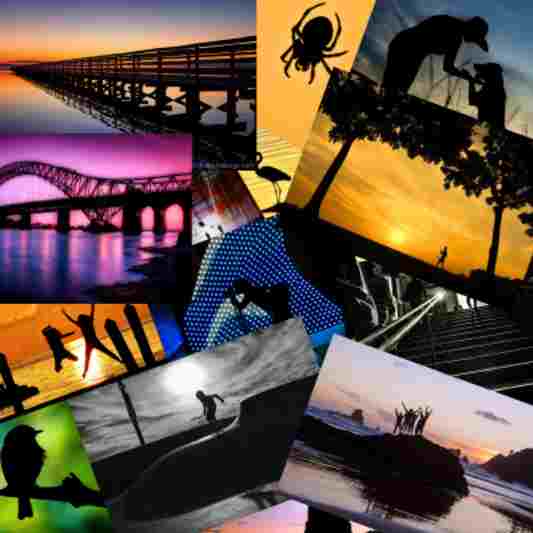} &
\includegraphics[width=\collagewidthhh]{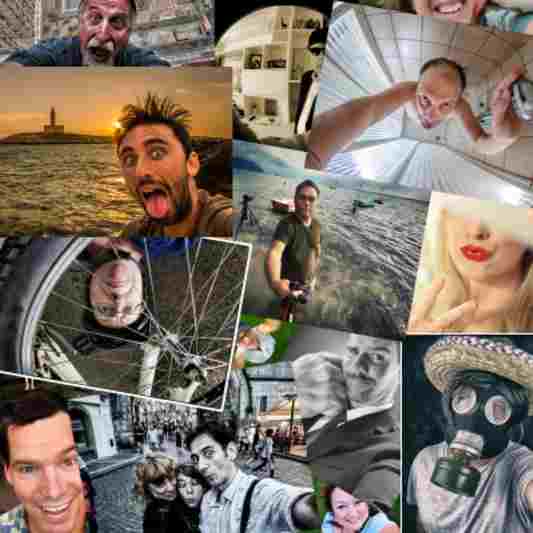} &
\includegraphics[width=\collagewidthhh]{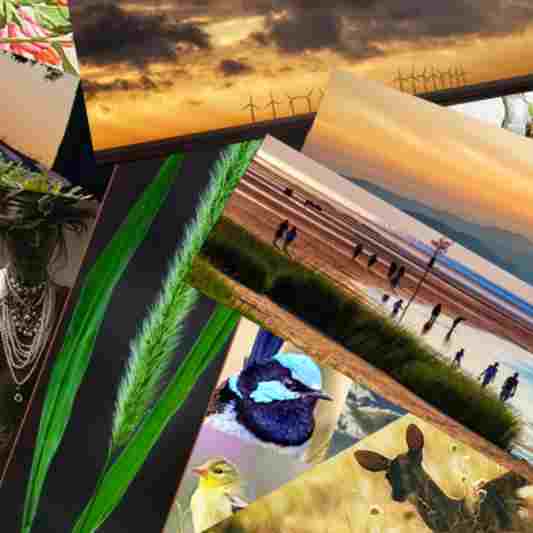} \\

\includegraphics[width=\collagewidthhh]{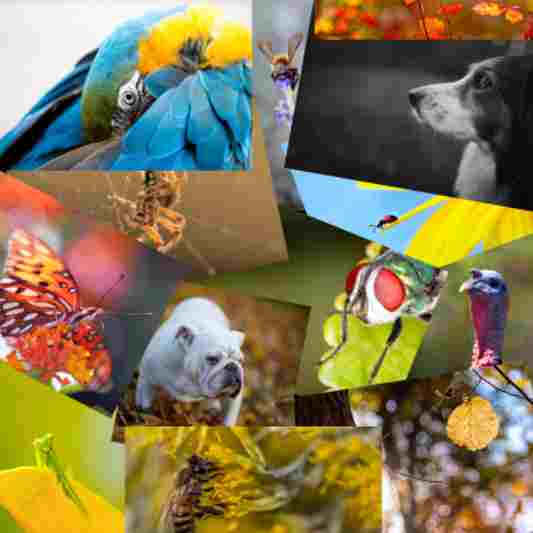} &
\includegraphics[width=\collagewidthhh]{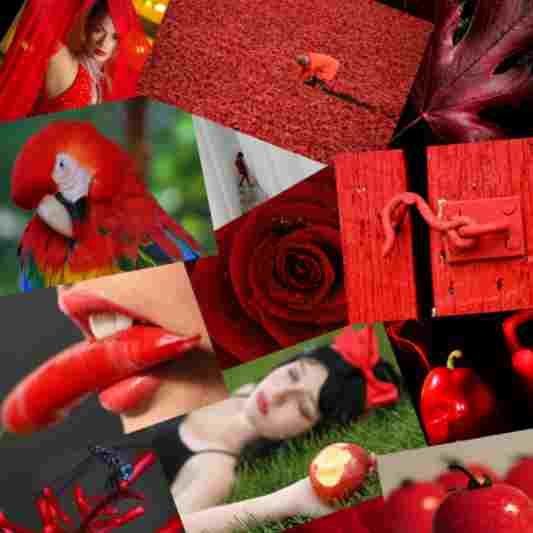} &
\includegraphics[width=\collagewidthhh]{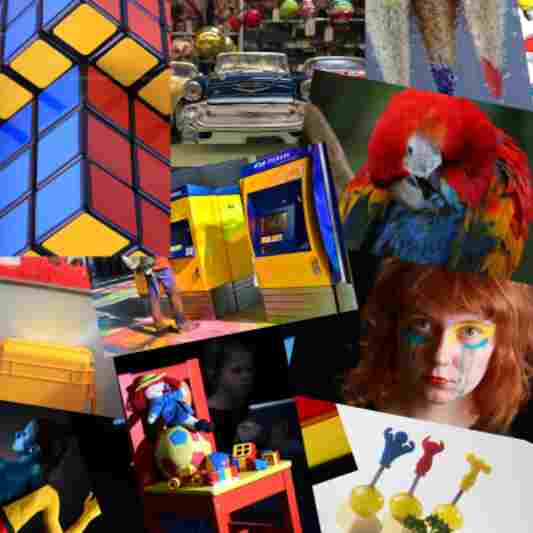} &
\includegraphics[width=\collagewidthhh]{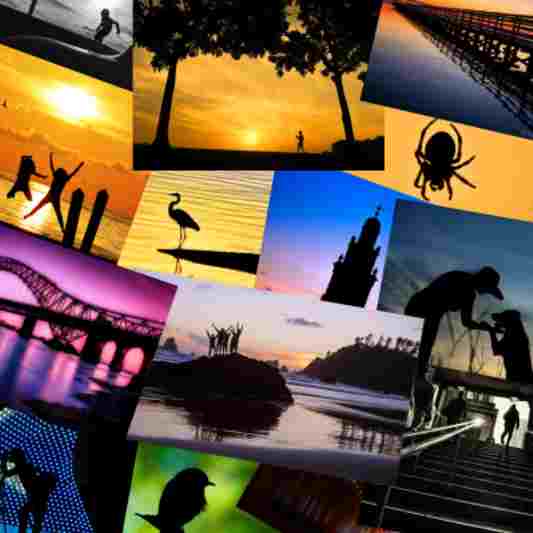} &
\includegraphics[width=\collagewidthhh]{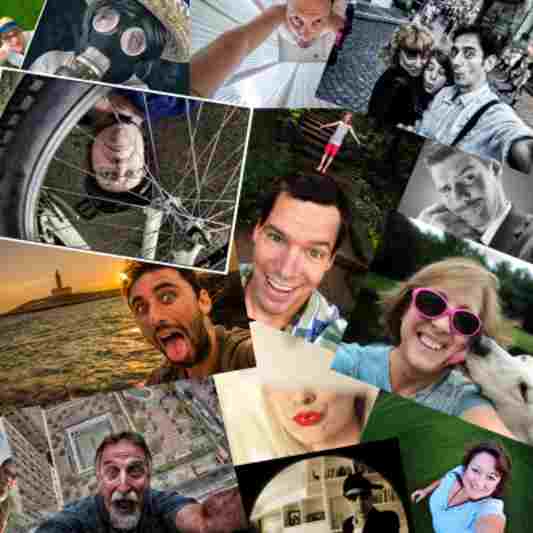} &
\includegraphics[width=\collagewidthhh]{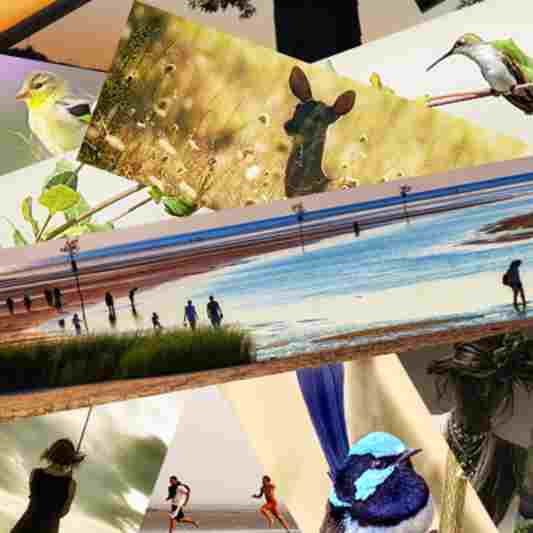}  \\
\end{tabular}
}

\normalsize{\begin{tabular}{|C{0.7in}|C{0.7in}|C{0.7in}|C{0.7in}|C{0.7in}|C{0.7in}|}
\hline
\dummytab $\vect{S}_{\shallow}^{sal}$    & $\vect{S}_{\red}^{sal}$    & $\vect{S}_{\primary}^{sal}$    & $\vect{S}_{\silhouettes}^{sal}$    & $\vect{S}_{\selfie}^{qua}$    & $\vect{S}_{\pixels}^{sal}$ \\ \hline
\dummytab $\vect{S}_{\shallow}^{\prime}$ & $\vect{S}_{\red}^{\prime}$ & $\vect{S}_{\primary}^{\prime}$ & $\vect{S}_{\silhouettes}^{\prime}$ & $\vect{S}_{\selfie}^{\prime}$ & $\vect{S}_{\pixels}^{\prime}$ \\ \hline
\end{tabular}}
\end{center}
\caption{\ADD{Generalization to new photo themes. These collages are created using the proposed framework with the single importance map (top) and the same learned fitness and weights used to create the collages in Figure \ref{fig:collages2} (bottom). Color and full size images can be found at {\pirlingurl}.}}
\label{fig:collages3}
\end{figure*}

The experiment performed is a pairwise subjective test similar to the one used in Experiment II. For each dataset, users were presented with the two collages in Figure \ref{fig:collages3} and they were asked to choose the preferred one. Following the results of Experiment I, one of the collage was generated using a single importance map; the other one was generated maximizing the learned fitness. A total of 42 subjects participated to this subjective experiment: 22 males and 20 females. 

Results from this experiment showed that on average 61\% of the subjects preferred the photo collages created using the learned definition of pleasantness. In particular, in four datasets (\shallow, \primary, \selfie and \pixels) these collages were preferred by over 66\% percent of the subjects. For the remaining two datasets (\red and \silhouettes) instead, the two collage versions tied.

\subsection{Comparing collages}
\label{sec:comparison}
\ADD{The different collage algorithms in the state of the art are based on different philosophies: keep images with the same size vs. allow image resize; allow image rotation vs. not; preserve image borders vs. blend image contents; allow images overlapping vs. not. We have run an experiment to compare our collage results with those of two algorithms belonging to the non-content preserving category (the same as ours) but using different philosophies: Shape Collage\footnote{\url{http://www.shapecollage.com/}} (a commercial software), and Autocollage \cite{Rother2006}.
The most relevant differences between the algorithms are that Shape Collage and our algorithm allow images to be rotated, while Autocollage does not. Moreover, Autocollage blends the images together to have a smooth transition between them, while Shape Collage and our algorithms do not.}

\begin{figure*}[!tb]
\small
\tabcolsep=2pt
\centering
\begin{tabular}{ccccccc}
 & Shallow & Red & Primary & Silhouettes & Selfie & Pixels \\
\begin{sideways}Shape collage\end{sideways} &
\includegraphics[width=\collagewidthhhh]{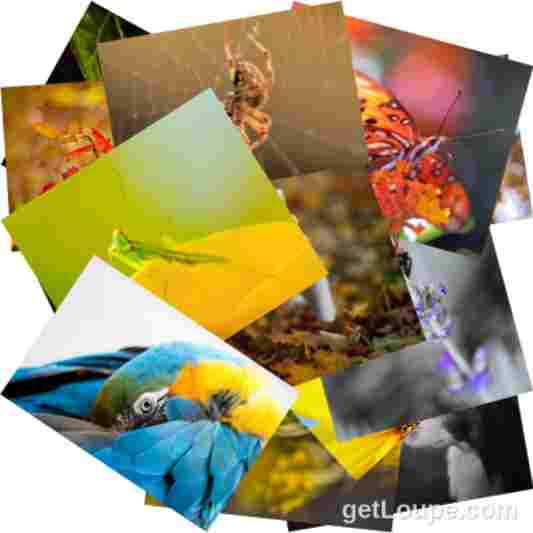} &
\includegraphics[width=\collagewidthhhh]{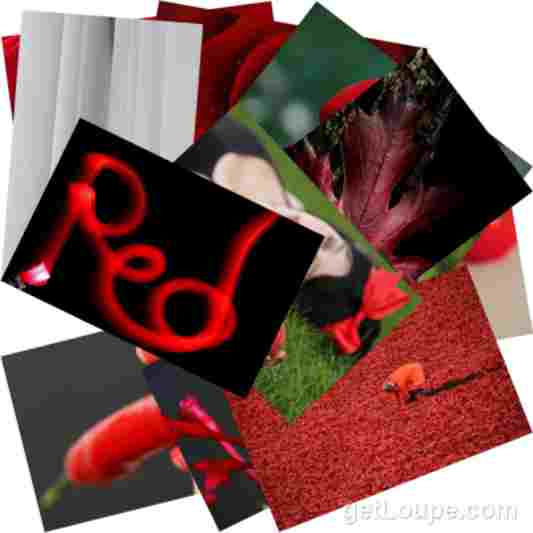} &
\includegraphics[width=\collagewidthhhh]{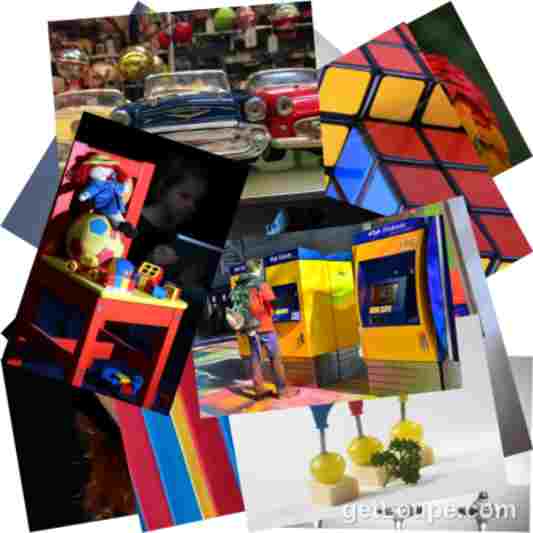} &
\includegraphics[width=\collagewidthhhh]{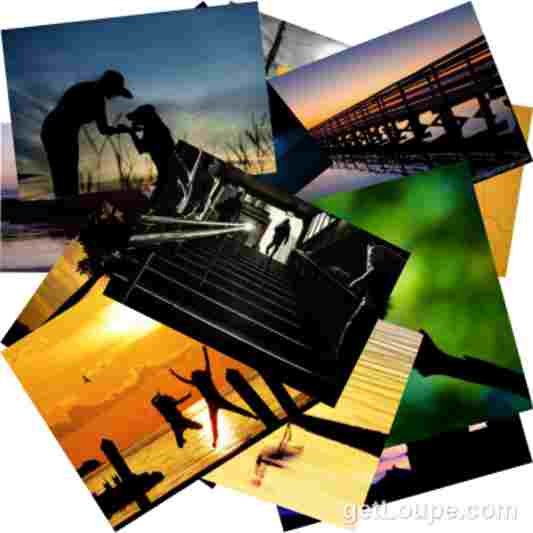} &
\includegraphics[width=\collagewidthhhh]{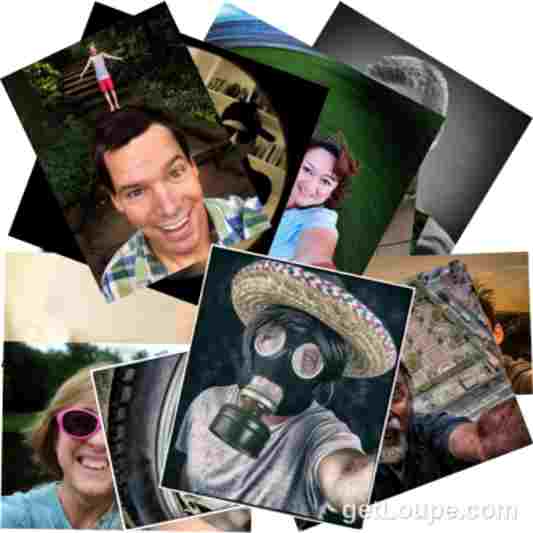} &
\includegraphics[width=\collagewidthhhh]{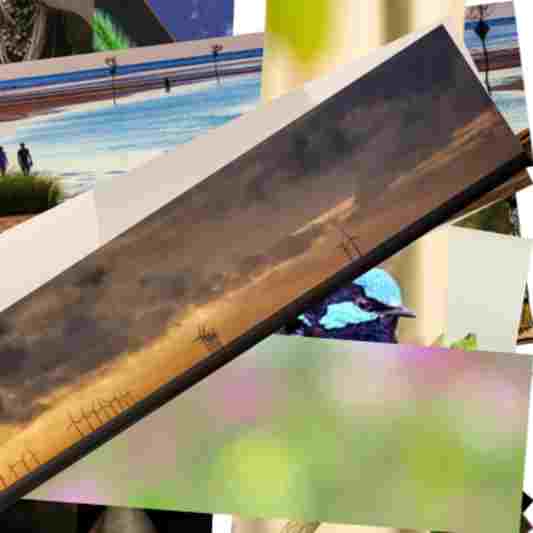} \\

\begin{sideways}Our method\end{sideways} &
\includegraphics[width=\collagewidthhhh]{collages3/Shallow_refinement_buchi3000_longer_4x.jpg} &
\includegraphics[width=\collagewidthhhh]{collages3/Red_refinement_buchi3000_longer_4x.jpg} &
\includegraphics[width=\collagewidthhhh]{collages3/Primary_refinement_buchi3000_longer_4x.jpg} &
\includegraphics[width=\collagewidthhhh]{collages3/Silhouettes_refinement_buchi3000_longer_4x.jpg} &
\includegraphics[width=\collagewidthhhh]{collages3/Selfie_refinement_buchi3000_longer_4x.jpg} &
\includegraphics[width=\collagewidthhhh]{collages3/Pixels_refinement_buchi3000_longer_4x.jpg}  \\

\begin{sideways}Auto collage\end{sideways} &
\includegraphics[width=\collagewidthhhh]{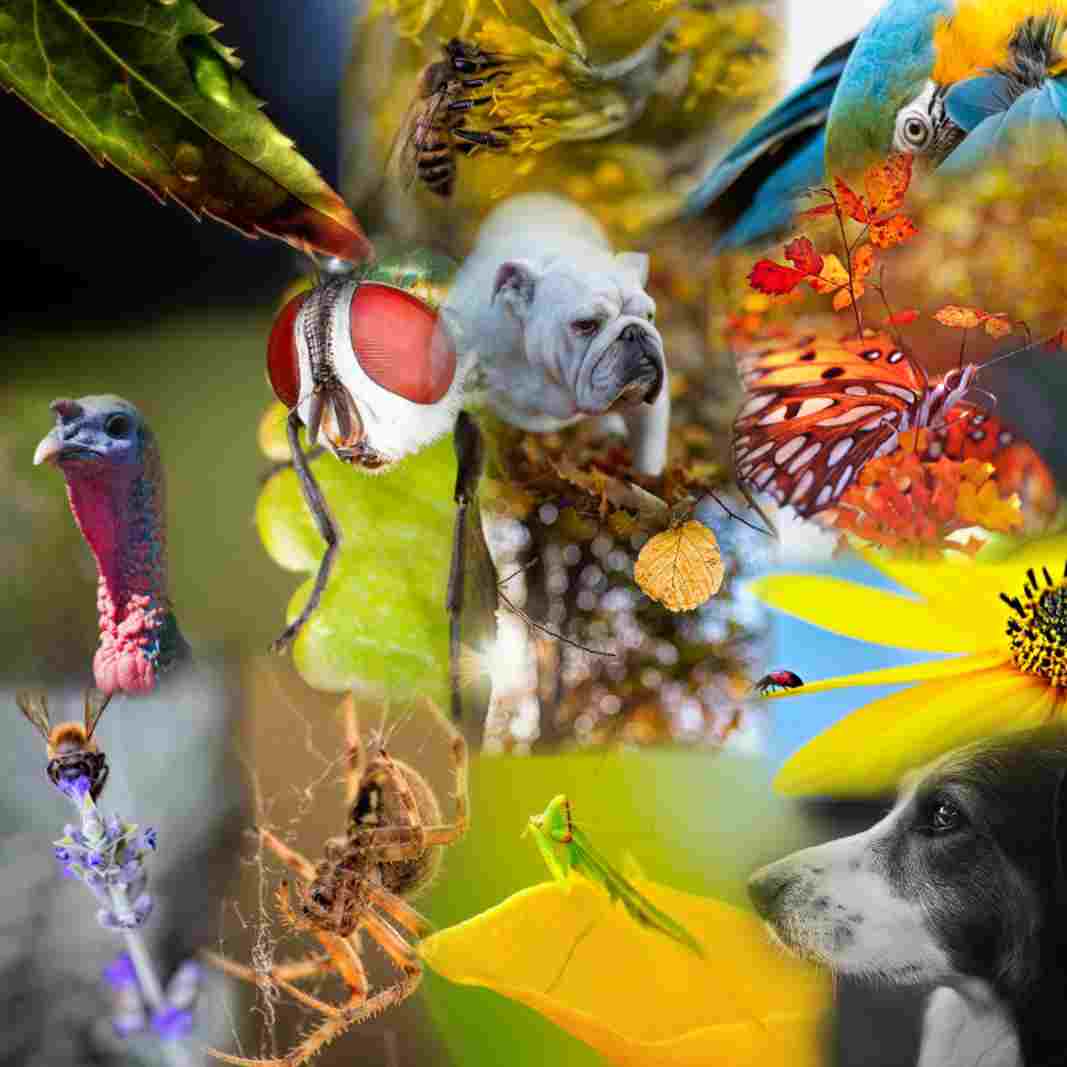} & 
\includegraphics[width=\collagewidthhhh]{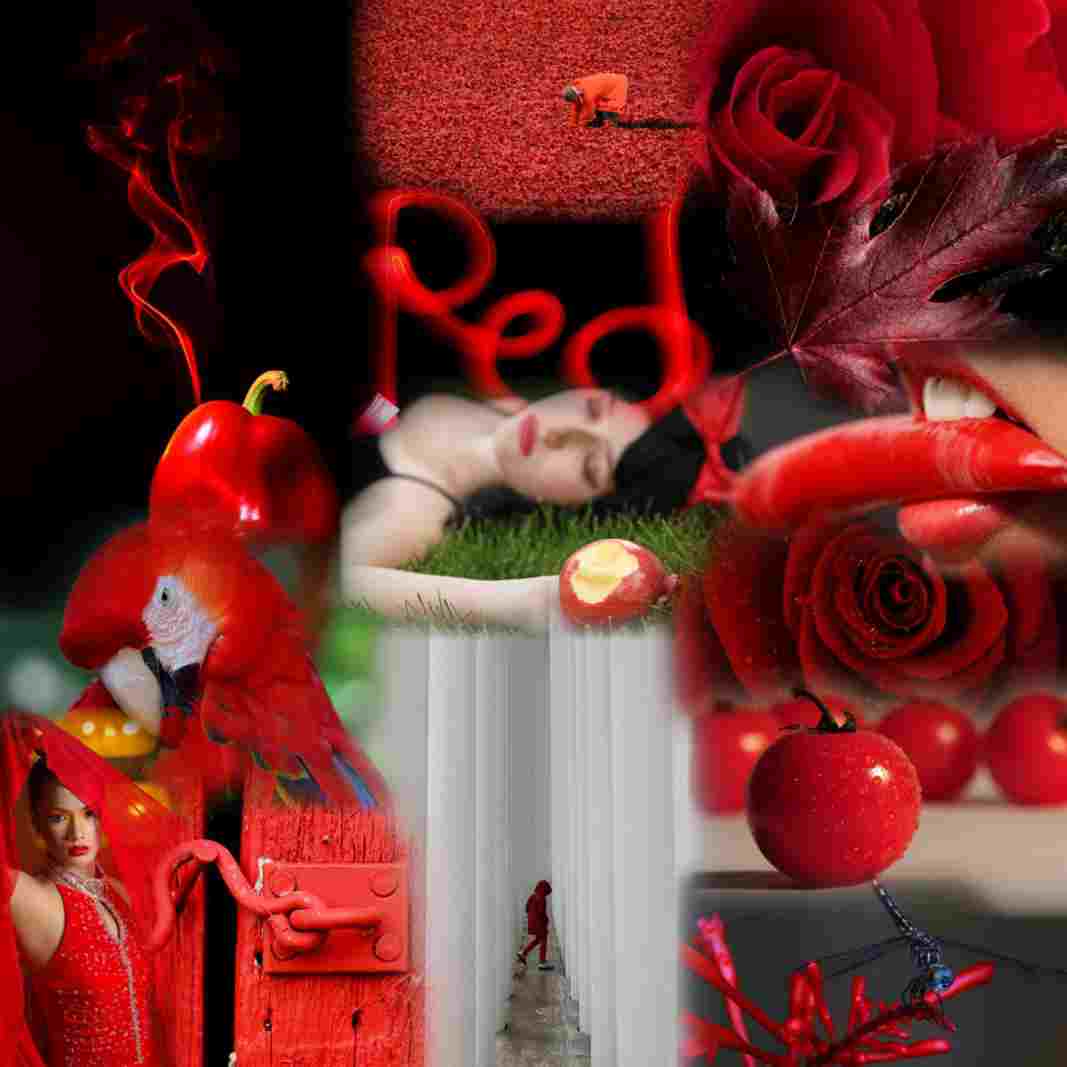} & 
\includegraphics[width=\collagewidthhhh]{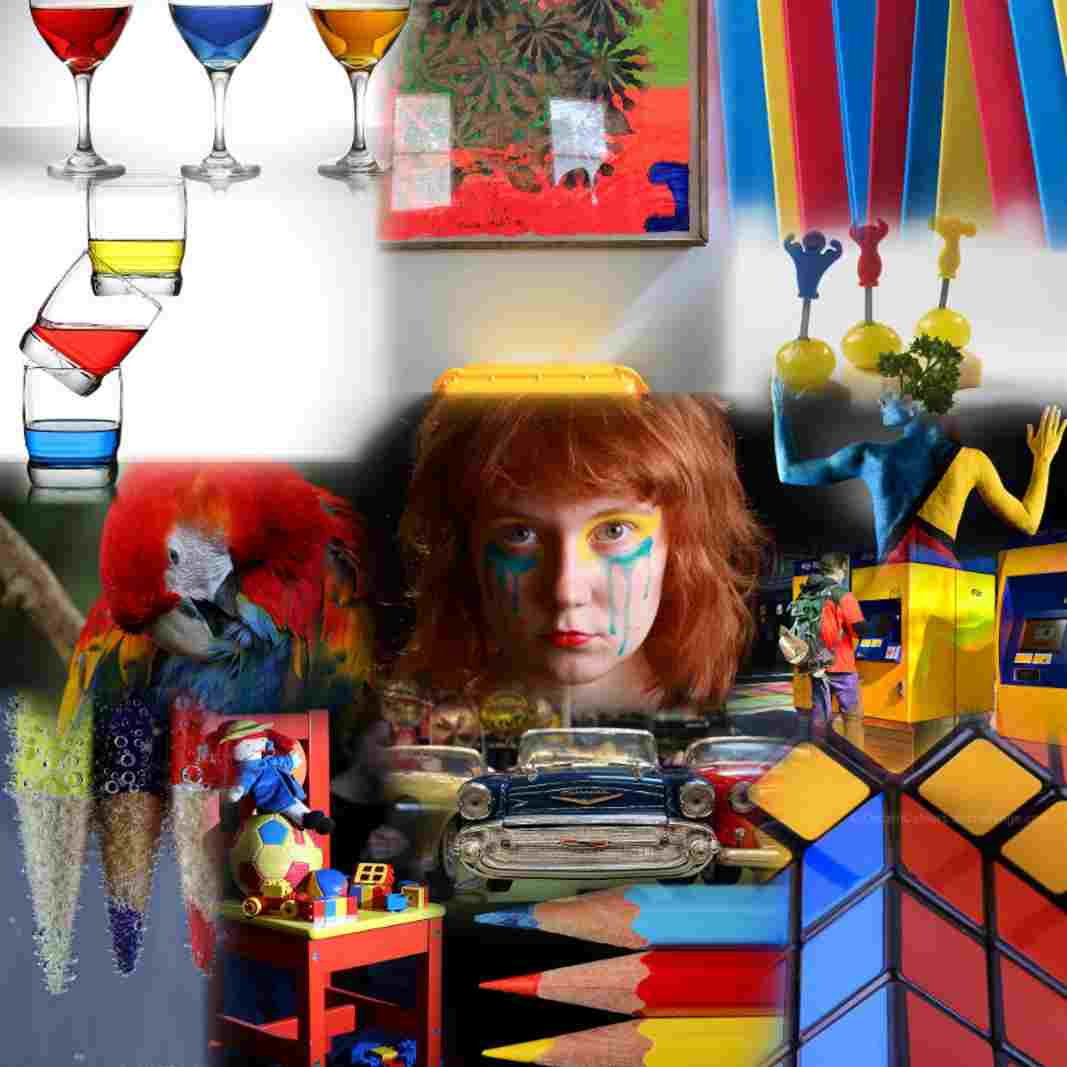} & 
\includegraphics[width=\collagewidthhhh]{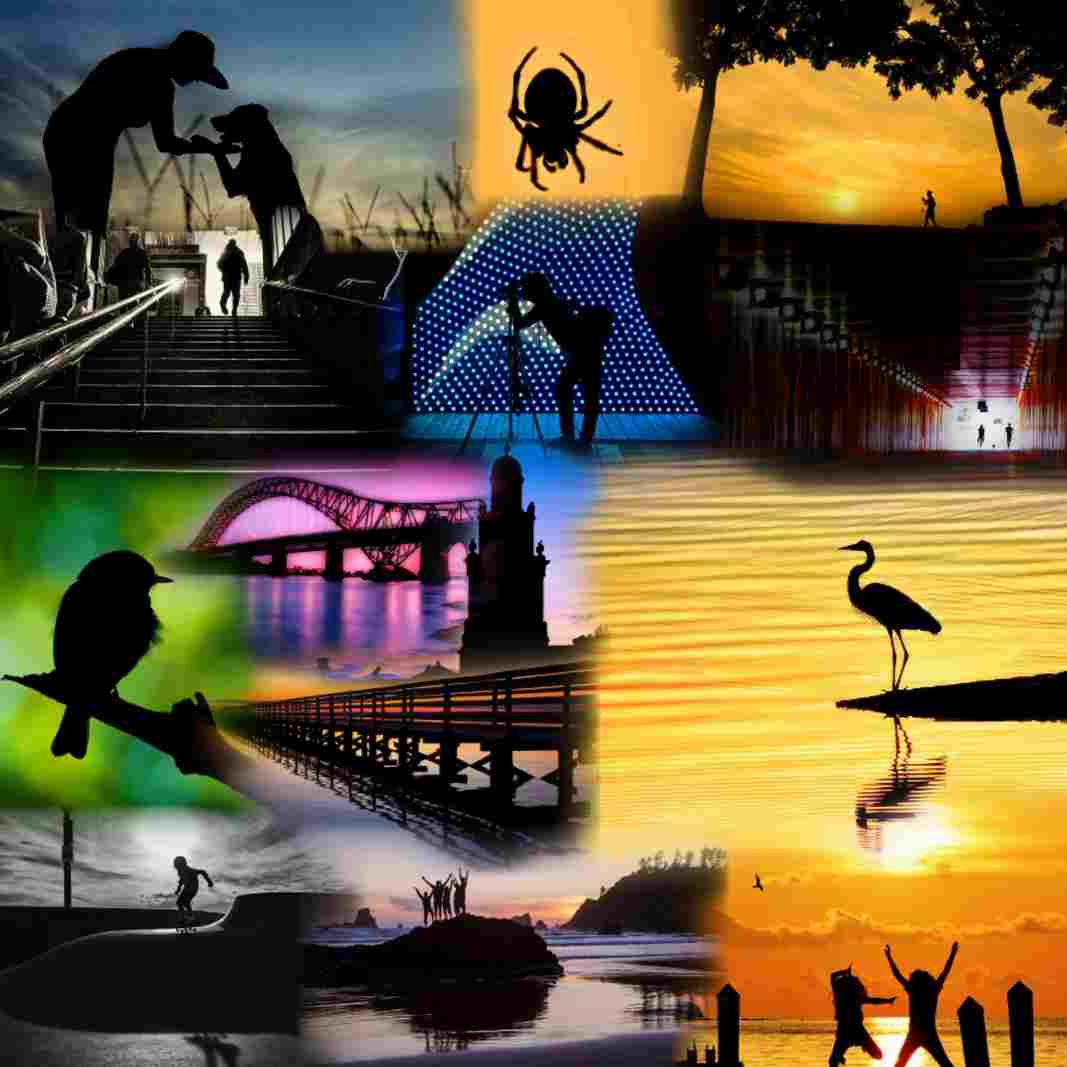} & 
\includegraphics[width=\collagewidthhhh]{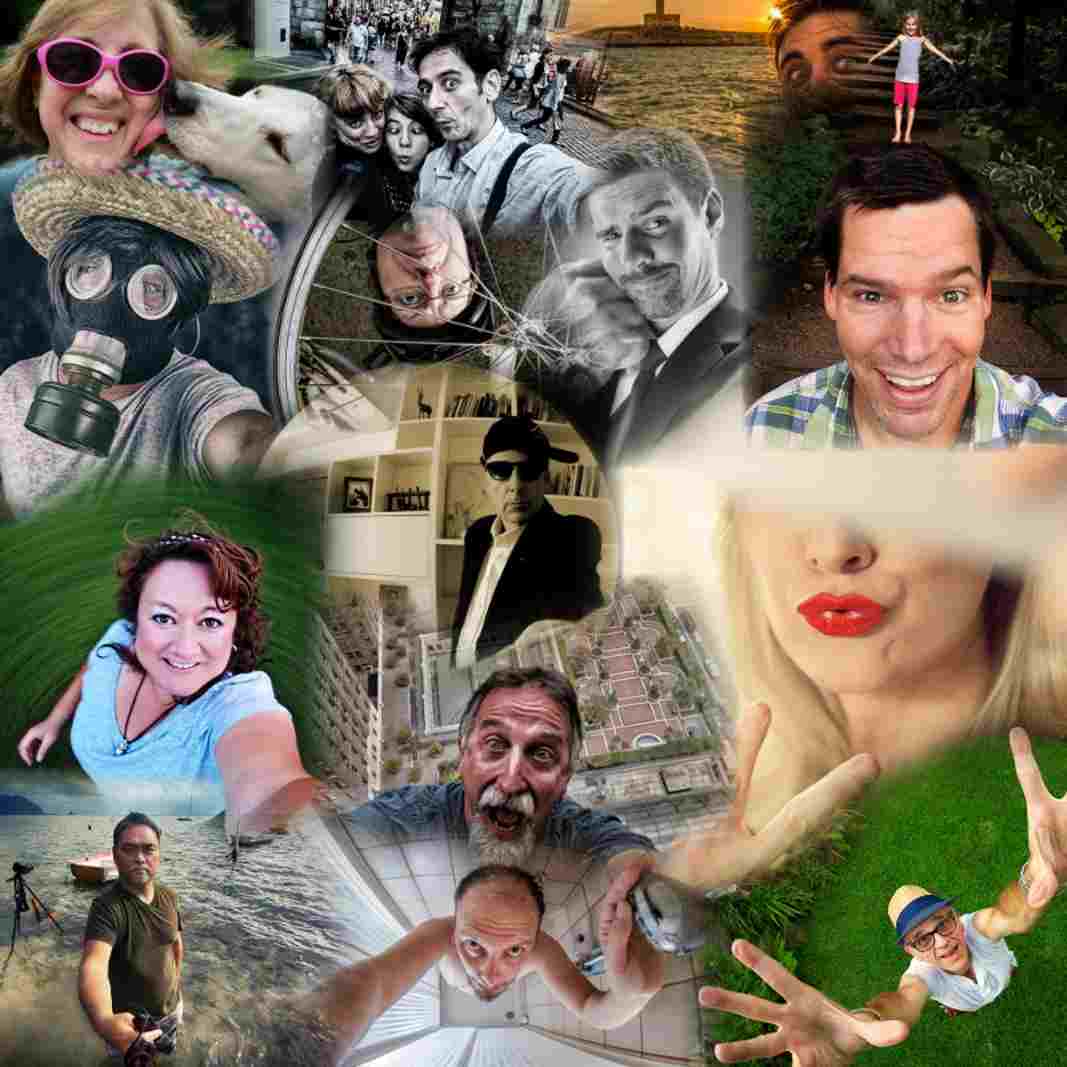} &
\includegraphics[width=\collagewidthhhh]{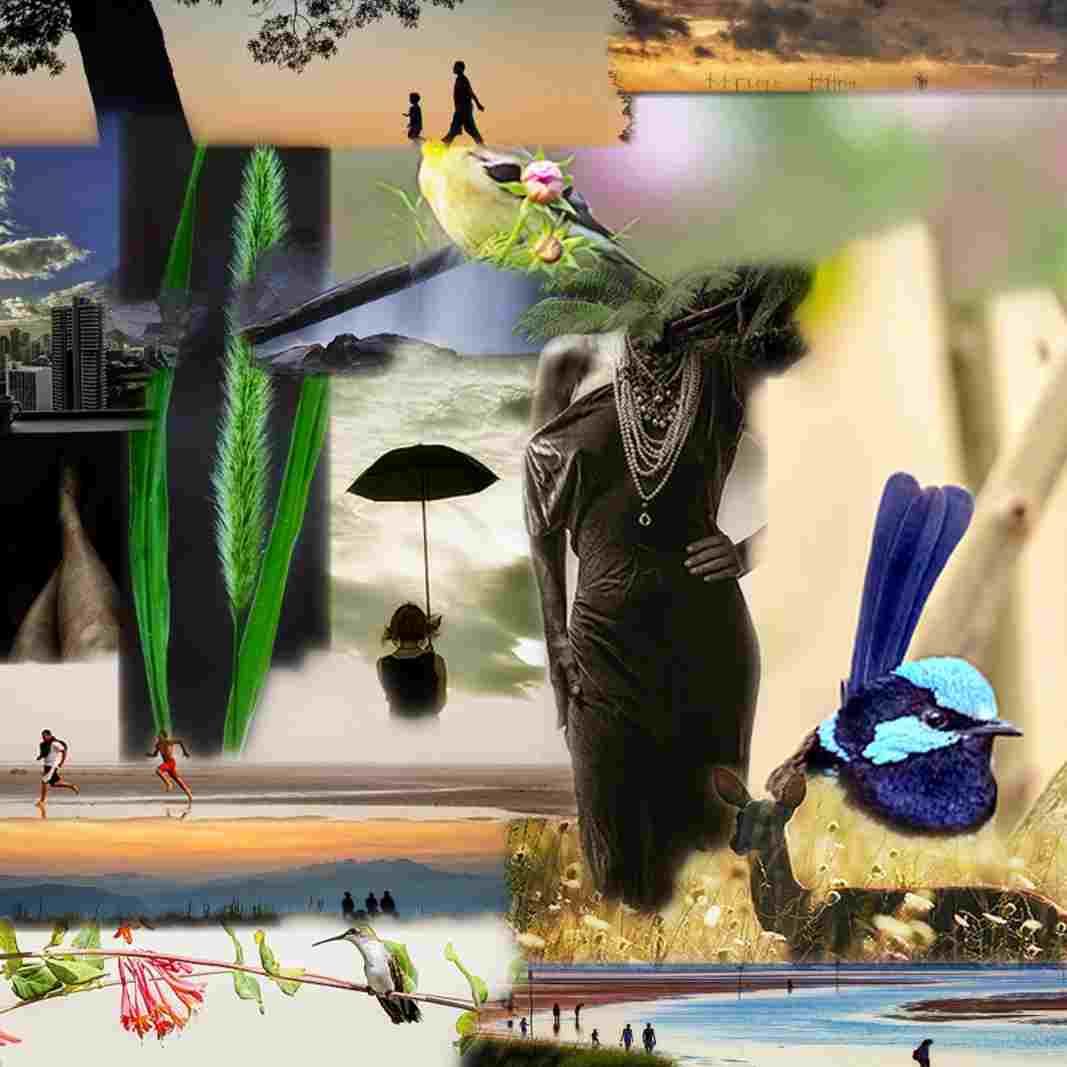} \\
\end{tabular}
\caption{\ADD{Comparison of our collage against Shape collage and Autocollage algorithms.}}
\label{fig:comparison}
\end{figure*}

\ADD{For this comparison, we used the six image datasets used in Section \ref{sec:generalization}. We set the parameters of the Autocollage and Shape Collage algorithms to generate collages of 14 images on a squared canvas and image to canvas ratio as similar as possible as in our set-up. The collages generated with the different methods are shown in Figure \ref{fig:comparison}.}

\ADD{The same subjects that participated in the previous experiment participated to this one. We asked them to choose among the three collages which they preferred. Results from this experiment showed that on average 56\% of the subjects preferred our photo collages; 42\% of the subjects preferred the Autocollage results and just 2\% preferred the Shape Collage results. In particular, in four datasets (Red, Primary, Selfie and 160 pixels) our collages were preferred by 65\% percent of the subjects. For the remaining two datasets (Shallow and Silhouettes) instead, 64\% of the subjects preferred the Autocollage results. This is due to the nature of the images used: in both categories the images contain a subject with an out-of-focus (Shallow) or almost uniform (Silhouettes) backgrounds. This makes easier for Autocollage to nicely blend image contents.}

\subsection{Varying collage sizes}
\label{sec:sizes}
\begin{figure*}[!tb]
\tabcolsep=2pt
\begin{center}
\small

\resizebox{\textwidth}{!}{
\setlength{\tabcolsep}{1pt}
\begin{tabular}{cccccc}
& 5 images & 10 images  & 14 images  & 25 images  & 50 images  \\
\begin{sideways} Our method\end{sideways} & 
\includegraphics[width=\collagewidthhh]{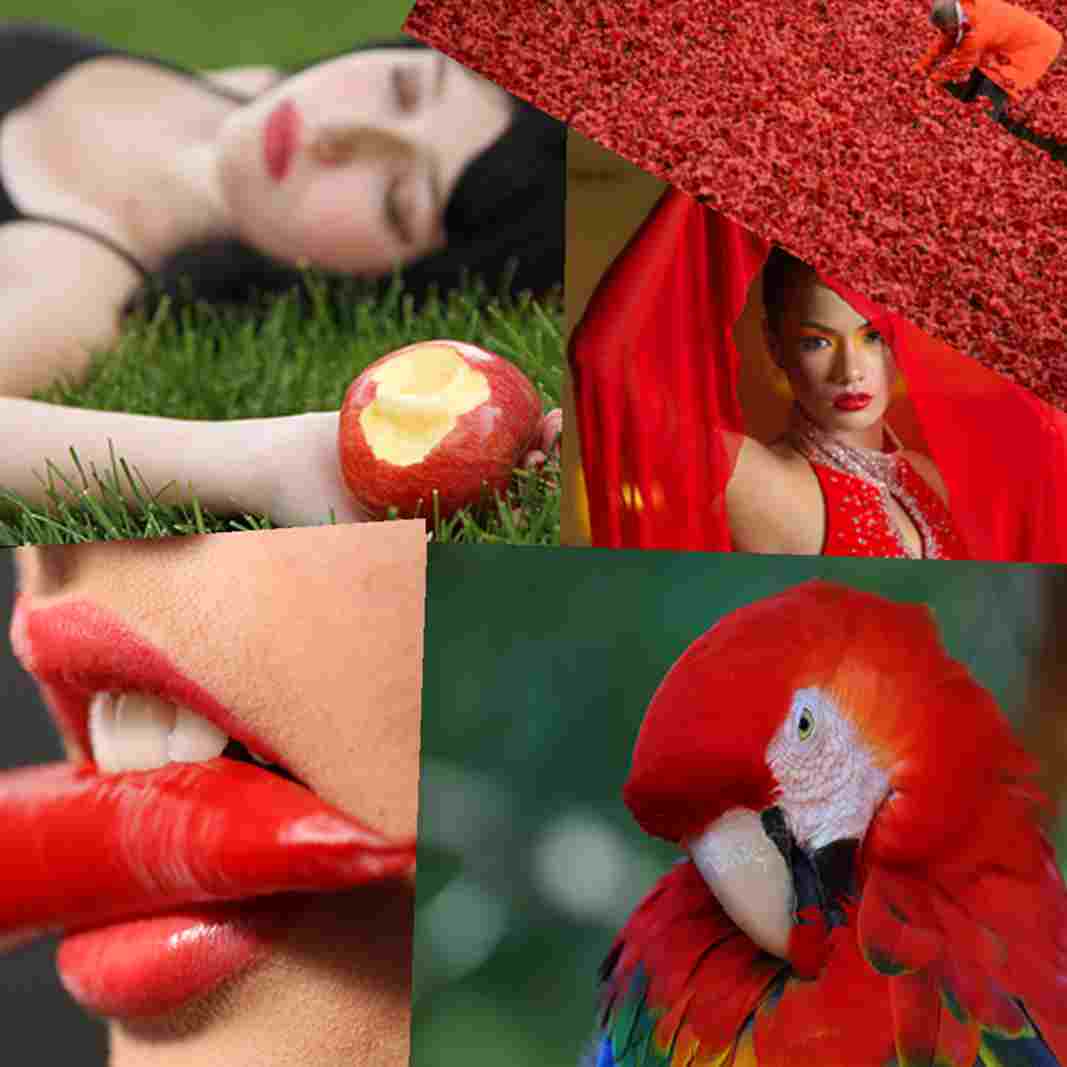} &
\includegraphics[width=\collagewidthhh]{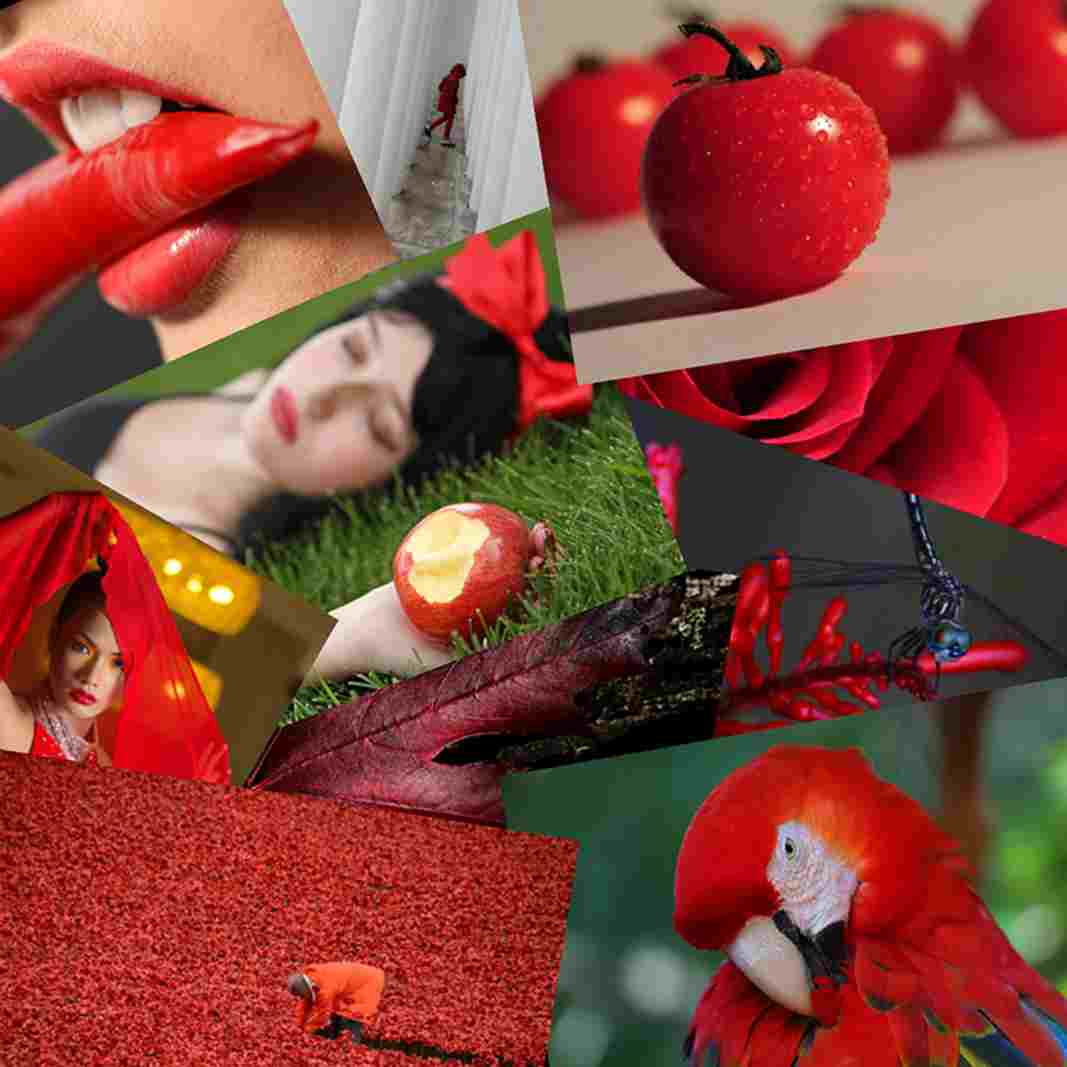} &
\includegraphics[width=\collagewidthhh]{collages3/Red_refinement_buchi3000_longer_4x.jpg} &
\includegraphics[width=\collagewidthhh]{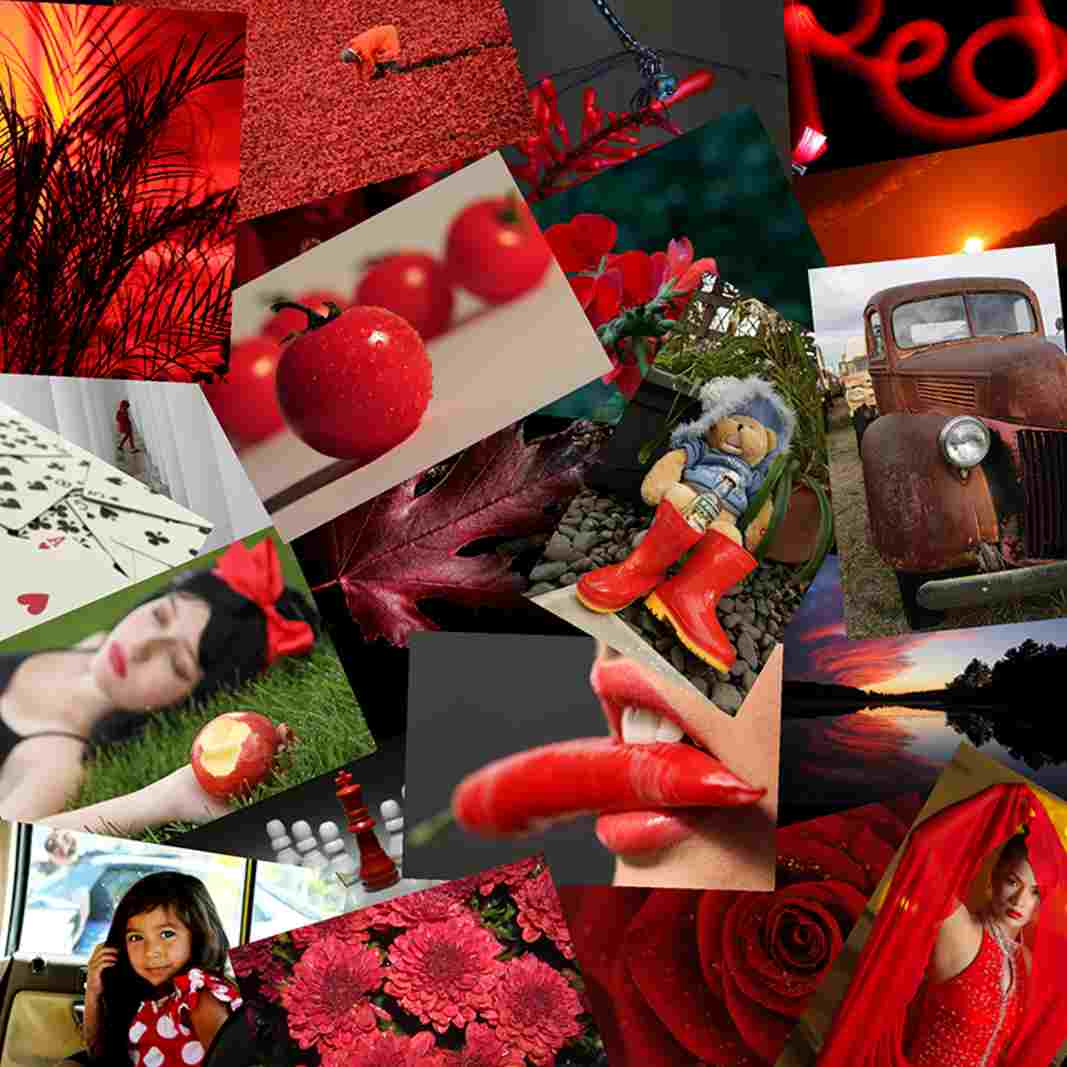} &
\includegraphics[width=\collagewidthhh]{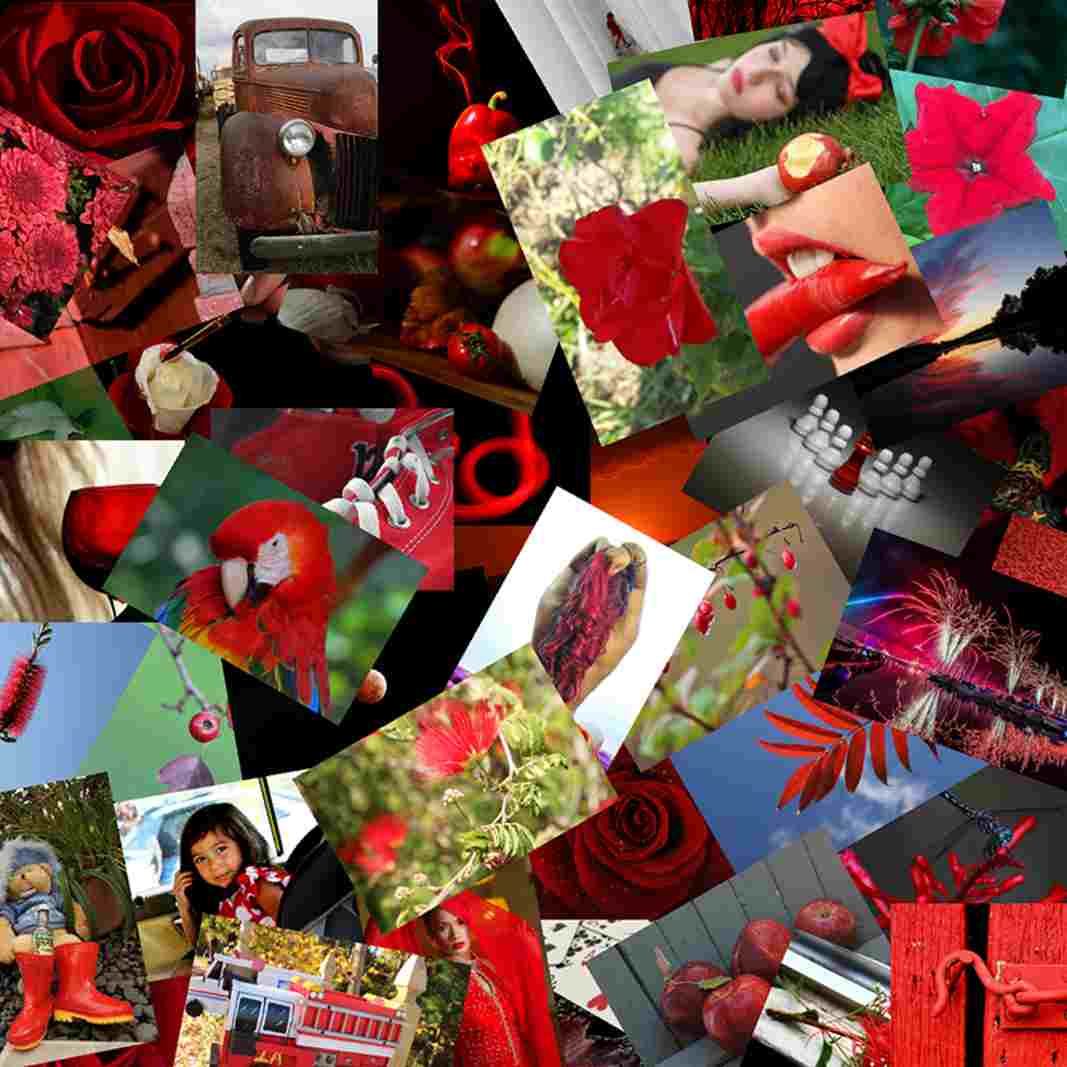} \\


\begin{sideways} Auto collage\end{sideways} & 
\frame{\includegraphics[width=\collagewidthhh]{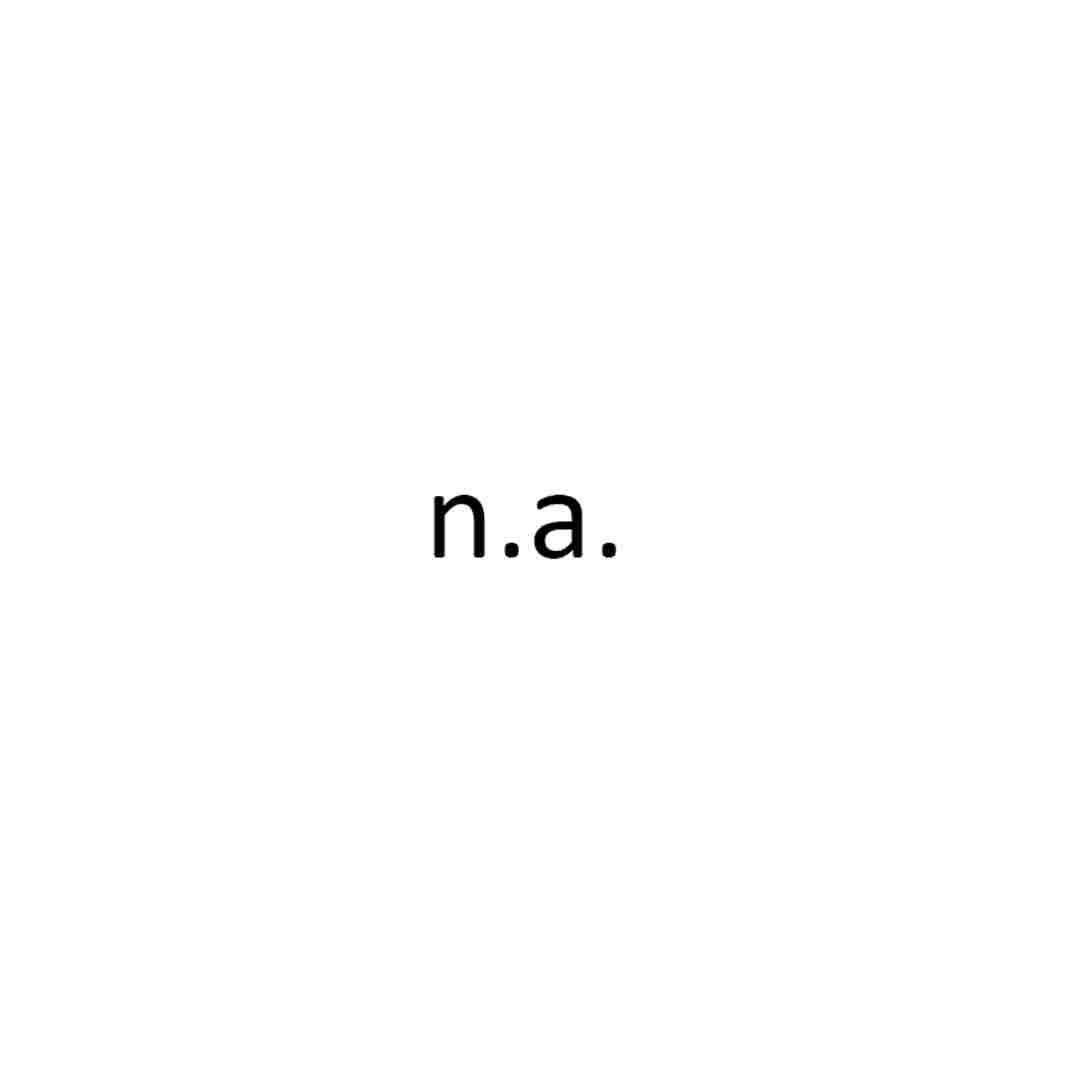}}&
\includegraphics[width=\collagewidthhh]{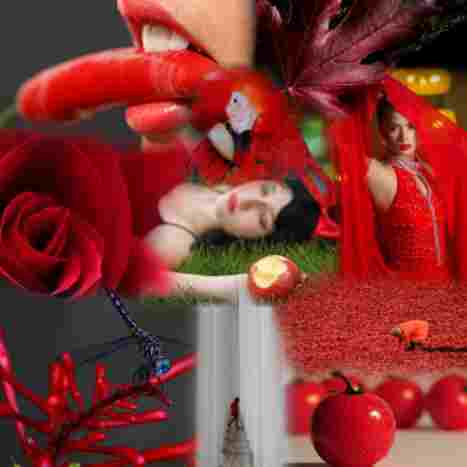} &
\includegraphics[width=\collagewidthhh]{comparison/red_autocollage_800x800.jpg} &
\includegraphics[width=\collagewidthhh]{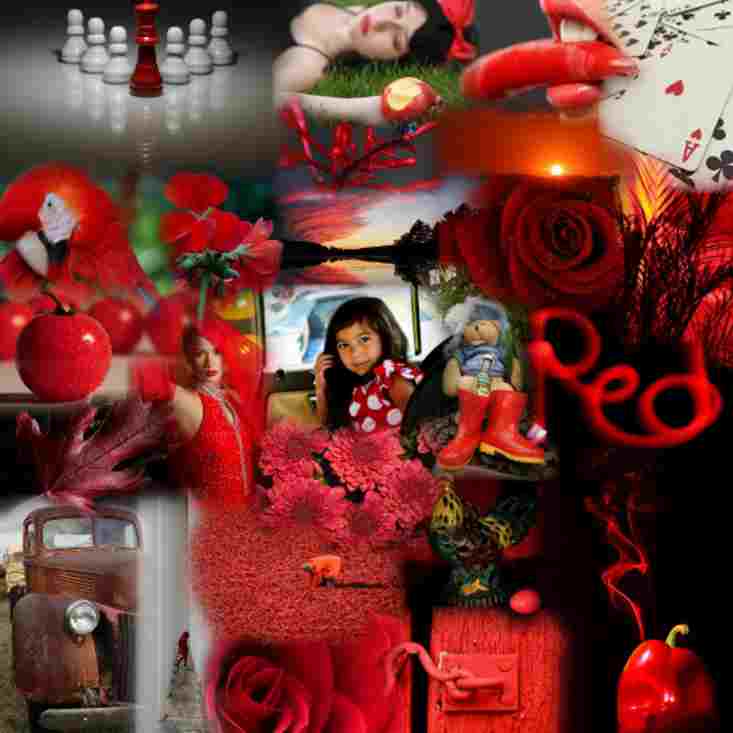} &
\includegraphics[width=\collagewidthhh]{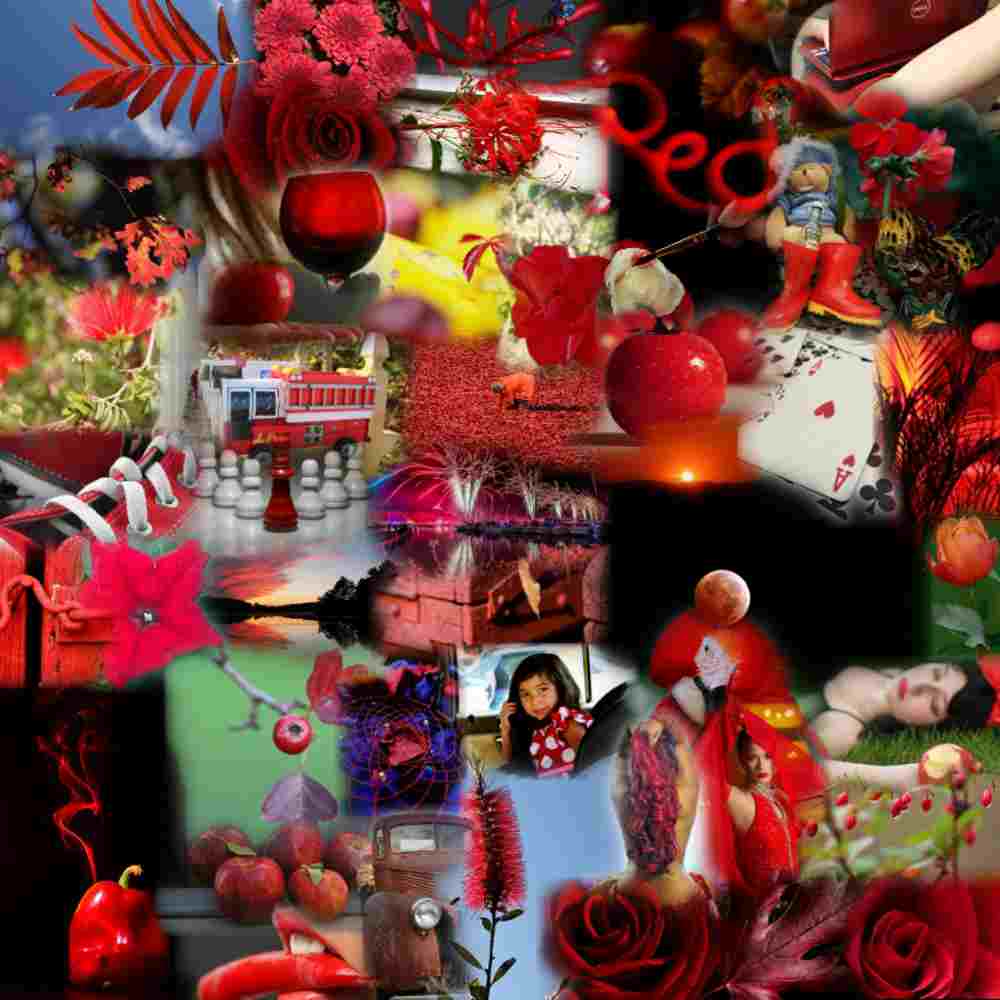} \\
\end{tabular}
}
\end{center}
\label{fig:mosaiciNimages}
\caption{\ADD{Generalization to datasets with different number of input images per collage (5, 10, 14, 25, 50) and different canvas sizes (250, 350, 550, and 750 pixels respectively). For each collage the canvas size is almost half of the area covered by all the images. The images have the same dimensions for visualization purposes.}}
\end{figure*}

\ADD{In this experiment we test how the learned definition of pleasantness generalizes to datasets with different number of input images per collage and different canvas sizes. Two smaller and two larger variants of the Red dataset have been considered, containing 5, 10, 25, and 50 images respectively. Canvas sizes have been chosen so that they are almost half of the total area covered by the images as in Section \ref{sec:generation}. Thus optimization has been performed on canvas having side length equal to 250, 350, 550, and 750. The results are reported in Figure \ref{fig:mosaiciNimages} and compared with the results obtained by Autocollage, which resulted in the best competing algorithm in the previous section. All the canvas have been resized to equal size for better visualization. The results of Autocollage in the case of 5 images is not available since the minimum number of images it can handle is 7. \ADD{The judgments of these collages have been performed by 20 subjects. We asked them to choose which collage among the two they prefer. After the test, the results that we collected are the following. For the three collage with few images, the majority of the subjects chose our collage with percentages of 100\%, 70\% and 65\% for the 5, 10 and 15 images respectively. In the case of 25 images, the difference between our collage and the Auto collage is reduced, with 55\% of the users choosing our collage and 45\% choosing the Auto collage. Finally, the gap between the two approaches further reduces in the case of 50 images to practically a tie (50\% of preferences). The interview with the users revealed that, when presented with the collages with 50 images, the limited canvas size made them paying less attention to the actual content of the images, while favoring the overall image distribution. In this case, the two collages were considered equally cluttered but the smooth carving of Auto collage made this collage more pleasing.} }

\section{Conclusion}
\label{sec:conclusion}

In this work we have considered the problem of creating pleasing photo collages by exploiting subjective experiments to model and learn user preferences. 
We designed an experimental framework for the identification of the criteria that need to be taken into account to generate a pleasing photo collage. 
Starting from collages created using state-of-the-art criteria, namely photo informativeness, canvas area coverage, and information ratio balance, we performed a subjective experiment involving several subjects on different thematic photo datasets. 
This experiment showed that different and more complex criteria are involved in the subjective definition of pleasantness. 
Inspired by the responses of the subjects, we have redefined the basic criteria and we have identified and implemented new global and local ones: face ratio, axis alignment, centrality, convexity, color similarity, orientation diversity and minimum orientation difference. 
The relative importance of all these criteria has been learned by exploiting user rankings. 
Moreover, with the proposed experimental framework we learned a composite photo informativeness description from saliency, quality and harmony.
A new set of collages has been generated using the identified criteria and evaluated in a pairwise comparison experiment against the previous best rated collages.
The new collages were preferred by the majority of the subjects for all the photo datasets considered, showing that the proposed framework is able to identify and combine the criteria at the basis of user preference, and to learn a computational model which effectively encodes an inter-user definition of pleasantness. A further experiment has been run, showing that the learned definition of pleasantness generalizes well to new thematic photo datasets not used in the training phase.

Photo informativeness has been described in terms of saliency, quality, and harmony maps, but other maps taking into account different image properties can be incorporated as well in our framework (e.g. photo memorability by Isola et al. \cite{isola2011makes,khosla2012memorability}). Furthermore, leveraging user preferences, the proposed framework permits to quantify the contribution of different  visual features to model new intrinsic properties of the images.  

\ADD{The proposed framework can benefit current collage generation algorithms in two different ways. The first regards its use to estimate the weights of the fitness function (also called energy function) in the different collage generation algorithms, e.g: weights associated to region importance, transition cost, object sensitivity and face presence in Autocollage \cite{Rother2006}; representativeness, compactness and transition smoothness in Video collage \cite{mei2009video}; salience visibility, salience ratio balance, penalty of severe occlusions, blank space presence, canvas shape constraint, spatially uniformity and orientation diversity in Picture collage \cite{Wang2006,liu2009picture}; image complexity and content distinctness in \cite{Yu2014}. \ADD{All these algorithms heuristically set the weights associated to the different terms in their fitness functions. With our framework, these weights can be systematically set using user preferences.} This way requires that a training data has to be generated in the form of multiple collages and the collection of user judgments about them. This operation has to be done only once and does not impact collage generation time. 
The second way in which existing algorithms can leverage our work regards the possibility of including the new criteria here defined inside their fitness/energy functions. This will not dramatically slow down the collage generation process, since the new criteria are fast to compute.}

As future work we plan to investigate if the learned definition of pleasantness changes when subjects and photos are linked. We plan also to expand the set of criteria by enlarging the number of subjects in the experiments, and by adding more thematic photo datasets.




\bibliographystyle{imsart-nameyear}
\bibliography{manuscript}


%
%
%
%
%
%

\end{document}